\documentclass[aip,reprint,jcp,amsmath,amssymb,showpacs,superscriptaddress]{revtex4-2}
\usepackage[utf8]{inputenc} %joris: for special characters
\usepackage{bm}
\usepackage{graphicx}
\usepackage{color}
\usepackage{appendix}
\usepackage{microtype}
\usepackage[breaklinks=true]{hyperref}
\usepackage{grffile} %joris: avoid "unknown graphics extension" error due to periods in filenames
\usepackage{lipsum}
\usepackage{booktabs} % joris: for tables

\newcommand{\beq}{\begin{equation}}
\newcommand{\eeq}{\end{equation}}
\definecolor{deepblue}{rgb}{0,0,0.5}
\definecolor{deepred}{rgb}{0.6,0,0}
\definecolor{snapyellow}{RGB}{204,188,41}
\definecolor{snapred}{RGB}{204,41,41}
\definecolor{apsblue}{rgb}{0.18,0.19,0.57}

\newcommand{\X}{\widetilde{X}}
\newcommand{\NiY}{Ni$_{33}$Y$_{67}$}
\newcommand{\CuZr}{Cu$_{64}$Zr$_{36}$}

\definecolor{orange}{rgb}{1,0.5,0}

\definecolor{mygreen}{rgb}{0.0,0.55,0.3}

\newcommand{\bvec}[1]{{\bm{#1}}}



\hypersetup{
colorlinks,%
citecolor=deepblue,%
filecolor=black,%
linkcolor=deepblue,%
urlcolor=deepblue
}

\graphicspath{{./sprouts/plots/paper/}:{./}}

\begin{document}

\title{Dimensionality reduction of local structure in glassy binary mixtures}

\author{Daniele Coslovich}
\email{dcoslovich@units.it}
\affiliation{Dipartimento di Fisica, Universit\`a di Trieste, Strada Costiera 11, 34151, Trieste, Italy}

\author{Robert L. Jack}
\affiliation{Yusuf Hamied Department of Chemistry, University of Cambridge, Lensfield Road, Cambridge CB2 1EW, United Kingdom}
\affiliation{Department of Applied Mathematics and Theoretical Physics, University of Cambridge, Wilberforce Road, Cambridge CB3 0WA, United Kingdom}

\author{Joris Paret}
\affiliation{Laboratoire Charles Coulomb, Universit\'e de Montpellier, Montpellier, France}

\date{\today}

\begin{abstract}
We consider unsupervised learning methods for characterizing the disordered microscopic structure of supercooled liquids and glasses. Specifically, we perform dimensionality reduction of smooth structural descriptors that describe radial and bond-orientational correlations, and assess the ability of the method to grasp the essential structural features of glassy binary mixtures. In several cases, a few collective variables account for the bulk of the structural fluctuations within the first coordination shell and also display a clear connection with the fluctuations of particle mobility. Fine-grained descriptors that characterize the radial dependence of bond-orientational order better capture the structural fluctuations relevant for particle mobility, but are also more difficult to parametrize and to interpret. We also find that principal component analysis of bond-orientational order parameters provides identical results to neural network autoencoders, while having the advantage of being easily interpretable.  Overall, our results indicate that glassy binary mixtures have a broad spectrum of structural features. In the temperature range we investigate, some mixtures display well-defined locally favored structures, which are reflected in bimodal distributions of the structural variables identified by dimensionality reduction.
\end{abstract}

\maketitle

\section{Introduction}

Short-range order in liquids and glasses is evidenced by a sharp peak in the radial distribution function, which defines the first coordination shell.
However, the spatial arrangement of atoms and molecules within this coordination shell is more difficult to characterize~\cite{royallRoleLocalStructure2015,Tanaka_Tong_Shi_Russo_2019}.
Conventional approaches to describe local order beyond the two-body level include higher order correlation functions~\cite{Coslovich_2013}, bond-orientational order parameters~\cite{steinhardt_bond-orientational_1983} and more general structural descriptors based on geometrical~\cite{Tanemura_Hiwatari_Matsuda_Ogawa_Ogita_Ueda_1977,gellatly_characterisation_1982,Honeycutt_Andersen_1987} or topological constructions~\cite{malinsIdentificationStructureCondensed2013,Lazar_Han_Srolovitz_2015}.
Over the last decades, local structure analysis provided evidence that some glassy colloidal suspensions and multi-component alloys display non-trivial local arrangements, known as locally favored structures (LFS)~\cite{royallRoleLocalStructure2015}.
These structures tend to be more symmetric and stable than the bulk, and also correlate to some extent with the local fluctuations of particle mobility~\cite{coslovichUnderstandingFragilitySupercooled2007a,malinsIdentificationLonglivedClusters2013,malinsLifetimesLengthscalesStructural2014,hocky_correlation_2014}, \textit{i.e.}, dynamic heterogeneity.
At present, however, there is no robust and generally accepted operational definition of a system's locally favored structure~\cite{Arceri2020}.

Unsupervised learning methods have recently emerged as a promising alternative to characterize materials' local structure~\cite{Mehta_Bukov_Wang_Day_Richardson_Fisher_Schwab_2019,Cheng_et_al_2020}.
Starting from a high-dimensional descriptor of the particles' arrangements, one searches for patterns and regularities in the statistics of the descriptor.
Unsupervised methods typically comprise two steps: (i) dimensionality reduction, to project the high-dimensional descriptor on a smaller subspace while retaining most of the original information, 
and (ii) clustering, to identify groups of points in the dataset that share similar values of the (reduced) descriptor.
This approach has been recently applied, for instance, to crystal structure identification in colloidal suspensions~\cite{boattini_unsupervised_2019,van_Damme_Coli_van_Roij_Dijkstra_2020} and the study of partially ordered systems~\cite{Cheng_et_al_2020, Becker_Devijver_Molinier_Jakse_2021}.
Recent studies have also tackled challenging problems of structural analysis in bulk disordered materials, such as simple models of supercooled liquids and glasses~\cite{boattini_autonomously_2020, paret_assessing_2020}, amorphous carbon~\cite{Deringer_Caro_2018,Cheng_et_al_2020}, and liquid water in normal and supercooled conditions~\cite{Monserrat_Brandenburg_Engel_Cheng_2020,Offei-Danso_Hassanali_Rodriguez_2022}.
However, some of these extensions employ high-dimensional structural descriptors, which are not always easy to interpret, and involve non-linear reduction methods, which mostly act as black boxes.

In this work, we focus on the first step of unsupervised learning, \textit{i.e.}, dimensionality reduction, and explore its role as a heuristic tool for structural analysis of binary glassy mixtures.
Our analysis is based on smooth structural descriptors that account for both density and bond-order fluctuations.
We apply principal component analysis (PCA), which is possibly the simplest linear dimensionality reduction method, to several descriptors and find evidence of a varying degree of structural heterogeneity across glassy mixtures.
The first few principal components, which capture the largest variance of the original descriptor's distribution, are closely connected to physically motivated measures of local order, as well as to dynamic heterogeneities.
Finally, we show that a neural network autoencoder (AE), which is a more complex dimensionality reduction method, provides nearly identical results to PCA.
Since linear methods have the advantage of being easily interpretable, they should be preferred when studying simple glassy systems.

The structure of the paper is as follows: Section~\ref{sec:methods} introduces our models and methods, including five different model systems, and several different descriptors that are used to characterize the local structure. Sections~\ref{sec:results}, ~\ref{sec:dynamics}, and~\ref{sec:pca_vs_ae} present our key findings. Section~\ref{sec:conc} gives a critical discussion and an outlook on the methodology.

\section{Methods}
\label{sec:methods}

We consider a pool of five computational models of binary glass-forming liquids, characterized by different types and degrees of local order: two canonical computer models, the Kob-Andersen~\cite{kob_scaling_1994} (KA) mixture and the Wahnström~\cite{wahnstrom_molecular-dynamics_1991} (Wahn) mixture, two Lennard-Jones models that mimic the structure of amorphous \NiY~\cite{dellavalle_microstructural_1994} and SiO$_2$~\cite{coslovich_dynamics_2009}, respectively, as well as a realistic embedded-atom model of \CuZr~\cite{Cheng_Sheng_Ma_2008}. We analyze statistically uncorrelated configurations, separated by at least one structural relaxation time $\tau_\alpha$, obtained from classical molecular dynamics simulations. As usual, $\tau_\alpha$ is defined by the condition $F_s(k^*,\tau)=1/e$, where $F_s(k^*,t)$ is the self intermediate scattering of the system at a wave-vector $k^*$ corresponding to the first peak of the structure factor. To provide a consistent comparison, all the systems are studied under equilibrium conditions at temperatures close to their respective mode-coupling theory (MCT) crossover temperatures, empirically identified by a power law fit of the structural relaxation time data~\cite{Berthier_Biroli_Coslovich_Kob_Toninelli_2012}.
Around the MCT crossover, the structural relaxation times are typically about 3 orders of magnitude larger than at the onset of slow dynamics.
Full details about the models are given in the Appendix~\ref{app:models}.

For each of these models, we analyze several structural descriptors, ranging from simple bond-order (BO) parameters and their smooth variants, to higher-dimensional descriptors that account for both orientational and radial correlations. To reduce the dimensionality of these descriptors, we use both linear (PCA) and non-linear (AE) methods. In this section, we present a synthesis of these methods and ideas. Readers already familiar with characterisations of local structure and dimensionality reduction may want to skip directly to Sec.~\ref{sec:results}.

\subsection{Structural descriptors}
\label{sec:descriptors}

\subsubsection{Bond order}\label{sec:bop}

Bond-order parameters are
standard measures of structure in the first coordination shell.%, we discuss bond order (BO) parameters.
~Let $\bvec{r}_i$ be the position of particle $i$ and define $\bvec{r}_{ij} = \bvec{r}_j - \bvec{r}_i$ and $r_{ij} = |\bvec{r}_{ij}|$.  Then consider the weighted microscopic density around particle $i$:
\beq
\rho(\bvec{r}; i) = \sum_{j=1}^{N_b(i)} w_j \delta(\bvec{r} - \bvec{r}_{ij})
\label{eq:density_micro}
\eeq
where $w_j$ is a particle-dependent weight and the sum involves a set of $N_b(i)$ particles, which defines the coordination shell of interest for particle $i$.

We project the microscopic density on a unit-radius sphere, that is, $\hat{\rho}(\hat{\bvec{r}}; i) = \sum_{j=1}^{N_b(i)} w_j \delta(\bvec{r} - \hat{\bvec{r}}_{ij})$,
where $\hat{\bvec{r}} = \bvec{r} / |\bvec{r}|$ and similarly $\hat{\bvec{r}}_{ij} = \bvec{r}_{ij}/|\bvec{r}_{ij}|$. % (\bvec{r}_j - \bvec{r}_i) / |\bvec{r}_j - \bvec{r}_i|$.
Expanding in
spherical harmonics yields
\beq
\hat{\rho}(\hat{\bvec{r}}; i) = \sum_{l=0}^\infty \sum_{m=-l}^l c_{l m}(i) Y_{l m}(\hat{\bvec{r}}) ,
\eeq
with coefficients
\beq
c_{l m}(i) =  \int d\bvec{r} \rho(\bvec{r}; i) Y_{l m}(\hat{\bvec{r}}) .
\eeq

In the conventional bond-order analysis, one sets the weights to unity and considers the normalized complex coefficients,
\begin{align}\label{eqn:bo}
q_{lm}(i) & = \frac{1}{N_b(i)} \int d\bvec{r} \rho(\bvec{r}; i) Y_{l m}(\hat{\bvec{r}}) 
\nonumber \\ & = \frac{1}{N_b(i)} \sum_{j=1}^{N_b(i)} Y_{l m}(\hat{\bvec{r}}_{ij}) .
\end{align}
The rotational invariants,
\beq\label{eqn:bo_inv}
Q_{l}(i) = \left( \frac{4\pi}{2l + 1}\sum_{m=-l}^l |q_{lm}(i)|^2 \right)^{1/2} ,
\eeq
provide a detailed structural description of the local environment around particle $i$.
By truncating the expansion to order $l_\mathrm{max}$, we obtain the simplest BO descriptor of an arbitrary particle $i$,
\beq
X^\textrm{BO}(i)
= (Q_0(i), \dots, Q_{l_\mathrm{max}}(i)) .
\label{equ:xBO}
\eeq
In the machine learning context, such a sequence is usually referred to as a ``feature vector''.

We note that the complex coefficients can be averaged over nearest neighbors, as suggested by Lechner and Dellago~\cite{lechner_accurate_2008},
% $$
% \bar{q}_{lm}(i) \equiv \langle q_{l m}(i) \rangle ,
% $$
% and then made invariant,
% $$
% \bar{Q}_{l}(i) \equiv \left( \frac{4\pi}{2l + 1}\sum_{m=-l}^l |\bar{q}_{lm(i)}|^2 \right)^{1/2} ,
% $$
to provide an improved descriptor for crystal structure detection.
We have tested this approach, but we do not use it in this work because we found that the additional average tends to smear the differences between disordered structural environments in the systems of our interest.

\subsubsection{Neighbor definition and smoothed BO parameters}\label{sec:neigbbors}

In the following, we mostly focus on the structural heterogeneity (or diversity of local arrangements) within the first coordination shell.
It is therefore important to discuss how this coordination shell is defined.

The most common approach is to define the neighbors on the basis of a fixed cutoff distance $r_{\alpha\beta}^c$, where $\alpha$ and $\beta$ are species indices. The cutoff distance is equal to the first minimum of the corresponding partial radial distribution functions $g_{\alpha\beta}(r)$.
Alternative definitions include setting the neighbors on the basis of a (radical) Voronoi tessellation~\cite{gellatly_characterisation_1982}, or using the solid angle nearest neighbor (SANN) approach~\cite{van_meel_parameter-free_2012}, which is parameter-free.

Irrespective of this choice, the coefficients $Q_l$ will change discontinuously whenever a particle leaves the coordination shell.
The fact that the descriptor's components are not smooth functions of particles' coordinates is a concern~\cite{Mickel_Kapfer_Schroder-Turk_Mecke_2013}.
The issue becomes particularly serious at low temperature, when thermal fluctuations are small and discontinuities may affect the distribution of the BO parameters.
An obvious approach to counter this issue is to use a smeared local density, in which the Dirac deltas are replaced by Gaussians~\cite{bartok_on-representing_2013,Behler_Parrinello_2007}, see Sec.~\ref{sec:soap}.
Alternatively, or even in addition to Gaussian smearing, one can consider a smooth version of the descriptor, in which the coefficients $q_{l m}$ are multiplied by a weighting function $f(r)$ that depends on the radial distance $r$ between the central particle and its neighbors,
\beq\label{eqn:sbo_coeff}
q_{lm}^{S}(i) = \frac{1}{Z(i)} \sum_{j=1}^{N} f({r}_{ij}) Y_{lm}(\hat{\bvec{r}}_{ij}) ,
\eeq
where $Z(i)=\sum_{j=1}^{N} f({r}_{ij})$ is a normalization constant and the superscript $S$ indicates the smooth nature of the descriptor.
In the following we use
\beq
f(r_{ij}) = \exp \left[- (r_{ij} / r_{\alpha\beta}^c)^\gamma \right] H(R_{\alpha\beta}^c - r_{ij}) ,
\label{eq:fr_sbo}
\eeq
where $r_{\alpha\beta}^c$ is the first minimum of the corresponding partial radial distribution function for the pair $(i,j)$ and $\gamma$ is an integer whose value is given in Sec.~\ref{sec:results}.
Also, $H$ is the Heaviside step function, which ensures, for efficiency reasons, that the descriptor only has contributions from particles within  a distance $R_{\alpha\beta}^c = 1.3 \times r_{\alpha\beta}^c$ from the central one.

The rotational invariants are defined similarly to Eq.~\eqref{eqn:bo_inv} and the corresponding ``smooth'' bond-order (SBO) descriptor of an arbitrary particle $i$ is given by 
\beq
X^\mathrm{SBO}(i) = (Q_0^S(i), \dots, Q_{l_\mathrm{max}}^S(i)) .
\label{equ:xSBO}
\eeq
We note that in network-forming systems, such as amorphous silica, the first coordination shell is sharply defined at low temperature.
In these systems, the breakage of a bond leads to a ``genuine'' discontinuity in the BO parameters.
This suggests to use a smoothing function that decays rapidly around $r_{\alpha\beta}^c$, \textit{i.e.}, a high value of $\gamma$, to avoid a too strong smearing of the local environment.
Moreover, in mixtures with strong chemical order due to covalent bonding, it is appropriate to compute separate descriptors depending on the species of the neighboring particles~\cite{Cheng_et_al_2020}.
For the silica model introduced in Sec.~\ref{sec:silica}, we therefore restrict the sum in Eq.~\eqref{eqn:sbo_coeff} to particles whose species is distinct from the one of the central particle $i$.

\subsubsection{Radial dependence of bond order: SOAP}\label{sec:soap}

By projecting the local density on a unit-radius sphere, the conventional BO order analysis treats the first coordination shell as a whole, and ignores any radial dependence.
Modern machine learning descriptors, used for instance to fit potential energy surfaces, provide instead a systematic expansion of the local density~\cite{Parsaeifard_De_Christensen_Faber_Kocer_De_Behler_von_Lilienfeld_Goedecker_2021}.
Here, we will focus on the smooth overlap of atomic positions (SOAP) descriptor~\cite{bartok_on-representing_2013}, which complements the spherical harmonics with a radial basis $g_n(r)$.
One can thus analyze the BO at different length scales, as well as correlations between BO orders at different distances~\cite{bartok_on-representing_2013}.
We will use the implementation of the SOAP descriptor provided by the \texttt{DScribe} package~\cite{dscribe}, whose documentation provides a wealth of related information.

Within the SOAP descriptor, the microscopic density is smeared with Gaussians of width $\sigma$,
\beq
\label{equ:soap-rho}
\rho(\bvec{r}; i) = \sum_{j=1}^{N_b(i)} w_j \exp{\left(-\frac{|\bvec{r} - \bvec{r}_{ij}|^2}{2\sigma^2}\right)} .
\eeq
and $w_j$ is set to 1.
Assuming that the radial basis functions $g_n(r)$ are orthonormal, the expansion reads
\beq
\rho(\bvec{r}; i) = \sum_{n=1}^{n_\textrm{max}} \sum_{l=0}^{l_\textrm{max}} \sum_{m=-l}^{-l} c_{nlm}(i)\, g_n(r) Y_{lm}(\hat{\bvec{r}}) ,
\eeq
with
\beq
c_{nlm}(i) = %\langle \rho | g_n Y_{l m}\rangle =
 \int d\bvec{r} \rho(\bvec{r}; i) g_n(r) Y_{l m}(\hat{\bvec{r}}) .
\eeq
Notice that, contrary to the conventional BO descriptor --see Eq.~\eqref{eqn:bo}--, the coefficients $c_{nlm}(i)$ are \textit{not} normalized by the number of neighbors.
This means that the descriptor explicitly accounts for the ``coordination number'' $N_b(i)$ around a particle, \textit{i.e.}, how dense is the local coordination shell.
We will further discuss this point in Sec.~\ref{sec:pc_vs_ql}.

The basic SOAP descriptor is then defined by the power spectrum $p_{nl}(i) = \sum_{m=-l}^{l} c_{nl m}^*(i) c_{nl m}(i)$~\cite{bartok_on-representing_2013}. However, this quantity is not very sensitive to angular correlations between particles at different distances, so that a faithful description of the local particle environment requires a more general descriptor that includes such correlations explicitly~\cite{bartok_on-representing_2013}.  A suitable choice is given in Ref.~\onlinecite{De_Bartok_Csanyi_Ceriotti_2016},
\beq
Q_{n n^\prime l}(i) = \left(\frac{8\pi^2}{2l + 1}\right)^{1/2} \sum_{m=-l}^{l} c_{nl m}^*(i) c_{n^\prime l m}(i) .
\eeq
This descriptor retains a lot more information than the single power spectrum $p_{ln}(i)$, which only includes the diagonal terms in the radial basis expansion, but is of course also computationally more expensive.

To sum up, the full SOAP descriptor of an arbitrary particle $i$ is defined by the following feature tensor
\beq
(\: \dots, Q_{nn'l}(i), \dots \:) ,
\label{eq:soap_tensor}
\eeq
with $0\leq l\leq l_\textrm{max}$, $1 \leq n\leq n_\textrm{max}$ and $n'\geq n$, \textit{i.e.}, dropping the terms that are identical by symmetry.
In the following, we will flatten this tensor as a vector to form the descriptor $X^\mathrm{SOAP}(i)$ of an arbitrary particle $i$.
We will use the original radial basis suggested in Ref.~\onlinecite{bartok_on-representing_2013},
\beq
g_n(r) = \sum_{n^\prime=1}^{n_\textrm{max}} \beta_{n n^\prime} (r - r_\textrm{cut})^{n^\prime + 2} ,
\eeq
where the coefficients $\beta_{n n^\prime}$ ensure orthonormality.
These basis functions are not associated with neighbors at specific distances: this helps to ensure a faithful description of the local environment,
at the cost of reducing the interpretability of the descriptor.
Note that choosing $n_\textrm{max}=1$ provides a smooth version of the conventional BO descriptor, which differs nonetheless from the SBO descriptor because of the Gaussian smoothing of the local density and the lack of normalization.

\subsubsection{Radial dependence of bond order: a simpler Gaussian basis}
\label{sec:rbo}

It is interesting to connect the SOAP descriptor to the one used by Boattini \textit{et al.}~\cite{boattini_averaging_2021} in a recent supervised learning study of dynamic heterogeneity.
We will refer to their descriptor as radial bond-order (RBO) descriptor, because it captures the radial dependence of bond order in the most straightforward way.
This descriptor does not involve any smoothing of the local density.
As a radial basis, Boattini \textit{et al.} used Gaussian functions of width $\delta$ centered on a grid of distances $\{d_n\}_{n=1 \dots n_\mathrm{max}}$,
\beq
G_n(r) = \exp{\left(-\frac{(d_n - r)^2}{2\delta^2}\right)} .
\eeq
As noted above, such descriptors are not very sensitive to angular correlations between particles at different distances~\cite{bartok_on-representing_2013}.
It is unclear \textit{a priori} to what extent this issue will affect supervised and unsupervised learning of structure and dynamics via this descriptor.

The complex radial bond-order coefficients are defined as
\beq
  q_{l m n}^{R}(i) = \frac{1}{Z(i)} \sum_{j=1}^{N}
  G_n(r_{ij}) Y_{l m}(\hat{\bvec{r}}_{ij}) ,
\eeq
where $Z(i) = \sum_{j=1}^N G_n(r_{ij})$ is a normalization constant and the superscript $R$ indicates the radial dependence of the descriptor.
In the following, we actually use
\beq
G_n(r_{ij}) = \exp{\left(-\frac{(d_n - r_{ij})^2}{2\delta^2}\right)} H(R_\mathrm{max} - r_{ij}) ,
\eeq
where $H$ is the Heaviside step function, which allows us once again to neglect the contributions of particles further than a distance $R_\mathrm{max} = d_{n_\mathrm{max}} + 2.5 \delta$ from the central particle, where $d_{n_\mathrm{max}}$ is the largest distance in the grid of points $\{ d_n \}$.
Then, only the diagonal coefficients of the power spectrum, namely 
\beq
Q_{ln}^R(i) = \left( \frac{4\pi}{2l + 1} \sum_{m=-l}^l |q_{l m n}^R(i)|^2 \right)^{1/2} ,
\label{eq:rbo}
\eeq
are retained to form the descriptor of particle $i$ as $(\dots, Q_{ln}^R(i), \dots)$, which is again flattened as a vector $X^\mathrm{RBO}(i)$, composed of $l_\textrm{max}\times n_\textrm{max}$ structural features.
The choice for the grid of points $\{ d_n \}$ is discussed in Sec.~\ref{sec:radial} and \ref{sec:propensity_fit_soap_rbo}.

One key advantage of the RBO descriptor is that it is easily interpretable: it describes the bond order of thin spherical shells at increasing distance from the central particle.
The radial basis considered within SOAP are much less interpretable and the descriptor involves the full power spectrum, $p_{nn^\prime l}$.
This provides a wealth of information, but it is unclear \textit{a priori} whether this is actually relevant for a specific problem.
We will come back to this issue in Sec.~\ref{sec:propensity}.

Finally, we note that the RBO descriptor does not explicitly account for the local coordination number (\textit{i.e.}, how dense is the shell of neighbors), because the projected density does not scale with the number of neighbors.
To include this information, Boattini\textit{ et al.} have complemented their structural descriptor with a measure of the local density in successive shells centered around a given particle.
Interestingly, however, we will find that the normalized BO descriptors are correlated with the local density of the first coordination shell (Sec.~\ref{sec:pc_vs_ql}).

\subsection{Dimensionality reduction}
\label{sec:dim-redux}

\newcommand{\dataX}{\textbf{X}}

We have described the construction of several structural descriptors $X(i)$ %$\vec{x}(i)$ 
that characterize the first coordination shell of particle $i$, as given in Eq.~\eqref{equ:xBO},~\eqref{equ:xSBO} and~\eqref{eq:rbo}. Let the dimension of $X(i)$ %$\vec{x}(i)$ 
be $M$. For a sample of $N_\textrm{tot} = N \times n_\textrm{conf}$ particles, this means that the coordination shells of all particles can be summarized in a matrix of size $N_\textrm{tot} \times M$, whose rows are the $X(i)$ % $\vec{x}(i)$:
\beq
\dataX
%X
 = \begin{pmatrix}
X(1) \\
\vdots \\
X(N_\textrm{tot})
\end{pmatrix}.
\eeq
In a liquid, each row of the matrix is different, which reflects the heterogeneity of the local structure.
However, correlations within the coordination shell mean that the row vectors have a non-trivial probability distribution, with correlations among their components.
The goals of dimensionality reduction are (i) to find out in a generic way to what extent the components are correlated, (ii) to extract a reduced descriptor of lower dimension, which embodies most of the ``relevant'' fluctuations, and (iii) to provide quick insight into structural heterogeneity, such as multimodality, non-Gaussianity, etc.

In simple systems and with simple descriptors, such as $X^\textrm{BO}$, one may always reduce the dimensionality of $X$ 
by focusing on selected components, \textit{e.g.}, $Q_4$ and $Q_6$, perhaps motivated by some physical intuition about the local structure.  One may then use the low-dimensional distributions of the chosen components to characterize structural heterogeneity.  
The aim of unsupervised learning is to avoid assumptions on the 
most important components~\cite{boattini_unsupervised_2019,van_Damme_Coli_van_Roij_Dijkstra_2020}, and to identify them using statistical methods.
This is particularly advantageous when $M$ is large, and even more so when applying unsupervised learning methods that require computing distances between datapoints~\cite{aggarwal_surprising_2001}.

Dimensionality reduction is a valuable tool to gather the relevant components into ``reduced descriptors''.
It involves mapping the high-dimensional descriptors into a lower-dimensional space
\beq
\mathcal{V}_\dataX : X \mapsto \widetilde{X},
\label{equ:mu}
\eeq
where $\mathcal{V}_\dataX$ is a function that maps the original $M$-dimensional descriptor (for example, $X(i)$) to a reduced $P$-dimensional descriptor (here $\widetilde{X}(i)$), with $P<M$.
The method exploits the full matrix $\dataX$ to construct the mapping, typically by optimizing some cost function, hence $\mathcal{V}_\dataX$ carries an implicit dependence on $\dataX$.
The main difference between the various dimensionality reduction methods lies in the nature of the mapping: we talk about \textit{linear} or \textit{non-linear} dimensionality reduction depending on the linearity of $\cal V_\dataX$, as will be discussed further in the next paragraphs.

Note that the dataset is often normalized before performing dimensionality reduction~\cite{Mehta_Bukov_Wang_Day_Richardson_Fisher_Schwab_2019}.
This normalization, known as feature scaling, is necessary if the dataset is heterogeneous, with different features having different physical dimensions.
In this work, the features have the same dimensions so scaling is not strictly necessary.  In some cases, we perform 
a simple Z-score scaling: each feature is normalized by subtracting the average and dividing by the standard deviation evaluated over all the particles.
We indicate whether or not feature scaling is applied, for each descriptor.

\subsubsection{Principal component analysis}
\label{sec:pca}

Principal component analysis (PCA)~\cite{jolliffe_principal_2016} is a linear dimensionality reduction method that aims to transform each descriptor $X$ into $\widetilde{X}$ by retaining directions associated to the largest variance of the dataset.
Since the transformation is linear, this is easily performed at the level of the data matrix $\mathbf{X}$, which is first transformed as
\beq 
\widetilde{\mathbf{X}} = \mathbf{X V},
\eeq
where $\mathbf{V}$ is a $M \times M$ matrix whose columns $\{V^{(j)}\}_{j=1 \dots M}$ are called the principal component (PC) directions of $\mathbf{X}$~\cite{Hastie_Tibshirani_Friedman_2016}.
They correspond to the eigenvectors of the features' covariance matrix, whose entry $(j,k)$ is
\beq
\mathrm{cov}(X_j, X_k) = \langle (X_j - \langle X_j \rangle) (X_k - \langle X_k \rangle) \rangle .
\eeq
where $\langle \dots \rangle$ denotes an average over the $N_\textrm{tot}$ particles of the sample.
This covariance matrix measures correlations between different elements of the descriptor.
The eigenvectors $\{V^{(j)}\}_{j=1 \dots M}$ correspond to the directions of the new feature space. 
Their associated eigenvalues $\{ \lambda_j \}_{j=1 \dots M}$ indicate the variance of the data along the corresponding eigenvector, and the explained variance ratio (EVR) of the $j$-th principal component is thus given by
\beq
\mathrm{EVR}(\mathrm{PC}_j) = \frac{\lambda_j}{\sum_{k=1}^M \lambda_k}.
\label{equ:EVR-PCA}
\eeq
By convention, the principal components are sorted in descending order according to their EVR.

Dimensionality reduction is performed by considering a truncated transformation, which retains only the first $P<M$ eigenvectors in the matrix $\mathbf V$.
Equivalently, we retain only the first $P$ columns of the transformed matrix $\widetilde{\mathbf X}$, yielding $P$ reduced features, $\widetilde{X}_1, \dots, \widetilde{X}_P$.
As a rule of thumb, one keeps as many PC directions as needed to explain to a significant fraction, say 80\%, of the total variance of the dataset, see also Sec.~\ref{sec:pc_vs_ql}.
Alternatively, one can inspect the EVR as a function of the component index and look for an inflection point, which defines empirically an optimal $P$ for dimensionality reduction.
The notion of intrinsic dimension~\cite{Goldt_Mezard_Krzakala_Zdeborova_2020,Mendes-Santos_Turkeshi_Dalmonte_Rodriguez_2021} of a dataset provides a more robust and objective criterion.
This kind of analysis is, however, beyond the scope of the present work and is left for a future study.
Note that within PCA, the reduced features are linearly related to the original ones.
Thus one can directly identify the main sources of variance in the data, \textit{e.g.}, which features are most responsible for the observed structural heterogeneity of the sample.

\subsubsection{Neural network autoencoder}
\label{sec:ae}

An autoencoder is an artificial neural network that learns an efficient coding of a dataset $\mathbf X$ in an unsupervised way~\cite{Hastie_Tibshirani_Friedman_2016}.
It is trained to reproduce its own $M$-dimensional input as output by forcing the input data through a lower $P$-dimensional bottleneck in the hidden layers.
An autoencoder is separated into two parts: (i) an \textit{encoder} (the first half of the hidden layers up to the bottleneck) that forces a compression of the input data, and (ii) a \textit{decoder} (the second half of the hidden layers), tasked to reconstruct the input $\mathbf X$ as output $\mathbf{X}^R$ with the highest accuracy.
This reduction method is non-linear since, in feed-forward neural networks, non-linear functions are generally used to control neuron activations between successive hidden layers.

Once the network has been optimized to minimize the reconstruction error, cutting it at the level of the encoder yields a mapping $\mathcal{V}_\dataX$ from the original high-dimensional data to a low-dimensional representation $\widetilde{\mathbf{X}}$.
The non-linearity of the method makes it possible to provide low-dimensional representations  of datasets with intricate structures.
However, an autoencoder also requires tuning a large number of parameters, such as the dimensions of the network (\textit{i.e.}, number and sizes of the hidden layers), choosing the activation functions, the cost function and regularization term, the solver, etc.
Moreover, training the neural network is a stochastic procedure that may yield very different outcomes and must be repeated until the optimal reconstruction error on $\mathbf{X}$ is reached.

The EVR can be determined by computing the mean squared error on the reconstruction rescaled by the mean squared deviation of the input vectors~\cite{boattini_autonomously_2020}, 
\beq
\mathrm{EVR} = 1 - \frac{ \sum_{i=1}^{N_\textrm{tot}} \lVert X(i) - X^R(i) \rVert ^2 }{ \sum_{i=1}^{N_\textrm{tot}} \lVert X(i) - \langle X \rangle \rVert ^2 },
\eeq	
where $X^{R}(i)$ is the $i$-th row of the reconstructed input, and $\langle X \rangle$ is the mean input vector. %  = (1/N_\textrm{tot}) \sum_{i=1}^{N_\textrm{tot}} X(i)
Note that this EVR is the fraction of the variance that is captured by the reduced dataset $\widetilde{\mathbf{X}}$ of the AE, while the corresponding quantity Eq.~\eqref{equ:EVR-PCA} for PCA is the variance captured along a single PC direction.

\section{Dimensionality reduction of local structure}
\label{sec:results}

In this section, we discuss some of the key features that emerge from dimensionality reduction of bond order in glassy binary mixtures.
The computation of the descriptors and most of the dimensionality reduction analysis reported in the following sections were performed using the \texttt{partycls} package~\cite{Paret2021}.
Data analysis has been carried out using a reproducible workflow, deposited in the Zenodo public repository~\cite{zenodo}.

We focus here on the local structure around the small particles of the ``closed-packed'' mixtures (Wahn, KA, \NiY, \CuZr) and on the Si particles of the SiO$_2$ model.
Local order around particles of these species becomes well-defined at low temperature.
We present some results for the other species in Appendix~\ref{app:sbo}.
We mostly limit ourselves to the first coordination shell and we set the weights $w_j$ to unity in Eq.~\eqref{eq:density_micro}.  We work at temperatures that correspond to moderate supercooling, \textit{i.e.}, close to the MCT crossover temperature, and we analyze
instantaneous configurations from dynamical simulations.
(At these temperatures, analysis of energy-minimized and instantaneous configurations differ only marginally.)

\subsection{Illustrative example: icosahedral local order in the Wahn model}

To illustrate the key features of the PCA approach to dimensionality reduction we start by analyzing the Wahn mixture.
This model serves as a useful benchmark, because its local structure has been characterized in-depth and is known to display strong icosahedral order~\cite{coslovichUnderstandingFragilitySupercooled2007a,malinsIdentificationLonglivedClusters2013}.
Its structural features should thus stand out most clearly.

\begin{figure}[!t]
  \includegraphics[width=.96\linewidth]{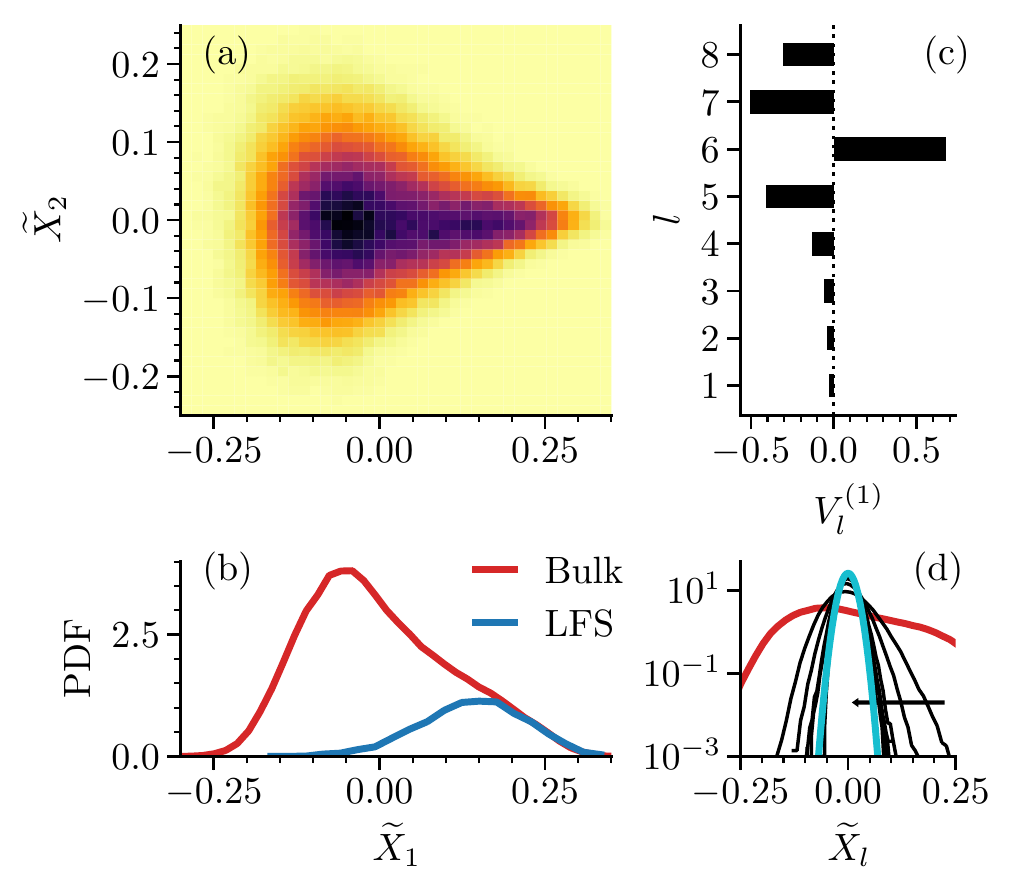}
\caption{\label{fig:bench} Overview of the results of PCA of the SBO descriptor for the small particles in the Wahn mixture. (a) Probability density $p(\X_1, \X_2)$. (b) Marginal distribution $p(\X_1)$ and $p(\X_1|\mathrm{LFS})$. (c) Components of the eigenvector $V^{(1)}$. (d) Distribution $p(\X_j)$ for $j>1$ (black) and $p(\X_1)$ (red). The light blue curve is a Gaussian with variance equal to the one of $\X_8$. The horizontal line indicates increasing PC index.}
\end{figure}

We consider the SBO descriptor, which avoids the discontinuous dependence of the features on the particle coordinates, taking $l_\mathrm{max}=8$ and $\gamma=8$ for the smoothing function $f(r)$ in Eq.~\eqref{eq:fr_sbo}.
Using PCA for dimensional reduction,
we write $\widetilde{X}_j(i)$ as the $j$-th element of $\widetilde{X}(i)$, which represents the projection of particle $i$'s descriptor on the $j$-th PC direction.
By considering all the particles in the system, we form a joint probability distribution of the PCs $(\widetilde{X}_1,\widetilde{X}_2,\dots)$.
Fig.~\ref{fig:bench}(a) shows the probability density function $p(\widetilde{X}_1, \widetilde{X}_2)$ of the projections on the two first PCs directions.
Since the values of $Q_l^S$ (and $Q_l$) are directly comparable in terms of both amplitudes and ranges, we do not perform feature scaling.

By visual inspection, we recognize two modes in the distribution: a central diffuse region and a more localized lobe for larger values of $\widetilde{X}_1$.
To illustrate the physical meaning of these modes,
we compare in Fig.~\ref{fig:bench}(b) the marginal distributions $p(\widetilde{X}_1)$ and $p(\widetilde{X}_1|\textrm{LFS})$, the latter being restricted to particles at the center of the LFS, as identified from a radical Voronoi tessellation~\cite{paret_assessing_2020}.
It is clear that the secondary lobe is due to local icosahedral order.
These observations suggest the presence of two distinct amorphous structural ``states'' in this model, similar to what found in polyamorphic materials~\cite{Tanaka_2012}.

Analyzing the results in more detail,
the first PC identifies the direction in feature space having the largest structural heterogeneity of the SBO parameters and accounts for about 62\% of the total variance.
One may expect that this component gives a strong weight to $Q_6$.
This is confirmed by Fig.~\ref{fig:bench}(c), which shows the individual components of the first eigenvector $V^{(1)}$ of the covariance matrix, see Sec.~\ref{sec:pca}.
We also see that $V^{(1)}$ couples several bond-order invariants and has large contributions also from $Q_5$, $Q_7$, $Q_8$, with a sign opposite to $Q_6$.
In other words, particles with large $Q_6$ (and icosahedral local environments) tend to have small projections on spherical harmonics of order $5,7,8$, while lower order invariants are statistically uncorrelated to $Q_6$.

Turning to higher principal components, 
we found that PC$_2$  gives strong and opposite weights to $Q_5$ and $Q_7$ (not shown) but the corresponding EVR is already fairly small (18\%).
Interestingly, principal components with even smaller EVR are characterized by marginal distributions $p(X_j)$ that are only slightly asymmetric and are rather well described by Gaussians, see Fig.~\ref{fig:bench}(d).
These observations suggest that it may be possible to schematically represent the feature space of the SBO descriptor of the Wahn mixture as having a single ``relevant'' direction, which captures the bulk of the structural heterogeneity, on top of a background of ``trivial'', nearly Gaussian directions.
Interestingly, this kind of simplified picture of high-dimensional datasets defines one of the simplest models of unsupervised statistical learning~\cite{Engel_Van_den_Broeck_2001} and is also amenable to analytical treatments.

\begin{figure*}[!t]
  \begin{tabular}{ccccc}
 %     \hskip-.5em\includegraphics[width=0.2\linewidth]{KA/T0.4500/config.xyz.sbo.scaling-False.redux-True.heatmap_marginal.species-2.pdf} &
 % \hskip-.5em\includegraphics[width=0.2\linewidth]{Ni33Y67/T0.55/config.xyz.sbo.scaling-False.redux-True.heatmap_marginal.species-2.pdf} &
 % \hskip-.5em\includegraphics[width=0.2\linewidth]{Wahn/T0.5800/config.xyz.sbo.scaling-False.redux-True.heatmap_marginal.species-2.pdf} &
 % \hskip-.5em\includegraphics[width=0.2\linewidth]{Cu64Zr36/T800/config.xyz.sbo.scaling-False.redux-True.heatmap_marginal.species-2.pdf} &
 % \hskip-.5em\includegraphics[width=0.2\linewidth]{SiO2/T0.3397/config.xyz.sbo.scaling-False.redux-True.heatmap_marginal.species-1.pdf} \\
 \hskip-.5em\includegraphics[width=0.2\linewidth]{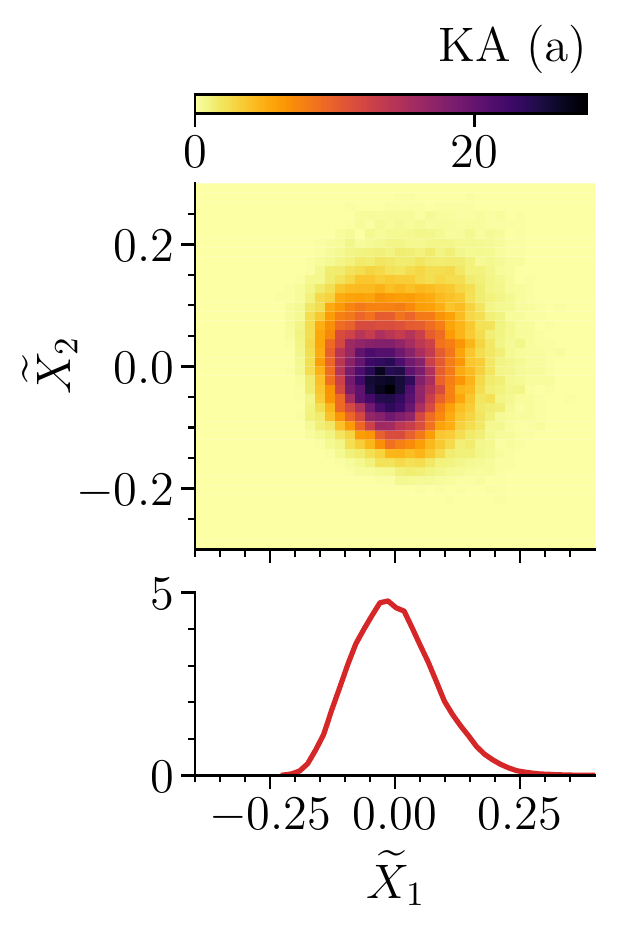} &
 \hskip-.5em\includegraphics[width=0.2\linewidth]{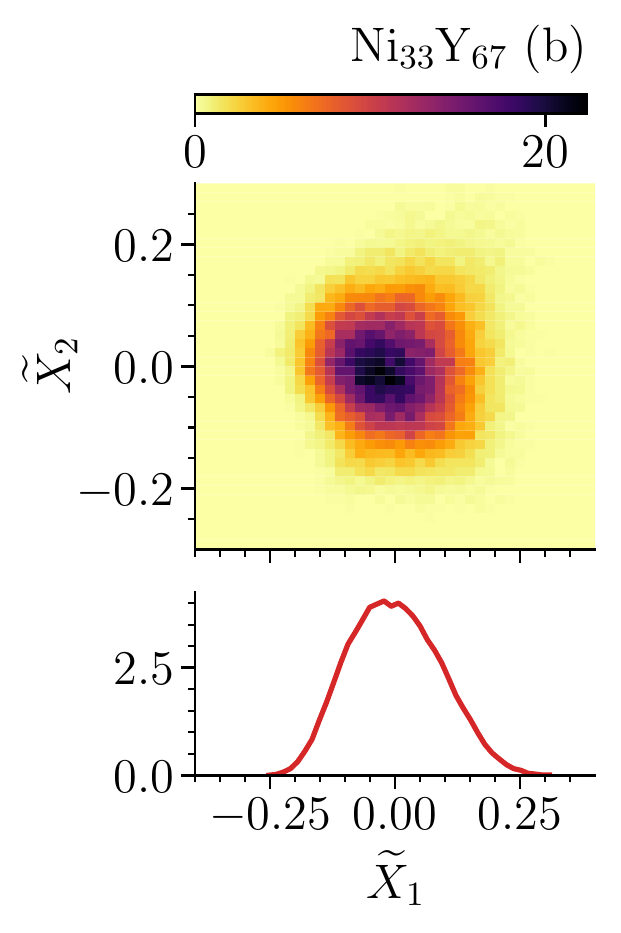} &
 \hskip-.5em\includegraphics[width=0.2\linewidth]{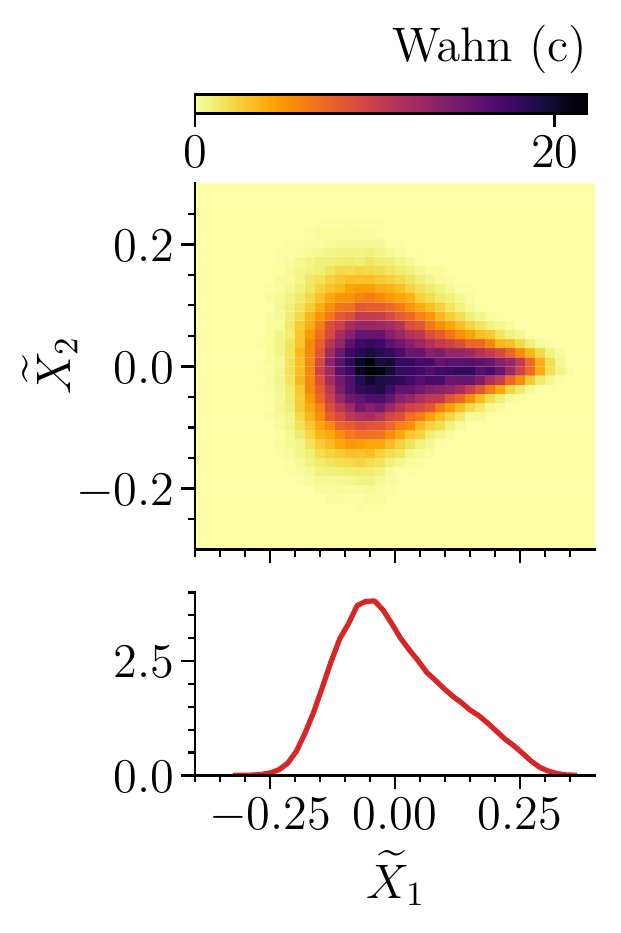} &
 \hskip-.5em\includegraphics[width=0.2\linewidth]{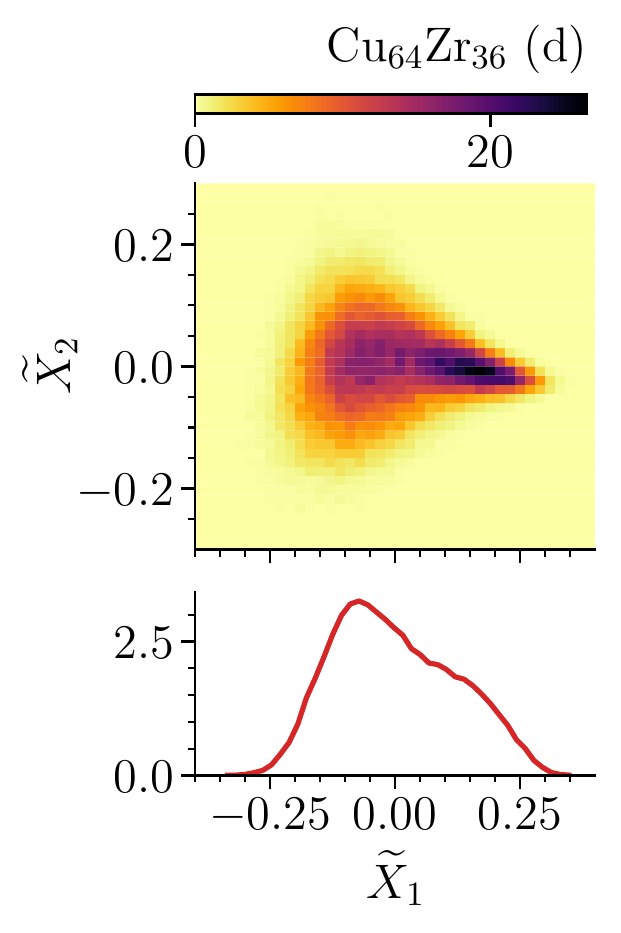} &
 \hskip-.5em\includegraphics[width=0.2\linewidth]{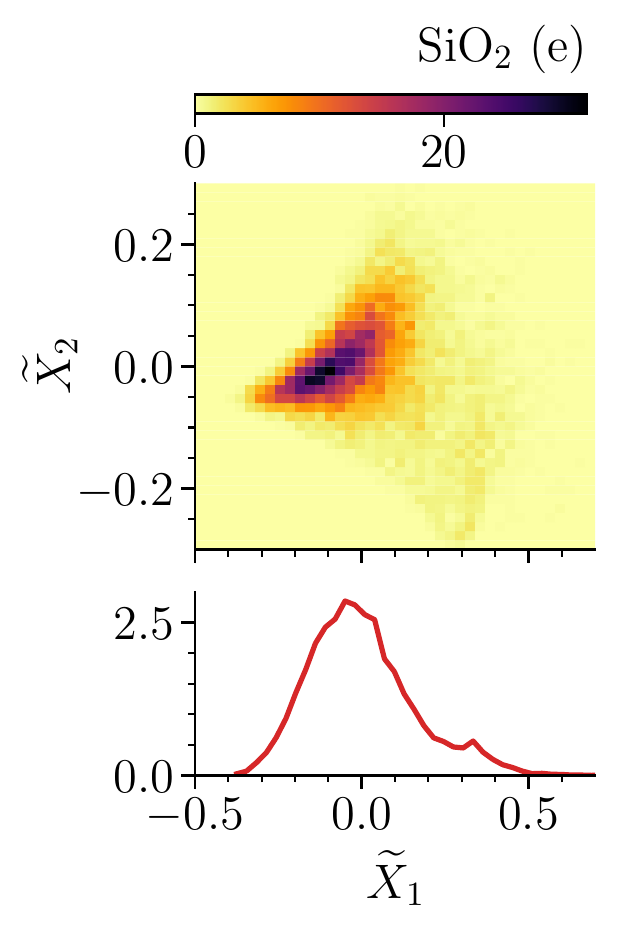} \\
  \end{tabular}
\caption{\label{fig:map_red} Probability density $p(\X_1, \X_2)$ from the PCA of the SBO descriptor in all the studied models. Results are shown for the small particles in panels (a)-(d) and for Si particles in (e). The bottom panels show the marginal distributions of $\widetilde{X}_1$.}
\end{figure*}

\subsection{From weak to strong structural heterogeneity in glassy binary mixtures}
\label{sec:weak_to_strong}

We now proceed with a more systematic comparison of the reduced structural features of glassy binary mixtures.
When analyzing bond order, the vast majority of previous studies have focused on rotational invariants of order 4 and 6, which are assumed to reflect the relevant symmetries of close-packed local structures~\cite{Leocmach_Tanaka_2012}.
The previous example shows that PCA identifies indeed $Q_6$ as the most relevant parameter in the Wahn mixture.
Here, however, we also consider systems for which the relevant symmetries are not obvious from physical intuition, so that unsupervised learning of local structure is useful.

As in the previous section, we use the SBO descriptor with $l_\textrm{max}=8$ and $\gamma=8$ for the smoothing function.
We checked that inclusion of higher order spherical harmonics ($l_\textrm{max}=16$) does not appreciably change the results.
We focus on the two PC directions with the largest eigenvalues, further analysis of the remaining components is given in  Sec.~\ref{sec:pc_vs_ql} below.
In Fig.~\ref{fig:map_red}, we show the probability density functions $p(\widetilde{X}_1, \widetilde{X}_2)$ for all the studied systems.
We start by analyzing the KA mixture and the \NiY~model.
In these systems, the first two PCs capture 37+31\% (KA) and 39+29\% (Ni$_{33}$Y$_{67}$) of the total variance.
The eigenvectors $V^{(1)}$ and $V^{(2)}$ of each model are given in Appendix~\ref{app:sbo}.
Even though the invariants $Q_4$ and $Q_6$ feature prominently in the first two PC directions, the fluctuations of other invariants are more relevant in some models.
For instance, in the KA mixture and \NiY~model, $Q_5$ has the largest contribution to PC$_1$.
As can be seen in Fig.~\ref{fig:map_red}(a,b), these two models display a broad but unimodal distribution of the reduced descriptor $(\X_1, \X_2)$.
These results can be contrasted to what we found in the Wahn mixture and the \CuZr~model, which display instead a nearly bimodal $p(\X_1, \X_2)$.

It is interesting to compare these results to those obtained without smoothing the descriptor, \textit{i.e.}, with the bare BO descriptor of Eq.~\eqref{equ:xBO}, see Fig.~\ref{fig:map_bare}.
This descriptor has been used in a recent unsupervised learning analysis of glassy systems~\cite{boattini_autonomously_2020}.
The PCA of $X^\mathrm{BO}$ reveals clear banding in the probability density of the reduced descriptor of all the studied models.
As shown in Fig.~\ref{fig:scatter_bare}, these bands are associated to different coordination numbers and disappear at high temperature, when thermal noise is large enough to smear the discontinuities of the descriptors~\cite{paret_thesis}.
Note that small signs of discontinuities are visible at the level of the distributions of the individual invariants, but becomes much more visible after dimensionality reduction.

The fact that the reduced BO descriptor is strongly modulated by coordination number may impact the cluster analysis of these structural descriptors.
However, we think these findings differ from ``structure from chance'' artifacts that may affect unsupervised learning~\cite{Engel_Van_den_Broeck_2001}.
The fact that bond order depends on the coordination number (or more generally local density) makes physical sense.
Interestingly, as we shall see in Sec.~\ref{sec:pc_vs_ql}, a strong correlation between the first few PCs and coordination number persists even after smoothing the descriptor.

Finally, we analyze the simple model of amorphous silica.
The local structure of this system can be easily characterized by the coordination number $z$, which equals $4$ for ideal tetrahedral structures around Si particles.
At the studied density and temperature, defects appear mostly in the form of undercoordinated structures, \textit{i.e.}, Si particles with $z=3$ and $z=2$.
We note that such defects would be almost completely removed by energy minimization at the studied temperature.
In Fig.~\ref{fig:map_red}(e) and Fig~\ref{fig:map_bare}(e) we show the distributions $p(\X_1, \X_2)$ obtained by applying PCA to $X^\mathrm{SBO}$ and $X^\mathrm{BO}$ of Si particles, respectively.
As anticipated in Sec.~\ref{sec:neigbbors}, we use a sharper $\gamma=18$ exponent to smear the SBO descriptor for this model, because the first coordination shell is more sharply defined than in the other mixtures, and we also restrict the calculation to oxygen neighbors.

As expected from the low coordination of the local structure, we find that the first two PC directions give strong weights to invariants with small $l$.
The PCs obtained from the bare BO descriptor display multiple sharp bands, along which $\widetilde{X}_1$ and $\widetilde{X}_2$ are strongly correlated.
The bands are associated to distinct coordination numbers, \textit{i.e.}, $z=4,3,2$ from left to right.
We found that, along a given band, the local environments around Si particles are also characterized by a marked gradient of tetrahedrality, \textit{i.e.}, how ideal is the local tetrahedral environment, see Sec.~\ref{sec:pc_vs_ql}.
On the other hand, the projections on the PCs obtained from $\widetilde{X}^\mathrm{SBO}$ have a much more diffuse character.
The scattered region at large $\X_1$ values is associated with defects in the network structure.
Thus, the relevant information about the local structure of this system is somehow washed out when using the smoothed descriptor.

\subsection{Connection of reduced structural features to other structural measures}
\label{sec:pc_vs_ql}

\begin{figure*}[!t]
  \begin{tabular}{ccccc}
% \hskip-.5em \includegraphics[width=0.2\linewidth]{KA/T0.4500/config.xyz.bo.scaling-False.redux-True.heatmap_marginal.species-2.pdf} &
% \hskip-.5em \includegraphics[width=0.2\linewidth]{Ni33Y67/T0.55/config.xyz.bo.scaling-False.redux-True.heatmap_marginal.species-2.pdf} &
% \hskip-.5em \includegraphics[width=0.2\linewidth]{Wahn/T0.5800/config.xyz.bo.scaling-False.redux-True.heatmap_marginal.species-2.pdf} &
% \hskip-.5em \includegraphics[width=0.2\linewidth]{Cu64Zr36/T800/config.xyz.bo.scaling-False.redux-True.heatmap_marginal.species-2.pdf} &
% \hskip-.5em \includegraphics[width=0.2\linewidth]{SiO2/T0.3397/config.xyz.bo.scaling-False.redux-True.heatmap_marginal.species-1.pdf} \\
 \hskip-.5em \includegraphics[width=0.2\linewidth]{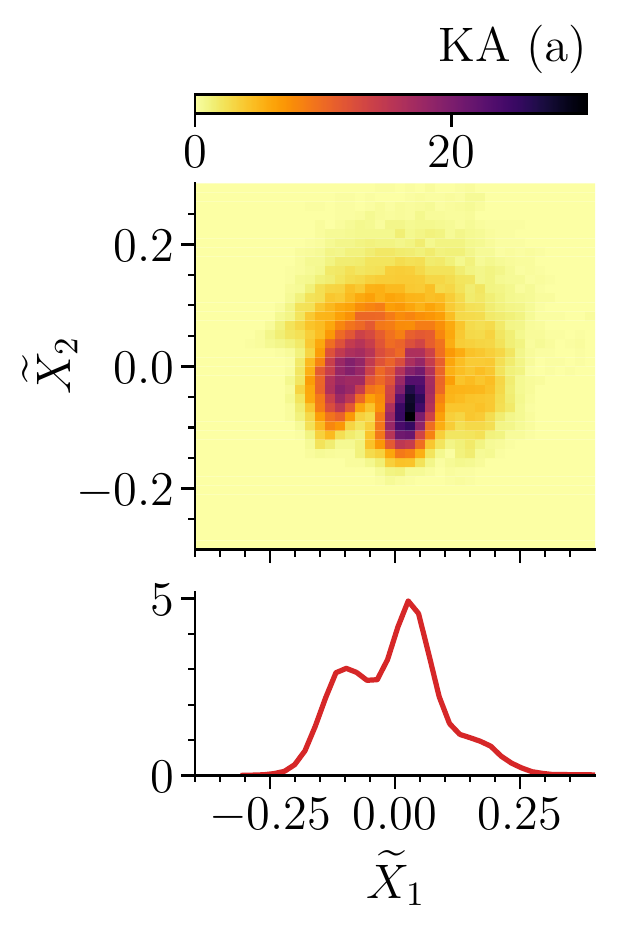} &
 \hskip-.5em \includegraphics[width=0.2\linewidth]{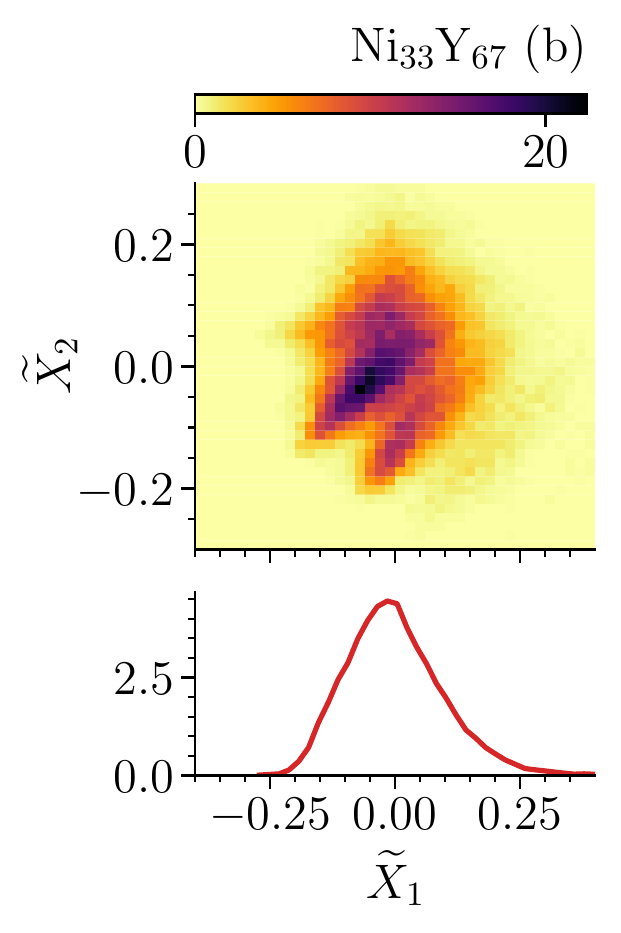} &
 \hskip-.5em \includegraphics[width=0.2\linewidth]{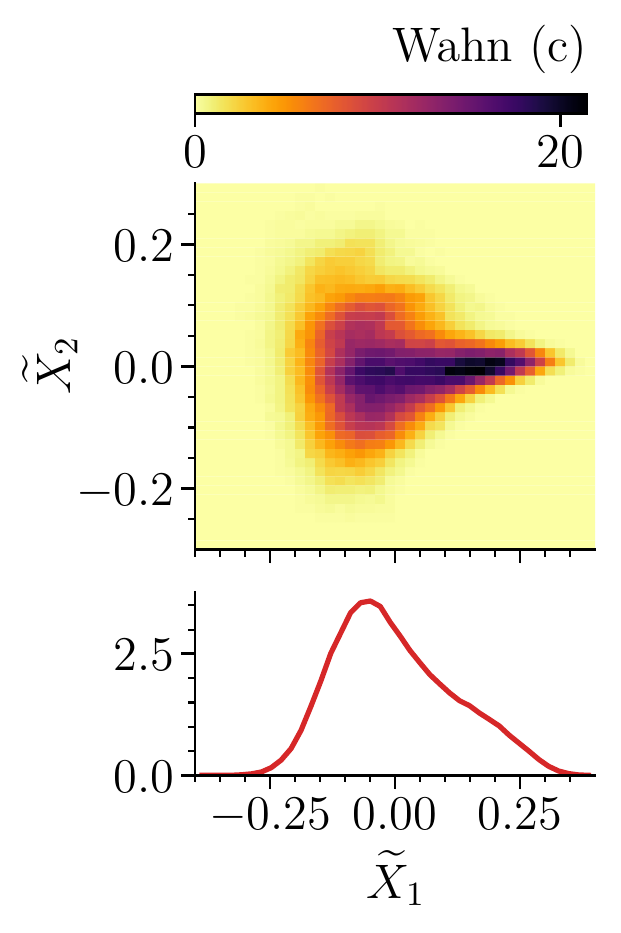} &
 \hskip-.5em \includegraphics[width=0.2\linewidth]{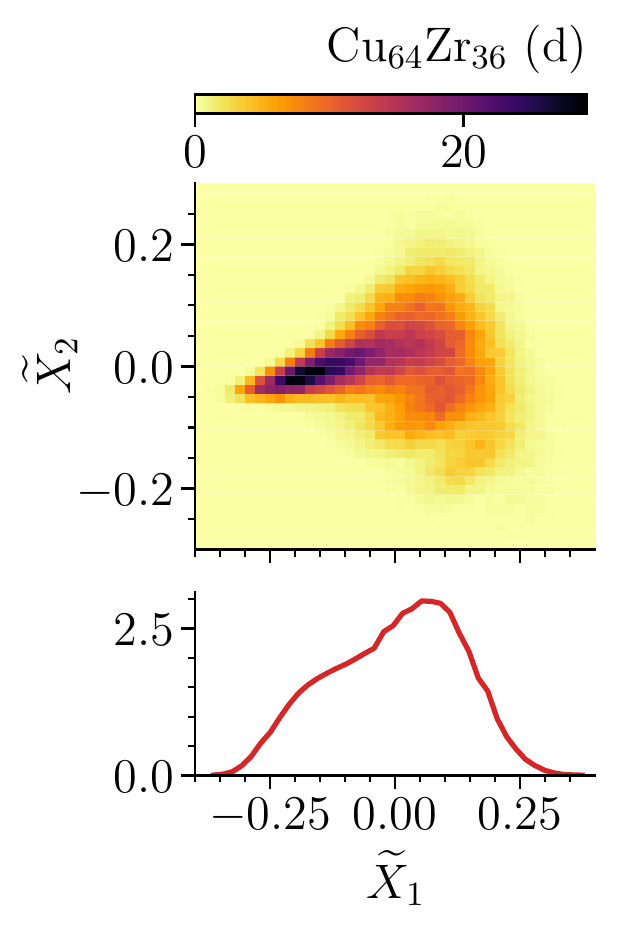} &
 \hskip-.5em \includegraphics[width=0.2\linewidth]{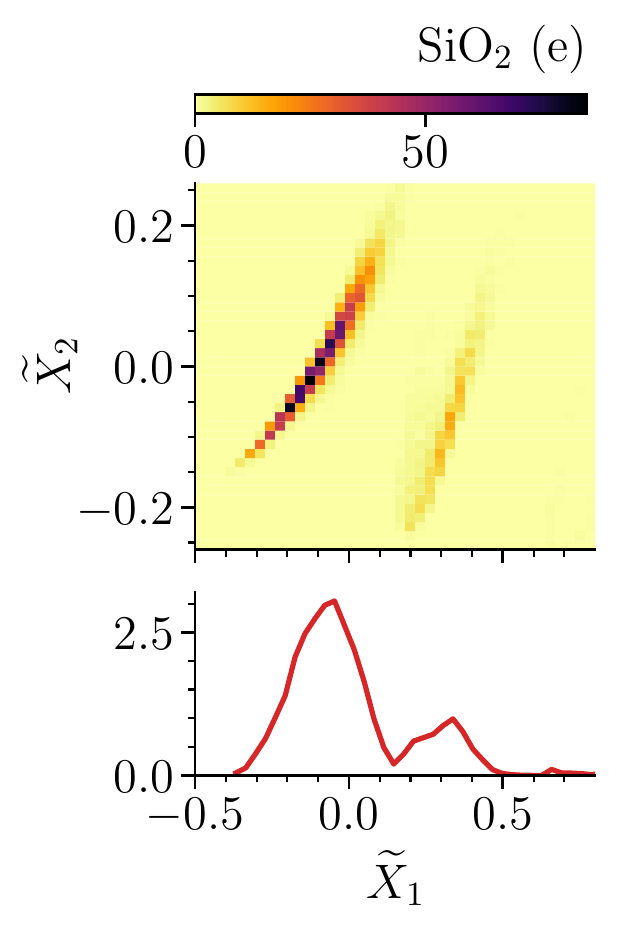} \\
  \end{tabular}
\caption{\label{fig:map_bare} Same as Fig.~\ref{fig:map_red} but for the BO descriptor.}
\end{figure*}

\begin{figure*}[!t]
    \begin{tabular}{ccccc}
 % \hskip-.5em      \includegraphics[width=.2\linewidth]{KA/T0.4500/config.xyz.bo.scaling-False.redux-True.scatter.species-2.pdf} &
 % \hskip-.5em       \includegraphics[width=.2\linewidth]{Ni33Y67/T0.55/config.xyz.bo.scaling-False.redux-True.scatter.species-2.pdf} &
 % \hskip-.5em       \includegraphics[width=.2\linewidth]{Wahn/T0.5800/config.xyz.bo.scaling-False.redux-True.scatter.species-2.pdf} &
 % \hskip-.5em       \includegraphics[width=.2\linewidth]{Cu64Zr36/T800/config.xyz.bo.scaling-False.redux-True.scatter.species-2.pdf} &
 % \hskip-.5em       \includegraphics[width=.2\linewidth]{SiO2/T0.3397/config.xyz.bo.scaling-False.redux-True.scatter.species-1.pdf}\\
 \hskip-.5em \includegraphics[width=.2\linewidth]{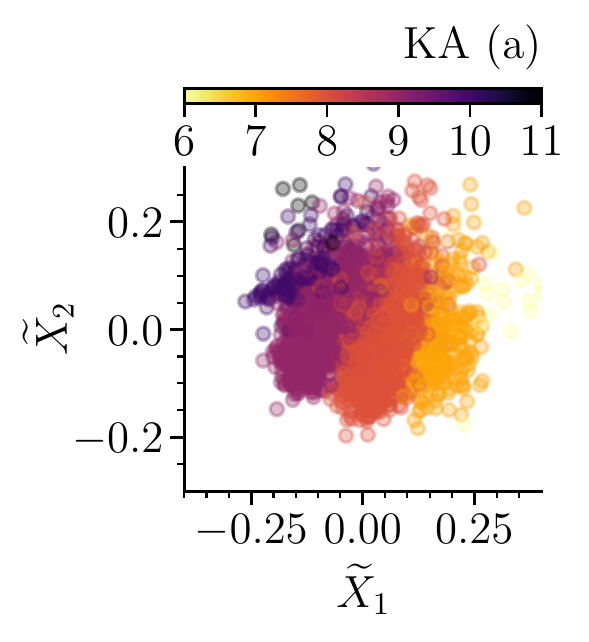} &
 \hskip-.5em \includegraphics[width=.2\linewidth]{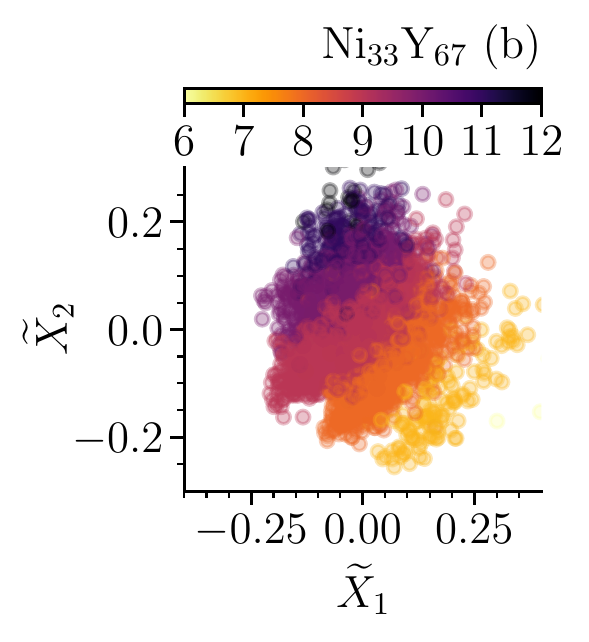} &
 \hskip-.5em \includegraphics[width=.2\linewidth]{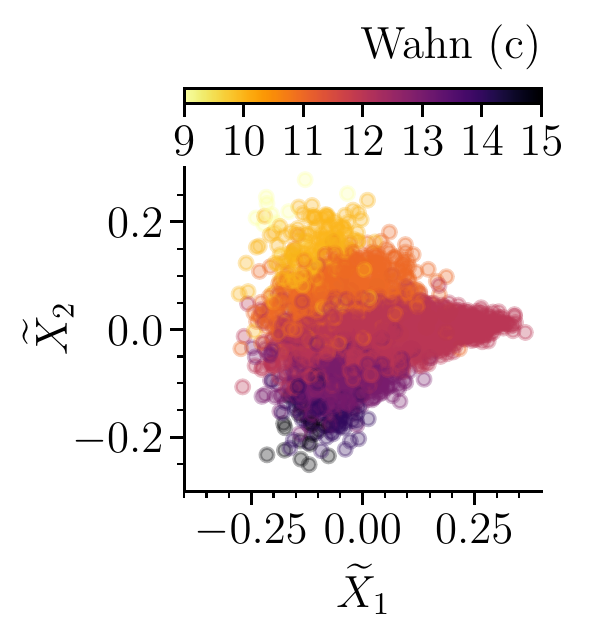} &
 \hskip-.5em \includegraphics[width=.2\linewidth]{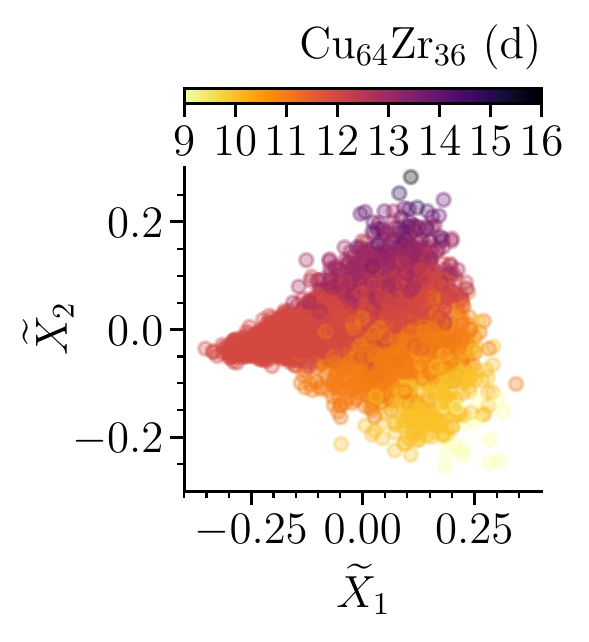} &
 \hskip-.5em \includegraphics[width=.2\linewidth]{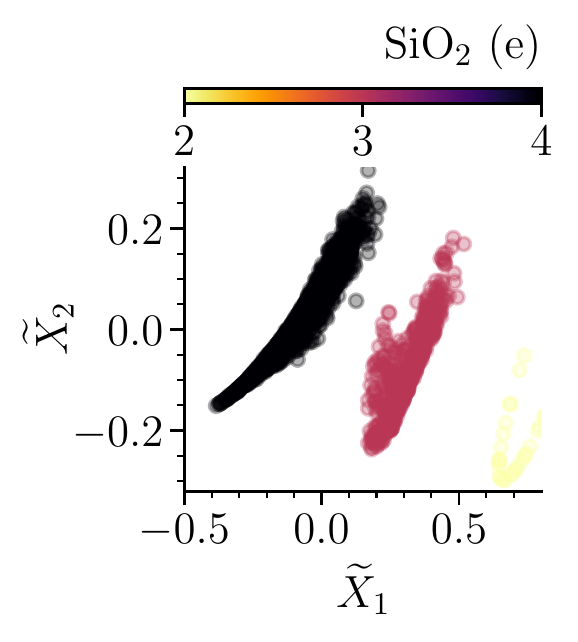}\\
    \end{tabular}
\caption{\label{fig:scatter_bare}Scatter plot of $(\X_1, \X_2)$ obtained from the PCA of the BO descriptor, color-coded by coordination number $z$. Results are shown for the small particles in panels (a)-(d) and for Si particles in (e). For clarity, only 3000 datapoints are shown for each system.}
\end{figure*}

The results discussed in Sec.~\ref{sec:weak_to_strong} provide some evidence of a varying degree of heterogeneity of the local structure of the studied systems.
The distribution $p(\widetilde{X}_1, \widetilde{X}_2)$ ranges from broad but unimodal (KA, \NiY) to nearly bimodal (Wahn, \CuZr) or even possibly multimodal (SiO$_2$).
In this section, we analyze the PCs of the SBO descriptor in detail and identify its connection with known measures of local order.
This analysis will corroborate the idea that the first few PCs are indeed sufficient to grasp relevant information about local order.

\begin{figure*}[!t]
  \begin{tabular}{ccccc}
 % \hskip-.5em\includegraphics[height=.258\linewidth]{KA/T0.4500/config.xyz.sbo.scaling-False.redux-True.species-2.overview_structure_color.pdf} &
 % \hskip-1.em\includegraphics[height=.258\linewidth]{Ni33Y67/T0.55/config.xyz.sbo.scaling-False.redux-True.species-2.overview_structure_color.pdf} &
 % \hskip-1.em\includegraphics[height=.258\linewidth]{Wahn/T0.5800/config.xyz.sbo.scaling-False.redux-True.species-2.overview_structure_color.pdf} &
 % \hskip-1.em\includegraphics[height=.258\linewidth]{Cu64Zr36/T800/config.xyz.sbo.scaling-False.redux-True.species-2.overview_structure_color.pdf} &
 % \hskip-1.em\includegraphics[height=.258\linewidth]{SiO2/T0.3397/config.xyz.sbo.scaling-False.redux-True.species-1.overview_structure_color.pdf} \\
 \hskip-.5em\includegraphics[height=.258\linewidth]{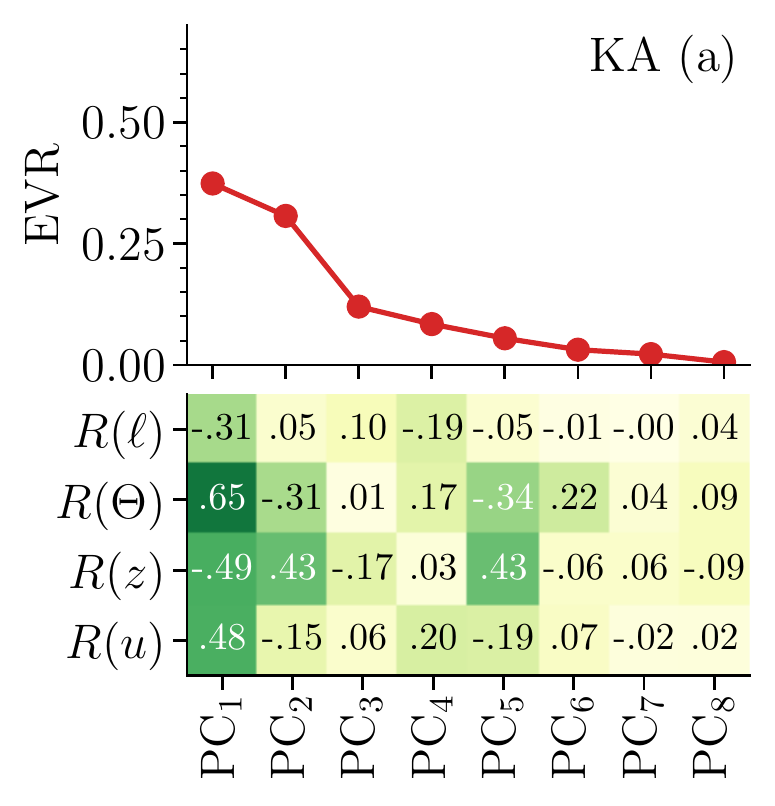} &
 \hskip-1.em\includegraphics[height=.258\linewidth]{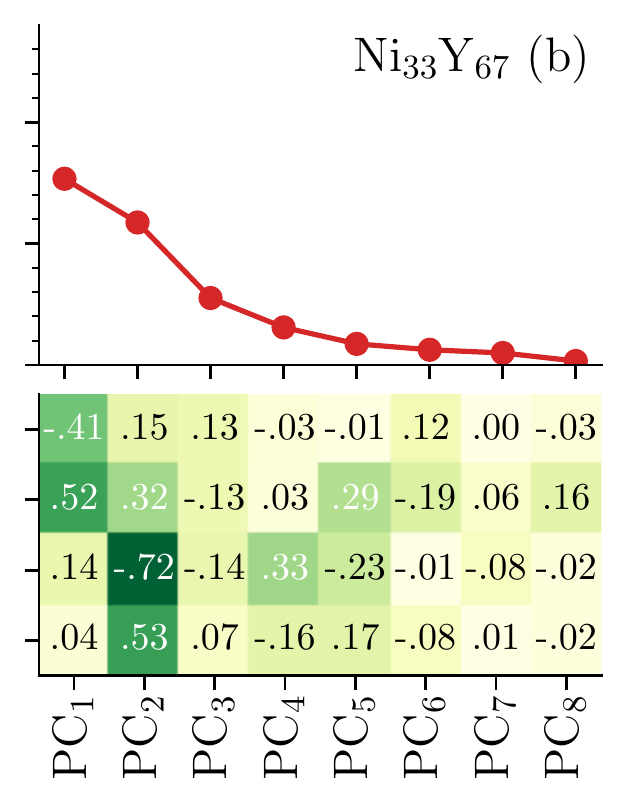} &
 \hskip-1.em\includegraphics[height=.258\linewidth]{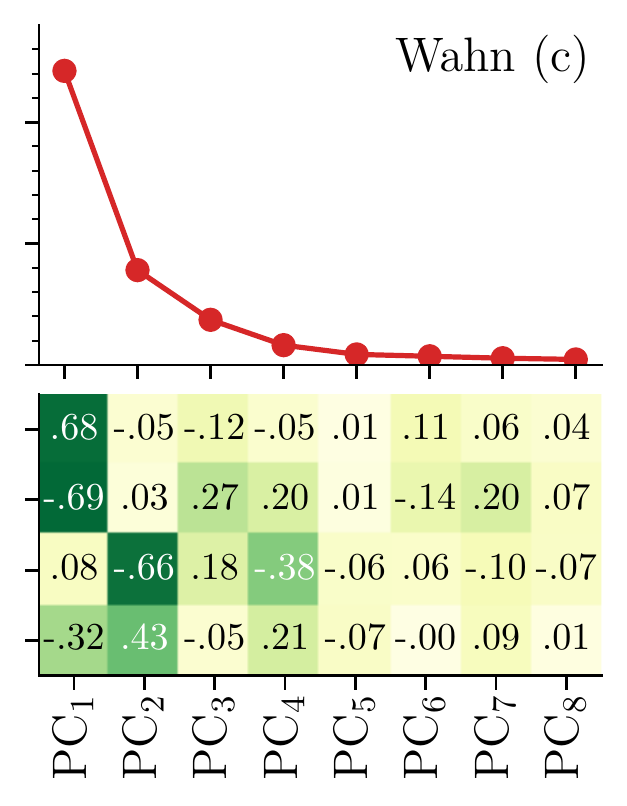} &
 \hskip-1.em\includegraphics[height=.258\linewidth]{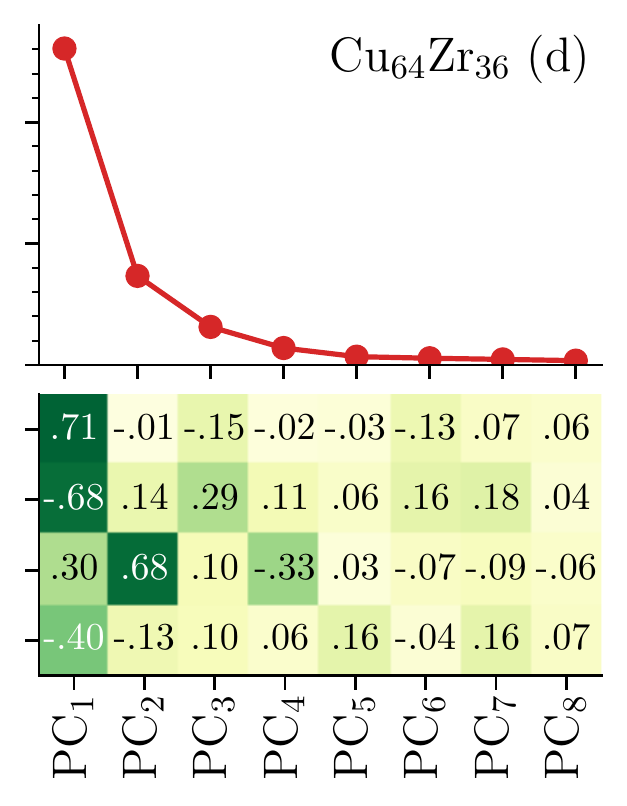} &
 \hskip-1.em\includegraphics[height=.258\linewidth]{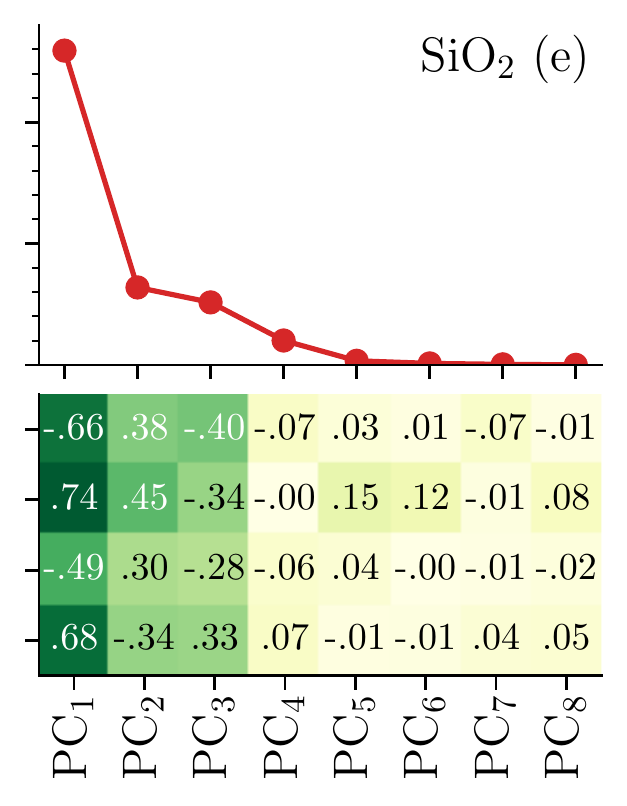} \\
  \end{tabular}
  \caption{\label{fig:overview_redux}Properties of the PC of the SBO descriptor for small particles in (a)-(d) and Si particles in (e). The top figures show the explained variance ratio (EVR) of each PC and the tables show the Pearson correlation coefficient $R$ between each $\X_j$ and other structural measures: LFS index ($\ell$), local compactness ($\Theta$), coordination number ($z$) and local potential energy ($u$). See the main text for a detailed description of these quantities.}
\end{figure*}

First, we analyze the EVR of all the PC directions in the top panel of Fig.~\ref{fig:overview_redux}.
All systems display a drop in the EVR after a few PCs: this occurs already after PC$_1$ in the Wahn, \CuZr~and SiO$_2$~models, while a little gap is visible after PC$_2$ in the more weakly heterogeneous systems.
This supports the view that the first two PCs alone are providing significant information about the underlying feature space.
Close inspection of Fig.~\ref{fig:overview_redux}(e) suggests the existence of an additional gap after PC$_3$ in the amorphous silica model, see also below.

We also confirmed that none of the low-EVR variables have strongly bimodal distributions.
We did so by employing the Hartigans' dip test~\cite{hartigan_dip_1985}, which provides a rather stringent criterion for the multi-modality of a distribution.
The output of the dip test is a value between 0 (unimodal distribution) and 1 (multimodal distribution).
We found that the marginal distributions $p(\widetilde{X}_1)$ of the SBO descriptor in the Wahn mixture and in the \CuZr~model have a significantly bimodal character, with dip test values close to 1.
For all the other marginal distributions the dip test value indicates a unimodal character.

We now connect the reduced structural variables $\widetilde{X}_j$ to selected measures of local order.
Namely, we compute the Pearson correlation coefficient $R$ between the PCs and (i) the coordination number $z$, measured by integrating the partial correlation function $g_{\alpha}(r)$, where $\alpha$ is the species of interest, up to its first minimum; (ii) the $\Theta$ parameter introduced by Tong and Tanaka~\cite{tongRevealingHiddenStructural2018}, which measures the compactness of local environment; as in the original implementation, we used radical Voronoi neighbors for this calculation~\cite{voro++} and we employed the positions of the first peak of the $g_{\alpha\alpha}(r)$ radial distribution functions as measures of the effective particle diameters; for the SiO$_2$ model, we use instead a simple measure of tetrahedrality, \textit{i.e.}, $\Theta$ is defined as the average deviation of the bond angles from the ideal tetrahedral angle $109.5^\circ$; (iii) the LFS index $\ell$, which equals 1 if the Voronoi polyhedron surrounding the particle corresponds to the system's LFS and 0 otherwise; the LFS has been identified from a radical Voronoi tessellation in previous work~\cite{coslovichUnderstandingFragilitySupercooled2007a, hocky_correlation_2014}; in the SiO$_2$ model model, $\ell$ is defined as 1 if the coordination number is $z=4$ for Si particles, and 0 otherwise; (iv) local potential energy $u$.

The bottom panels of Fig.~\ref{fig:overview_redux} provide an overview of these correlation coefficients for all the studied mixtures.
Interestingly, the first few PCs are quite strongly connected to the local order parameters ($\Theta$, $\ell$), local potential energy ($u$) or coordination number ($z$).
The degree of correlation depends on the PC and on whether the mixture is weakly or strongly heterogeneous.
In particular, we notice that the correlation between PC$_1$ and the LFS determined from the Voronoi tessellation is significant in systems with marked structural heterogeneity (Wahn, \CuZr, SiO$_2$) and less so in the other models (KA, \NiY).
Even in systems with weak structural heterogeneity, however, the projections on the first two PCs are correlated to measures of local structure, despite the more uniform spectrum of features.
Interestingly, PC$_2$ and PC$_3$ in the SiO$_2$ model display a higher correlation than the remaining low-EVR components, confirming the above observations that these variables are structurally relevant.

\begin{figure*}[!t]
  \begin{tabular}{ccccc}
 \hskip-.5em\includegraphics[height=.248\linewidth]{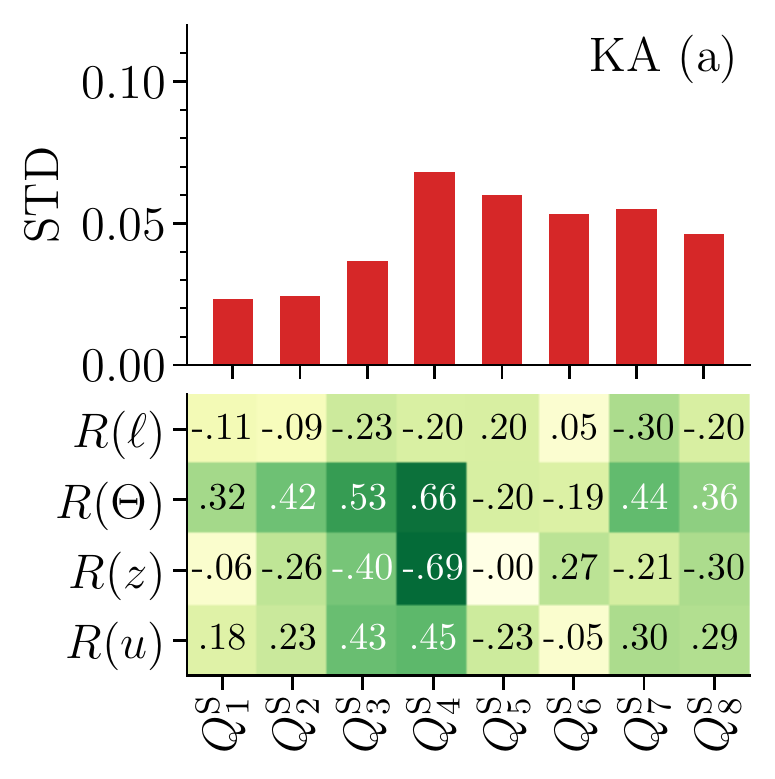} &
 \hskip-1.em\includegraphics[height=.248\linewidth]{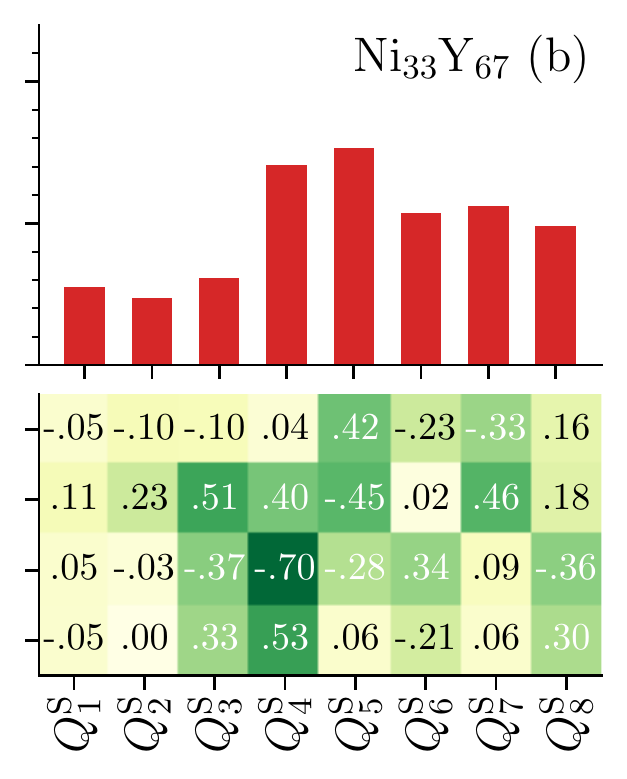} &
 \hskip-1.em\includegraphics[height=.248\linewidth]{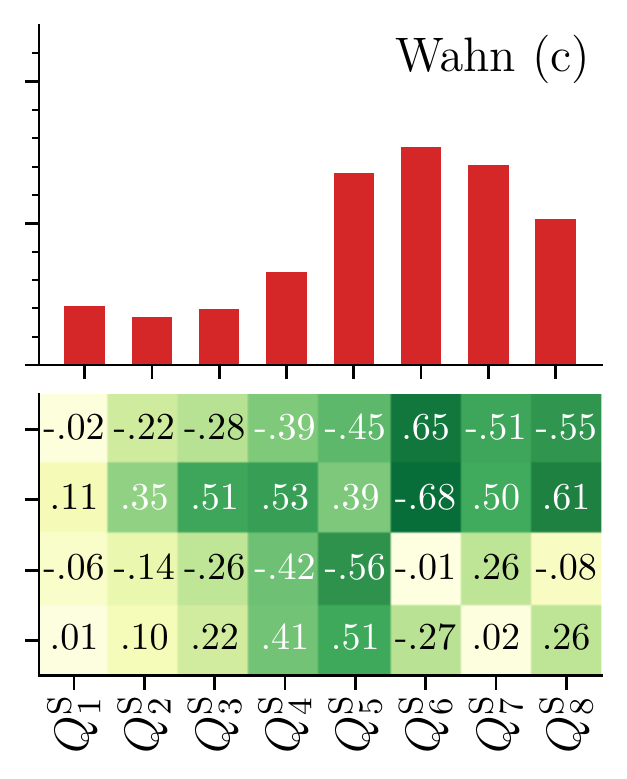} &
 \hskip-1.em\includegraphics[height=.248\linewidth]{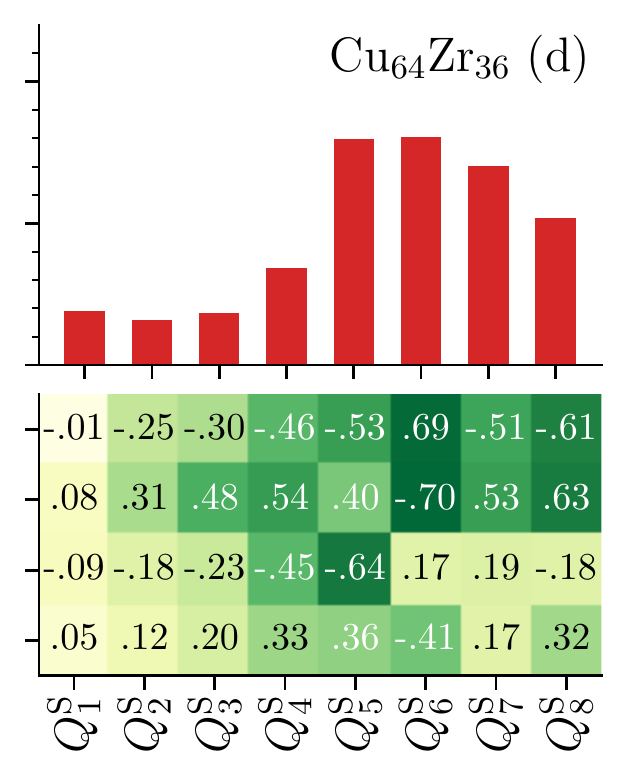} &
 \hskip-1.em\includegraphics[height=.248\linewidth]{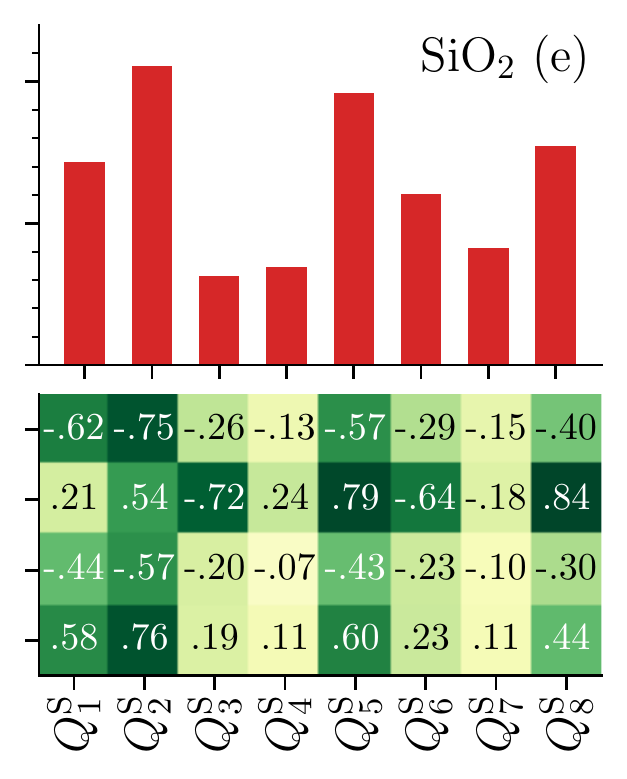} \\
 % \hskip-.5em\includegraphics[height=.248\linewidth]{KA/T0.4500/config.xyz.sbo.scaling-False.redux-False.species-2.overview_structure_color.pdf} &
 % \hskip-1.em\includegraphics[height=.248\linewidth]{Ni33Y67/T0.55/config.xyz.sbo.scaling-False.redux-False.species-2.overview_structure_color.pdf} &
 % \hskip-1.em\includegraphics[height=.248\linewidth]{Wahn/T0.5800/config.xyz.sbo.scaling-False.redux-False.species-2.overview_structure_color.pdf} &
 % \hskip-1.em\includegraphics[height=.248\linewidth]{Cu64Zr36/T800/config.xyz.sbo.scaling-False.redux-False.species-2.overview_structure_color.pdf} &
 % \hskip-1.em\includegraphics[height=.248\linewidth]{SiO2/T0.3397/config.xyz.sbo.scaling-False.redux-False.species-1.overview_structure_color.pdf} \\
  \end{tabular}
\caption{\label{fig:overview_bare}Properties of the individual SBO parameters, $Q_l^\mathrm{S}$, for small particles in (a)-(d) and Si particles in (e). The top panels show the standard deviation on the distributions of each parameter $Q_l^\mathrm{S}$ and the tables show the Pearson correlation coefficients $R$ between the individual values of $Q_l$ and the same structural measures featured in Fig.~\ref{fig:overview_redux}.}
\end{figure*}

We also found that these conventional structural measures are all somewhat correlated to one another (with cross-correlations in the range 0.2-0.6), capturing slightly different aspects of structural heterogeneity.
The PCA of $X^\mathrm{SBO}$ allows one to capture all these features through a few collective variables, defined on the basis of structural heterogeneity alone.
This suggests that the intrinsic dimension of the SBO feature space is indeed low -- a more precise determination would require additional work~\cite{Mendes-Santos_Turkeshi_Dalmonte_Rodriguez_2021}.
A somewhat striking feature is that, even though the bond-order coefficients do not scale with the number of neighbors [see Eq.\eqref{eqn:bo}], the first PCs are still strongly correlated to the local coordination number, $z$.
Thus, smoothing the coefficients removes the discontinuities in the standard BO descriptor, but does not wash out the correlation with local density evidenced in Fig.~\ref{fig:map_bare}.

To further illustrate the advantage of considering a few PCs instead of selected SBO parameters $\{ Q_l^\mathrm{S} \}$, we computed the same structural indicators as in Fig.~\ref{fig:overview_redux} for the bare SBO descriptor.
These results are shown in Fig.~\ref{fig:overview_bare}, along with the standard deviation of each marginal distribution $p(Q_l^\mathrm{S})$.  They demonstrate that, for any given system, it is difficult to pinpoint a small subset of SBO parameters that capture the bulk of structural heterogeneity.
Instead, all the SBO parameters are to some extent correlated or anti-correlated to measures of local order and to local density.
We note that a trivial source of correlation between $z$ and $X^\mathrm{BO}$ (or $X^\mathrm{SBO}$) arises from the reduction (in magnitude) of $Q_l$ as the number of neighboring particles increases.
This leads to negative correlations between most of the $Q_l$ and $z$.
However, in mixtures with strong icosahedral local order,  $Q_6$ has a non-monotonic dependence with $z$: it peaks at $z=12$ before dropping off at larger $z$.
Thus, $Q_6$ and $Q_l^\mathrm{S}$ display a very weak correlation with $z$.
Finally, we note that in the SiO$_2$ model, the first two PCs are both strongly and positively correlated to $\Theta$.
Indeed, we found a marked gradient of $\Theta$ along each of the bands visible in Fig.~\ref{fig:map_bare}(e) (not shown).

To summarize, PCA applied to the SBO descriptor provides a few collective structural variables that aggregate information on local order.
These variables are simple linear combinations of smooth BO parameters and capture the largest fraction of structural fluctuations.
Our analysis has been restricted to the small particles of the close-packed mixtures, building on the insight that local order is less pronounced around the big particles (see Figs.~\ref{fig:map_red_big} and~\ref{fig:overview_structure_big} in Appendix~\ref{app:sbo} for analogous results for these particles).
For the small particles, the first few PCs are strongly connected to well-known measures of local order and gather a signal that would otherwise be scattered across almost all the individual BO parameters.
Even though some low-variance components may retain relevant information (see also Sec.~\ref{sec:propensity}),
our results suggest that the first PCs suffice to characterize structural heterogeneity in glassy binary mixtures, within the first coordination shell.

\subsection{Fine-grained local structure: using the SOAP descriptor to analyze radial dependence of bond order}\label{sec:radial}

Can we gain additional insight into the local structure by considering a finer-grained expansion of the local density?
The SOAP descriptor provides a natural framework for this extended analysis, since it performs a systematic expansion of the local density up to a distance $r_\mathrm{cut}$ from a central particle.
The cost is a reduced interpretability of the descriptor and a larger hyper-parameter space to explore.
Indeed, structural analysis using the SOAP descriptor must be combined with chemical and physical intuition to grasp the key results~\cite{Offei-Danso_Hassanali_Rodriguez_2022}.
In this section, we summarize the few robust trends we have identified, as well as the shortcomings of this kind of analysis.

The SOAP descriptor depends on two hyper-parameters: the smearing parameter $\sigma$ in Eq.~\eqref{equ:soap-rho}, which sets the length scale on which the particles are localized, and the cutoff distance $r_\mathrm{cut}$, used to define the coordination shells of interest.
In addition, it depends on the number of radial and angular basis functions $n_\mathrm{max}$ and $l_\mathrm{max}$, which control the radial and angular resolution respectively, and on feature scaling.
Unsurprisingly, we found that all these hyper-parameters have a significant impact on the outcome of dimensionality reduction, see also Ref.~\onlinecite{Cheng_et_al_2020} for a discussion on SOAP hyper-parameter tuning.
In this section, we again limit ourselves to the first coordination shell and we set $r_{\alpha\beta}^c = r_\mathrm{cut}=1.6$, slightly larger than the minimum of the total $g(r)$.
This effectively ignores slight differences between coordination shells defined by particles of different species. 
As in the previous sections, we use $l_\textrm{max}=8$ orientational basis functions and varied the largest number of radial basis functions from $n_\mathrm{max}=1$ to $n_\mathrm{max}=8$.
The qualitative features discussed below are rather insensitive to this parameter for large enough $n_\mathrm{max}$.
We use $n_\mathrm{max}=6$ in the plots that follow.
As a guideline, $\sigma$ should be around $\sigma \approx r_\textrm{cut} / n_\textrm{max}$ and, in thermal systems, larger than the typical vibrational amplitude of the particles.
We chose $\sigma=0.2$. Note that perturbations around $\sigma=0.2$ may change the outcomes of dimensionality reduction, we quantify these effects later in this section.

\begin{figure}[!t]
\includegraphics[height=0.65\linewidth]{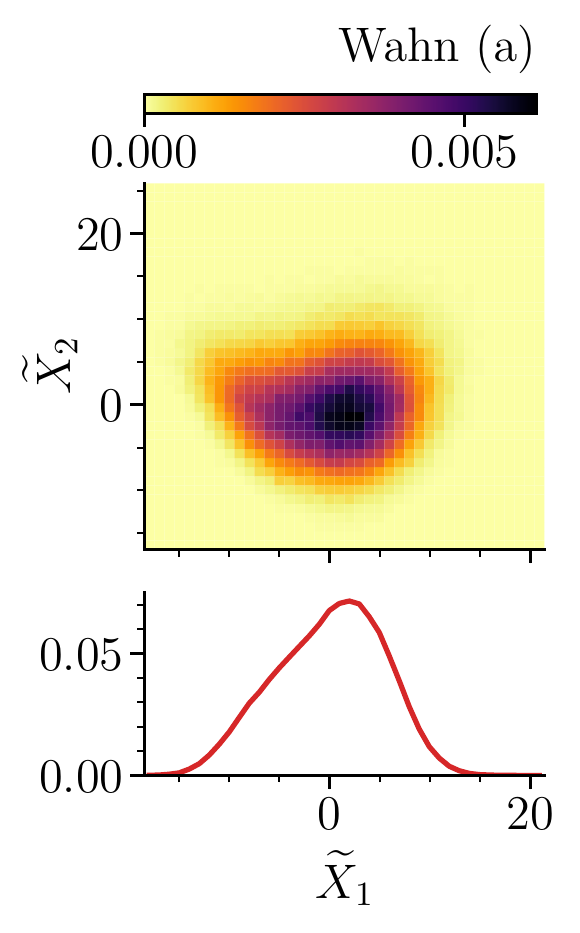}
\includegraphics[height=0.65\linewidth]{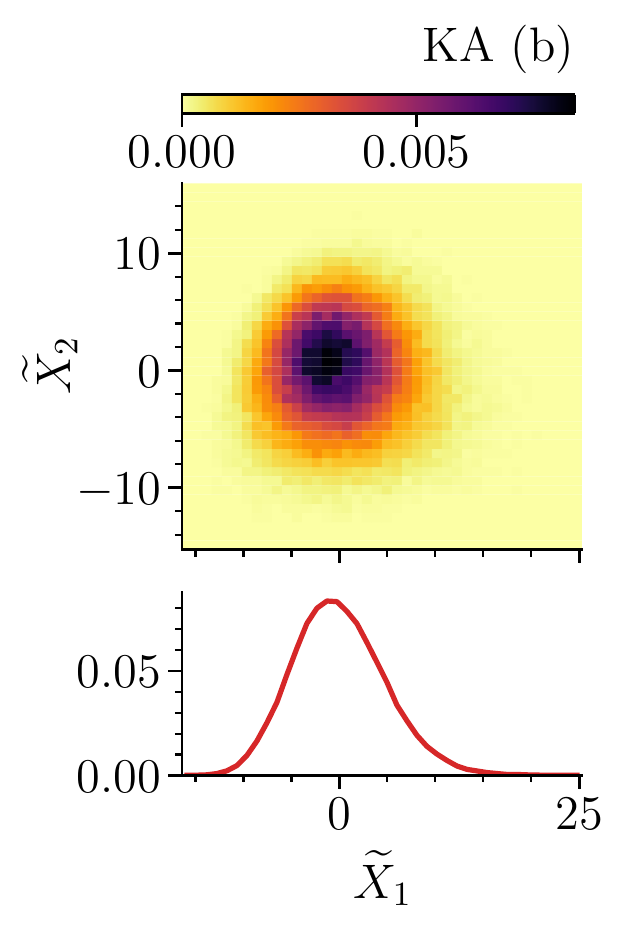}
\caption{\label{fig:map_soap_1}Probability density $p(\X_1, \X_2)$ from the PCA of the SOAP descriptor for the small particles of (a) the Wahn mixture and (b) the KA mixture. The bottom panels show the marginal distribution $p(\X_1)$.}
\end{figure}

\begin{figure}[!t]
\includegraphics[width=0.49\linewidth]{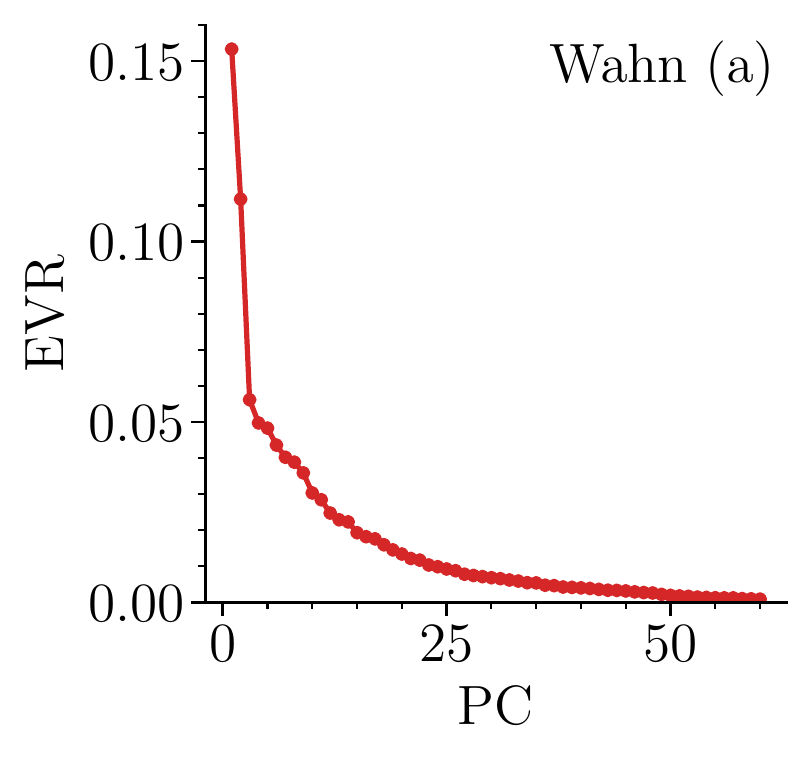}
\includegraphics[width=0.49\linewidth]{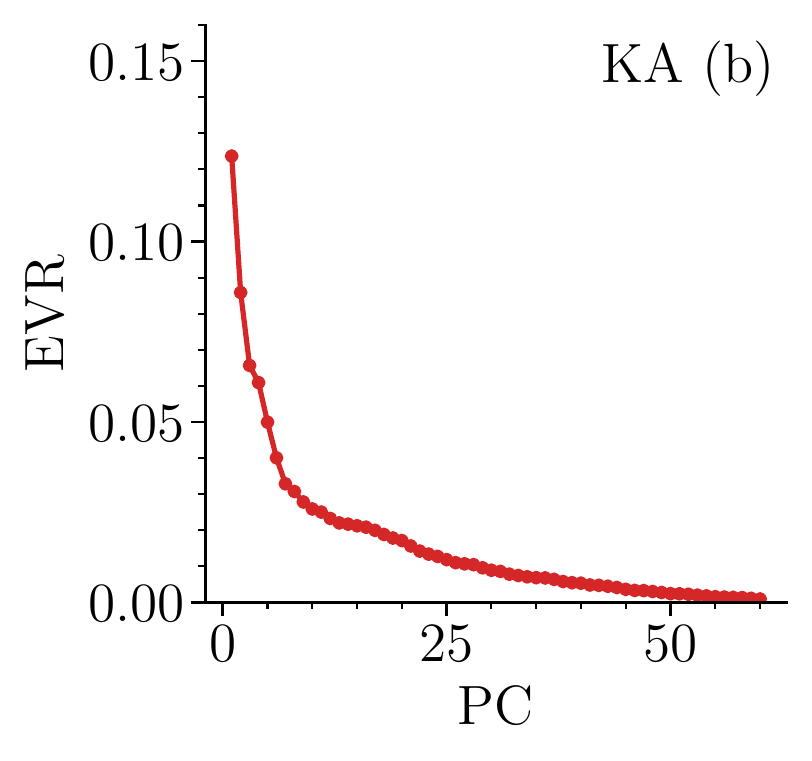}
\caption{\label{fig:evr_soap_1}EVR along the first 60 PC directions of the SOAP descriptor for the small particles of (a) the Wahn mixture and (b) the KA mixture.}
\end{figure}

We consider the KA and Wahn mixtures, which are representative of close-packed mixtures with weak and strong structural heterogeneity, respectively.
For this analysis, we focus on a smaller dataset composed of 10 configurations for each model system.
In Fig.~\ref{fig:map_soap_1} we show the probability density $p(\X_1, \X_2)$ of the first two PCs of the SOAP descriptor for these two systems.
In both cases, feature scaling is applied to adsorb the lack of normalization of the power spectrum\footnote{The range of the components in the reduced SOAP descriptor scales with $n$ [see Eq.~\eqref{eq:soap_tensor}], which creates a trivial increase of the variance across features. The features with the largest variance would thus dominate the PCA, hence the need for feature scaling to bring all the features to the same range.}.
The qualitative features observed in Fig.~\ref{fig:map_soap_1} are similar to those discussed in Sec.~\ref{sec:weak_to_strong}: $p(\X_1, \X_2)$ is nearly bimodal in the Wahn mixture, while it is unimodal in the KA mixture.
Compared to the SBO descriptor, however, the EVR is now distributed more evenly among the PC directions and the first two capture a much smaller fraction of the total variance, see Fig.~\ref{fig:evr_soap_1}.
This may be expected in view of the larger feature space of SOAP\footnote{Note that without feature scaling the PC$_1$ direction gathers instead a \textit{large} variance, which further increases with increasing $\sigma$. Very likely, this reflects the lack of normalization of the descriptor, which gives a strong weight to $Q_0$, \textit{i.e.}, the local coordination number.}.
However, we can still appreciate from Fig.~\ref{fig:evr_soap_1} the presence of gaps between the EVRs of the first two PCs and the bulk in the Wahn mixture: beyond PC$_2$, the shape of the EVR as a function of component index evolves smoothly.
In the KA mixture, a similar change is observed after 4-5 PCs, which still suggest a relatively low intrinsic dimension~\cite{Mendes-Santos_Turkeshi_Dalmonte_Rodriguez_2021}.
However, compared to the low-dimensional SBO descriptor, these gaps are small enough to be perturbed by noise and small changes in the descriptor's hyper-parameters.

\begin{figure}[!t]
\includegraphics[width=0.48\linewidth]{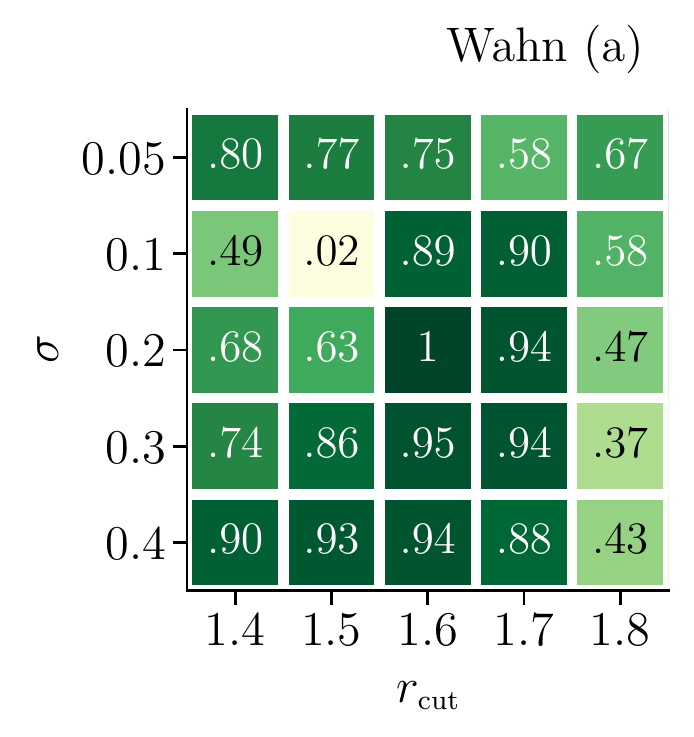}
\includegraphics[width=0.48\linewidth]{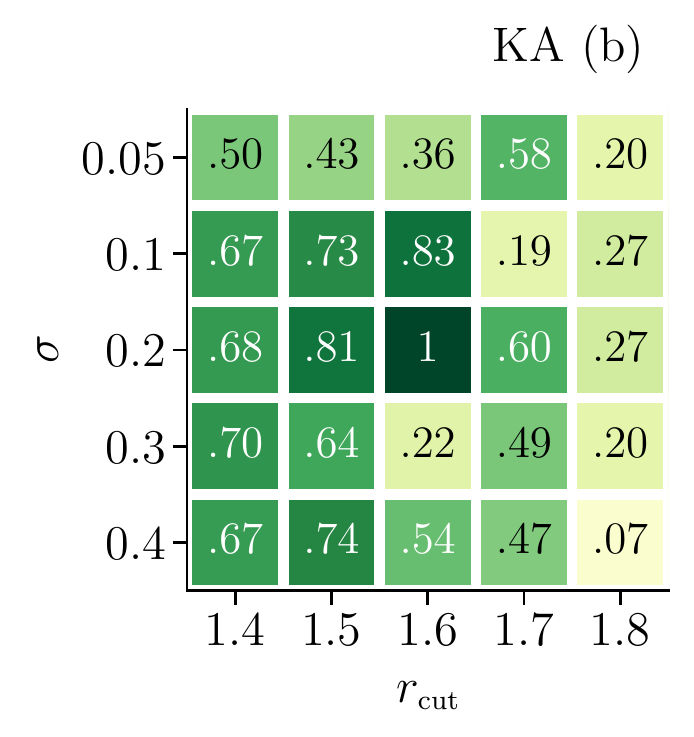}
\caption{\label{fig:sim_soap_1} Pearson correlation coefficient between the feature $\X_1$ of the SOAP descriptor (obtained for $\sigma=0.2$, $r_\mathrm{cut}=1.6$) and $\X_1(\sigma, r_\mathrm{cut})$ (obtained for different values of the parameters $\sigma$ and $r_\mathrm{cut}$, as indicated in the figure). Results are shown for the small particles of (a) the Wahn mixture and (b) the KA mixture.}
\end{figure}

To address this point more systematically, we assessed the similarity between the reductions obtained using different choices of hyper-parameters, $\sigma$ and $r_\textrm{cut}$.
We computed the Pearson correlation coefficient $R$ between the variable $\X_1$ obtained using the reference values, $\sigma=0.2$, $r_\textrm{cut}=1.6$, and the ones obtained for small perturbations of these hyper-parameters.
From the results collected in Fig.~\ref{fig:sim_soap_1}, we observe that the results are relatively stable for the Wahn mixture, as expected, but the outcome of the reduction deviates significantly from the reference when $r_\textrm{cut}$ is increased.
Even more dramatic are the changes observed for the KA mixture: in this case, $\X_1$ may change significantly and erratically as a function of $\sigma$ and $r_\textrm{cut}$.
We attribute part of this effect to the dependence of nearest neighbor distances on the chemical composition of the first coordination shell of the KA model.  It is also a general feature that smaller gaps in EVR between PCs leads to increased sensitivity of PC directions to small perturbations in the parameters, or the data.  This is consistent with the trend between Wahn and   KA mixtures.
The second PC displays an even stronger variability for both models, as expected (not shown).

Note that across the range of hyper-parameters spanned in Fig.~\ref{fig:sim_soap_1}, the distributions $p(\X_1, \X_2)$ maintain the same qualitative differences between the two models observed in Fig.~\ref{fig:map_soap_1}.
However, specific choices of the parameters may enhance or suppress the bimodal character of $p(\X_1, \X_2)$ for the Wahn mixture.
Given the large hyper-parameter space associated to the descriptor, our results call for a principled approach to hyper-parameter tuning for unsupervised learning of glassy materials, where the relevant length scales cannot be determined in a simple way~\cite{Cheng_et_al_2020}.

We note one possible shortcoming of the radial basis for the present analysis: as is clear from Fig.~3 of Ref.~\onlinecite{bartok_on-representing_2013}, the first radial functions are non-zero over a range of distances that are depleted in our range of temperatures and they vary only mildly over the first coordination shell\footnote{The closest approach distance in our mixtures is typically half of the cut-off distance $r_\textrm{cut}$}.
This basis may thus require a large value of $n_\textrm{max}$ to properly account for correlations within the first coordination shell.
To test whether this could be an issue, we also used a different basis formed by Gaussian-type orbitals~\cite{Jager_Morooka_Federici_Canova_Himanen_Foster_2018} which effectively suppress contributions from close-contact distances.
We found qualitatively similar results to those already discussed in this section.
An alternative route to provide a fine-grained description of correlations within the first shell is to use the RBO descriptor, Eq.~\eqref{eq:rbo}, which we tailored to cover the the first coordination shell using $n_\textrm{max}=5$ Gaussians of width $\delta=0.2$ centered at $(0.9, 1.0, 1.1, 1.2, 1.3)$, and $l_\textrm{max}=8$.
This choice of hyper-parameters leads to PCA maps that are again qualitatively similar to those of Fig.~\ref{fig:map_red}(a,c) and \ref{fig:map_soap_1}, but the underlying PC directions are again sensitive to addition or removal of one Gaussian (not shown here).

To sum up, SOAP has proven to be a good descriptor for supervised and unsupervised learning studies of materials and molecules~\cite{Cheng_et_al_2020}, in particular for fitting macroscopic properties and potential energy surfaces~\cite{Ceriotti2018}.
As we shall see in the next section, the SOAP descriptor also provides a good fit to dynamic heterogeneity, from purely structural information.
Its application in the context of unsupervised learning of disordered materials is more recent and focused so far on systems with highly directional correlations, such as amorphous carbon~\cite{Deringer_Caro_2018, Cheng_et_al_2020}, or normal and supercooled liquid water~\cite{Monserrat_Brandenburg_Engel_Cheng_2020,Offei-Danso_Hassanali_Rodriguez_2022}.
Even in such systems with low coordination numbers and sharp orientational correlations, interpreting the results requires some physical and chemical intuition.
In our opinion, the utility of this descriptor for unsupervised learning of the structure of glassy materials with close-packed local structures, like metallic glasses, remains to be clarified.
Our results indicate that, for the first coordination shell of these systems, the fine radial basis of SOAP does not reveal qualitatively new geometrical features, compared with the much simpler SBO descriptor.
We also found that even the first few PC directions are sensitive to the choice of hyper-parameters and also to thermal noise in the particle configurations\footnote{Namely, we found that energy-minimized configurations provide somewhat more robust results upon changes of hyper-parameters.}.
We expect that studying the mixtures at much lower temperature and combining separate descriptors for neighbors of different species, as done in~\cite{Offei-Danso_Hassanali_Rodriguez_2022}, will provide more insight into this issue.

\section{Correlations of structure with dynamics}
\label{sec:dynamics}

Having identified a few collective variables that capture structural heterogeneity, one obvious question is whether they are related to \textit{dynamic} heterogeneity.
To address this point, we follow the usual approach of measuring the particles' square displacements in the iso-configurational ensemble~\cite{widmer-cooperHowReproducibleAre2004}.
Given an initial equilibrium configuration sampled at a temperature $T$, the propensity for motion of particle $i$ after a time interval $t$ is 
$$
\mu_i(t)=\langle|\bvec{r}_i(t) - \bvec{r}_i(0)|^2\rangle_{ic} \, ,
$$
where $\langle \dots \rangle_{ic}$ denotes an average over an ensemble of independent trajectories that start from the initial positions $\{\bvec{r}_i(0)\}_{i=1,\dots,N}$ with initial velocities drawn from the Maxwell-Boltzmann distribution at temperature $T$.  
Hereafter, we refer to the propensity for motion simply as the propensity.
Data for the propensity in this section were obtained in Ref.~\onlinecite{paret_assessing_2020}, to which we direct the reader for further details.

Spatial fluctuations of the propensity are directly connected to the structure of the initial configuration.
Thus, by construction, the propensity captures the structural component of dynamic heterogeneities~\cite{berthierStructureDynamicsGlass2007}.
The key issue is then to identify \textit{which aspects} of structure correlate to the propensity.
This challenge is closely related to the task of predicting localized plastic events in sheared amorphous solids, using only structural data~\cite{Richard_Ozawa_et_al_2020}.

\subsection{On the relationship between structural and dynamic heterogeneity: insights from PCA}\label{sec:propensity}

To assess this, we computed the absolute value of the Pearson correlation coefficient between the propensity and each of the reduced features obtained from PCA.
Our analysis focuses on two reference models, the KA and Wahn mixtures.
Since the small and big particles of these systems have quite different dynamics, the calculation is carried out separately for each species, yielding distinct correlation coefficients $R_A$ and $R_B$ for big and small particles, respectively.
We emphasize that this correlation is computed at the single-particle level as a correlation between $\mu_i$ and $\widetilde{X}_j$, contrary to the majority of recent studies on this topic where the structural descriptor is averaged over some length scale~\cite{hocky_correlation_2014,tongRevealingHiddenStructural2018,Tong_Tanaka_2019,paret_assessing_2020,boattini_averaging_2021}.  Since there is no physically-motivated choice for this length scale yet, our approach allows a simple comparison between different structural observables, even though larger absolute values for the correlation coefficient would be obtained if local averaging were performed. This increase is particularly strong at long times, due to a coupling with the local density~\cite{jackInformationTheoreticMeasurementsCoupling2014}.

\begin{figure}[!t]
  \begin{tabular}{cc}
% \includegraphics[width=.52\linewidth]{Wahn/T0.5800/all/config.lammps.xyz.propensity.sbo.scaling-False.redux-True.fit-linear_regression.species-B.propensity-pcs.pdf} &
% \hspace{-1em}\includegraphics[width=.52\linewidth]{KA/T0.4500/all/config.lammps.xyz.propensity.sbo.scaling-False.redux-True.fit-linear_regression.species-B.propensity-pcs.pdf} \\[-.5em]
% \includegraphics[width=.52\linewidth,clip,trim=0 0 0 4.5em]{Wahn/T0.5800/all/config.lammps.xyz.propensity.sbo.scaling-False.redux-True.fit-linear_regression.species-A.propensity-pcs.pdf} &
% \hspace{-1em}\includegraphics[width=.52\linewidth,clip,trim=0 0 0 4.5em]{KA/T0.4500/all/config.lammps.xyz.propensity.sbo.scaling-False.redux-True.fit-linear_regression.species-A.propensity-pcs.pdf}
\includegraphics[width=.52\linewidth]{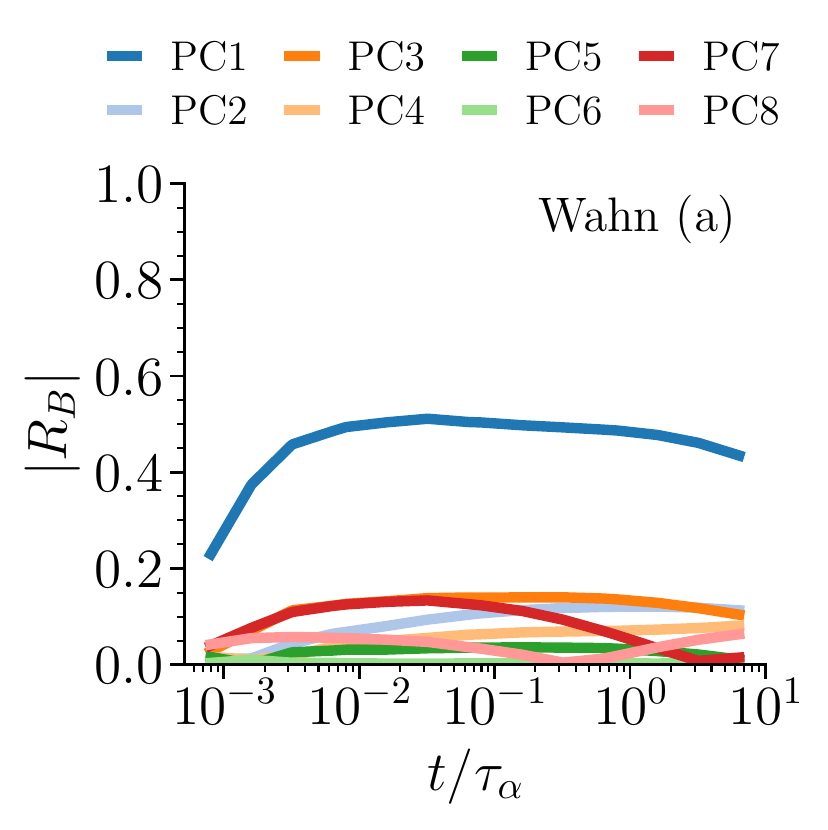} &
\hspace{-1em}\includegraphics[width=.52\linewidth]{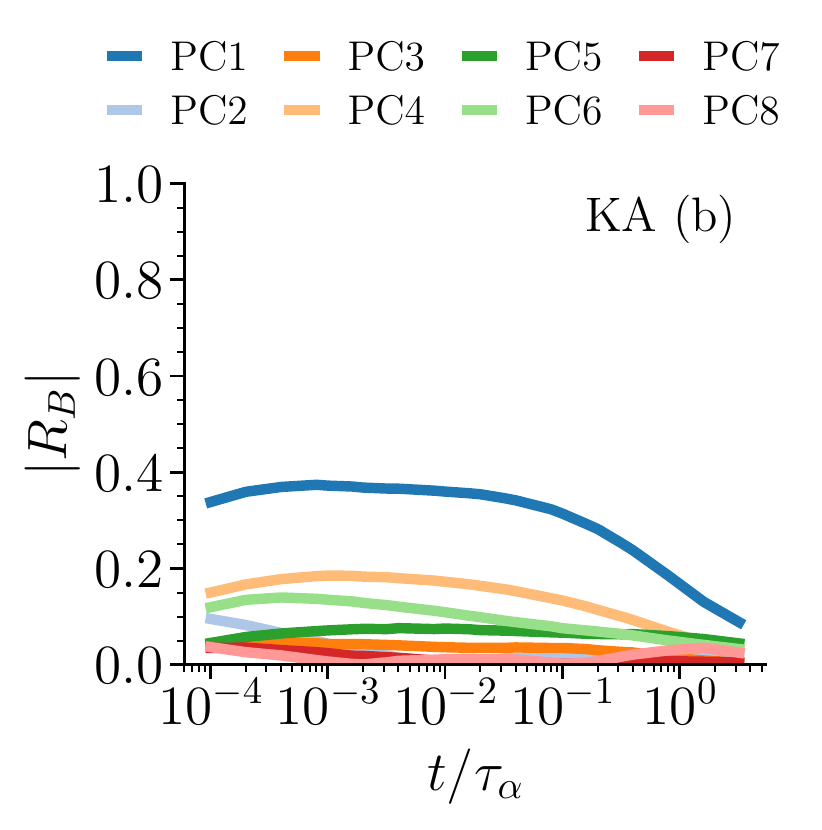} \\[-.5em]
\includegraphics[width=.52\linewidth,clip,trim=0 0 0 4.5em]{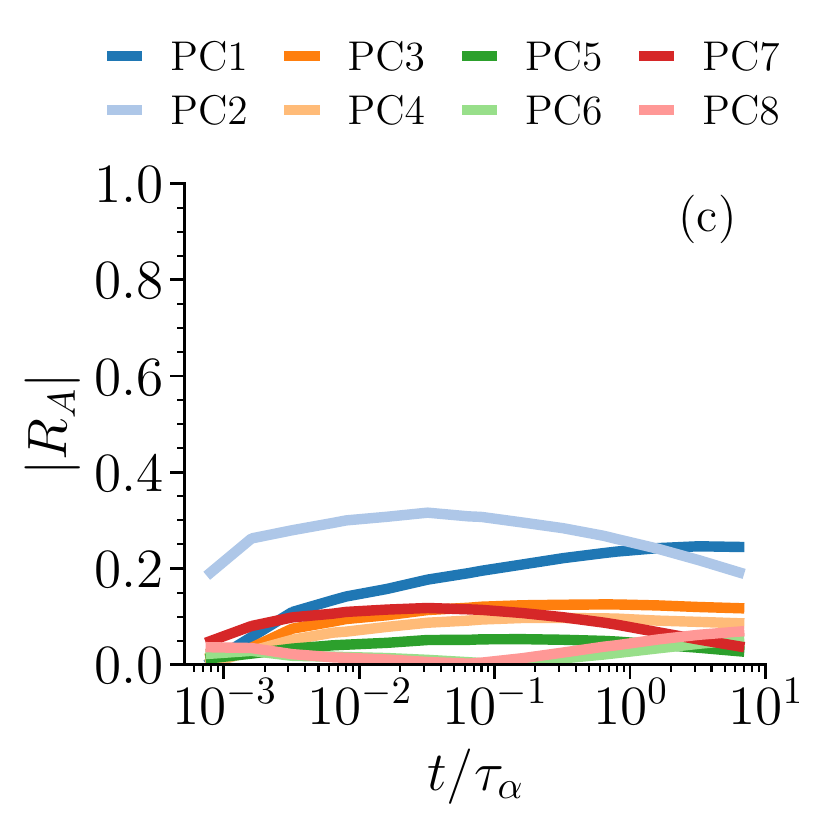} &
\hspace{-1em}\includegraphics[width=.52\linewidth,clip,trim=0 0 0 4.5em]{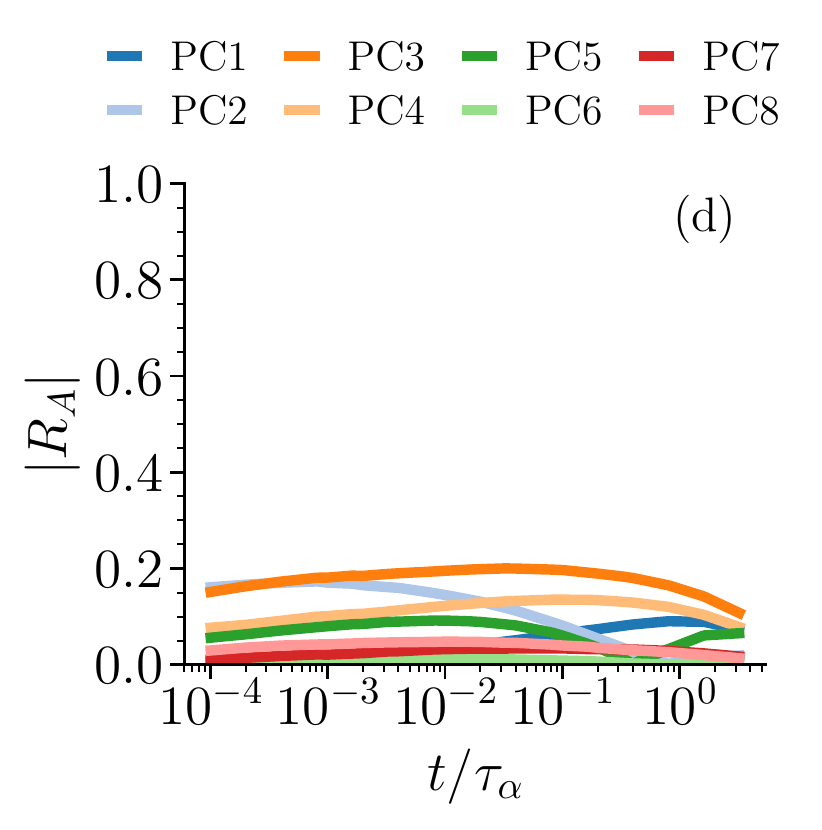}
  \end{tabular}
  \caption{\label{fig:K_1}Pearson correlation coefficients between the propensity at time $t$ and each of the reduced feature $\X_j$ obtained from the PCA of the SBO descriptor, for the Wahn mixture [(a) and (c)] and the KA mixture [(b) and (d)]. Results are shown separately for the small particles [$|R_B|$, (a) and (b)] and the big particles [$|R_A|$, (c) and (d)].}
\end{figure}

We start by analyzing the small particles, on which our analysis has focused so far.
In the top panels of Fig.~\ref{fig:K_1}, we show the absolute values of the correlation coefficient $|R_B|$ between the propensity at time $t$ and each of the reduced features $\widetilde{X}_j$ obtained from the PCA of the SBO descriptor.
The first PC gives the strongest correlation. 
For both models, the correlation displays a flat maximum and starts decaying around the structural relaxation time $\tau_\alpha$, which is measured from the decay of the total self-intermediate scattering function~\cite{paret_assessing_2020}.
The values of $\tau_\alpha$ are $1.6\times 10^2$ and $9\times 10^2$ for the Wahn and KA mixtures, respectively.
The maximum of the correlation coefficient is around 0.5 in in the Wahn mixture.  
For the KA mixture, the first PC still carries significant information about the dynamics, but the correlation is weaker (max. $\approx 0.35$).   
In both models, these results show that a single structural variable, $\widetilde{X}_1$, obtained without any supervision, has a significant correlation with the dynamics of the small particles.  
Comparing the two models, correlations between structure and dynamics are stronger in the Wahn mixture, consistent with previous work~\cite{hocky_correlation_2014,paret_assessing_2020}.
For the higher PCs, individual correlations with dynamics are weaker than PC$_1$.

Corresponding results for big particles are shown in the bottom panels of Fig.~\ref{fig:K_1}.
Compared to the small particles, the correlations are much weaker, and there is no clear separation between the first PC and the rest.
This points to a more complicated relationship between structure and dynamics, in this case.

\subsection{Multiple linear regression of the propensity of motion using structural data}\label{sec:propensity_fit}

To characterize more complex structure-dynamics relationships, one must go beyond correlations between individual PCs and propensity.
Supervised learning methods provide a data-driven route to identify such relationships in supercooled liquids and they have been used quite intensively in recent years~\cite{Cubuk_Schoenholz_Rieser_Malone_Rottler_Durian_Kaxiras_Liu_2015, Schoenholz_Cubuk_Sussman_Kaxiras_Liu_2016, bapst_unveiling_2020, boattini_averaging_2021, Alkemade_Boattini_Filion_Smallenburg_2022}.
In the absence of a physically motivated framework for such an analysis, we follow the simplest approach and perform a multiple linear regression (LR) to connect the propensity to structural data~\cite{boattini_averaging_2021}.

Consider a generic set of $P$ structural features: $X_1, \dots, X_P$.
We least-square fitted the propensity after time $t$ to a linear combination of structural features 
by defining, for each particle $i$,
\begin{equation}
\mu^X_i(t; \beta_1, \dots, \beta_P) = \beta_1 X_1(i) + \dots + \beta_P X_P(i) ,
\end{equation}
where $\beta_1,\dots,\beta_P$ are fitting parameters.
The optimal parameters $\hat{\beta}_1, \dots, \hat{\beta}_P$ are determined by minimizing the cost function
\begin{equation}
\chi = \frac{1}{N_\alpha} \sum_{i=1}^{N_\alpha} |\mu^X_i(t; \beta_1, \dots, \beta_P) - \mu_i(t)|^2, 
\end{equation}
where the sum is restricted to all the particles of species $\alpha$ in the sample.
The fit was performed using the \texttt{LinearRegression} function of the \texttt{scikit-learn} package~\cite{scikit-learn}.
We also tested the effect of weight regularization in the cost function (the Ridge method~\cite{Hastie_Tibshirani_Friedman_2016}), as done in Ref.~\onlinecite{boattini_averaging_2021}, but this did not lead to any significant change in the results.

\begin{figure}[!t]
  \begin{tabular}{cc}
% \includegraphics[width=.52\linewidth]{Wahn/T0.5800/all/config.lammps.xyz.propensity.fit-linear_regression.species-B.propensity-fit-struct.pdf} &
% \hspace{-1em}\includegraphics[width=.52\linewidth]{KA/T0.4500/all/config.lammps.xyz.propensity.fit-linear_regression.species-B.propensity-fit-struct.pdf} \\[-.5em]
% \includegraphics[width=.52\linewidth,clip,trim=0 0 0 4em]{Wahn/T0.5800/all/config.lammps.xyz.propensity.fit-linear_regression.species-A.propensity-fit-struct.pdf} &
% \hspace{-1em}\includegraphics[width=.52\linewidth,clip,trim=0 0 0 4em]{KA/T0.4500/all/config.lammps.xyz.propensity.fit-linear_regression.species-A.propensity-fit-struct.pdf}
\includegraphics[width=.52\linewidth]{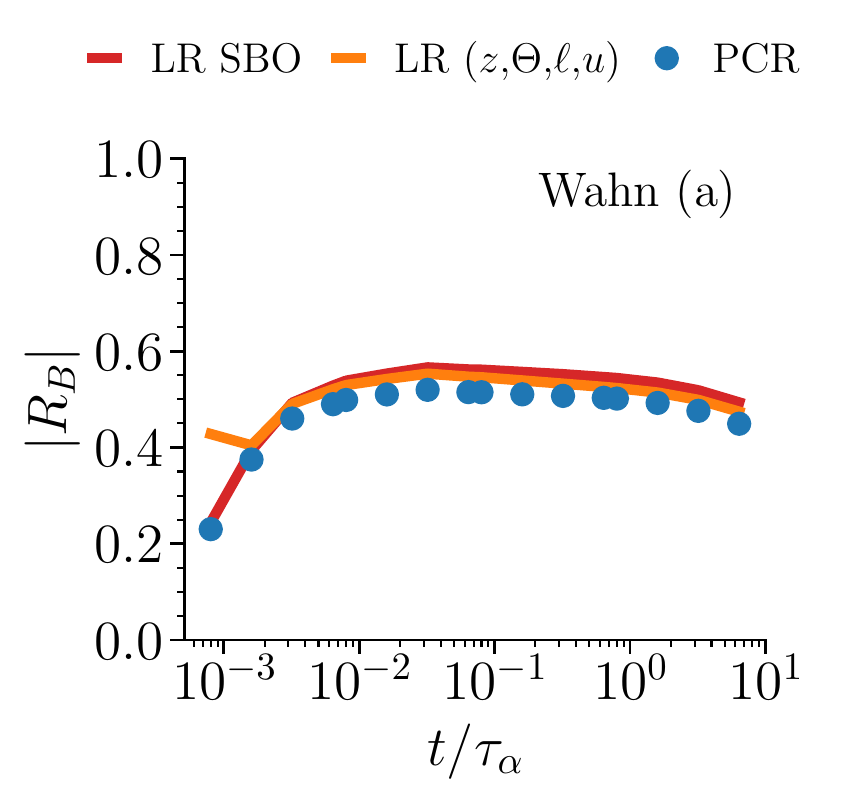} &
\hspace{-1em}\includegraphics[width=.52\linewidth]{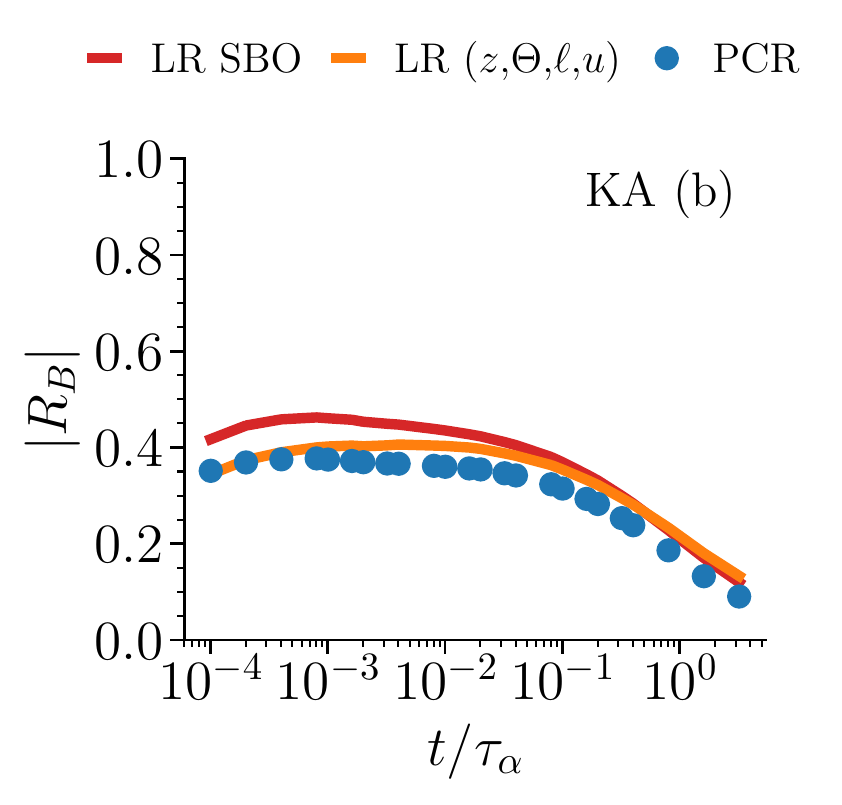} \\[-.5em]
\includegraphics[width=.52\linewidth,clip,trim=0 0 0 4em]{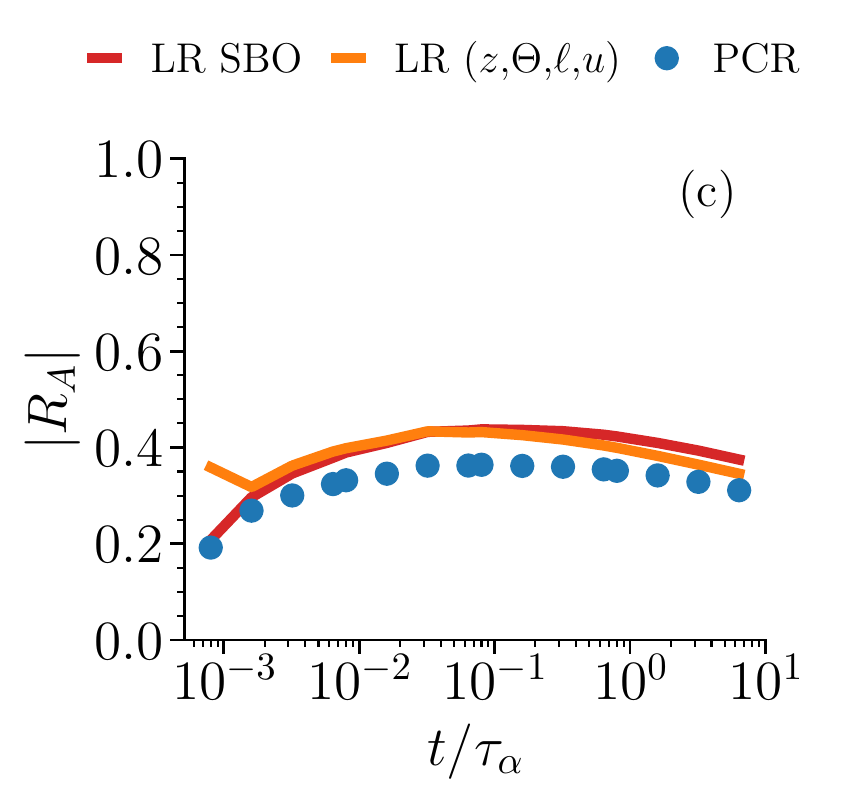} &
\hspace{-1em}\includegraphics[width=.52\linewidth,clip,trim=0 0 0 4em]{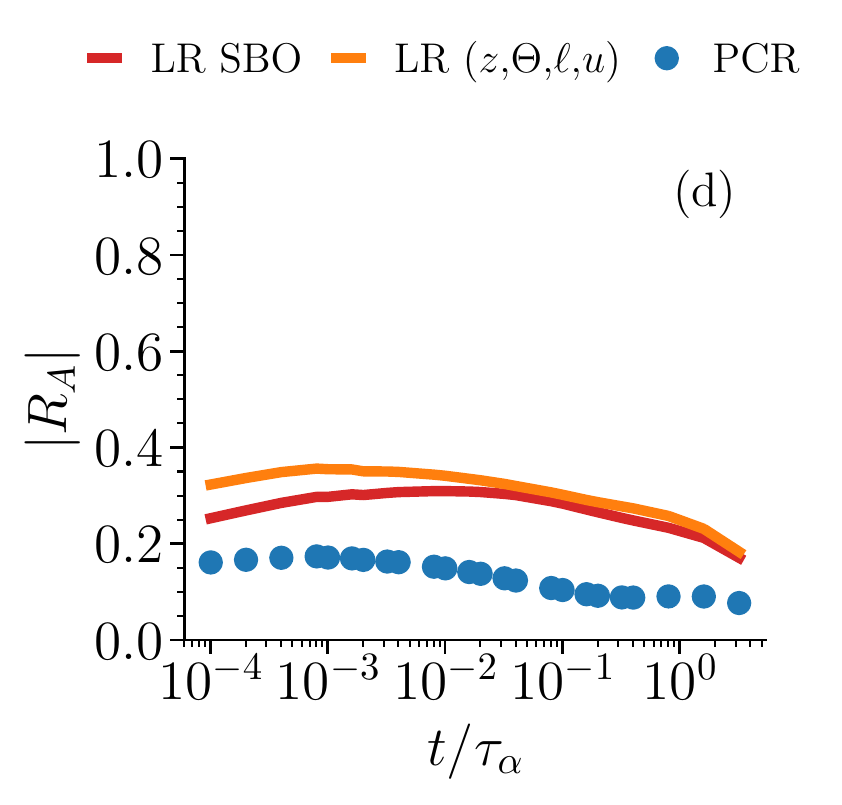}
  \end{tabular}
  \caption{\label{fig:K_3} Pearson correlation coefficients between the propensity at time $t$ and a multiple linear regression of different structural measures, for the Wahn mixture [(a) and (c)] and the KA mixture [(b) and (d)]. Results are shown separately for the small particles [$|R_B|$, (a) and (b)] and the big particles [$|R_A|$, (c) and (d)]. The three sets of features used for fitting are described in the main text.}
\end{figure}

For each particle $i$, the fit yields an interpolated value of the propensity $\hat{\mu}_i = \mu^X_i(t; \hat{\beta}_1, \dots, \hat{\beta}_P)$,
which can be interpreted as a data-driven prediction for the propensity, based on the structure.
To assess the quality of these predictions,
we use Pearson correlation coefficients $R$ between the interpolated and the actual propensity, which can be obtained by normalizing and shifting the optimal cost $\chi$.

We performed this test for three different structural descriptors.
As a reference, we first consider the full SBO descriptor: $X_j = X_j^{\rm SBO}$ with $j=1,\dots, 8$.
Second, we consider principal component regression (PCR)~\cite{Hastie_Tibshirani_Friedman_2016}, in which we fit a linear combination of $\widetilde{X}_1^{\rm SBO}$ and $\widetilde{X}_2^{\rm SBO}$.
Third, we considered the physically-motivated structural measures introduced in Sec.~\ref{sec:pc_vs_ql}: $X_1=z$, $X_2=\ell$, $X_3=\Theta$, $X_4=u$ (coordination, LFS membership, tetrahedrality parameter, local potential energy).

Results are shown in Fig.~\ref{fig:K_3}.  For small particles, very similar results are obtained for all three descriptors.
We conclude that the two most relevant PCs capture almost all of the correlations between the SBO descriptor and the dynamics, and that
these descriptors predict the dynamics just as well as the structural measurements.
That is, two PCs are enough, in these cases, to capture the bulk of the correlations.
For big particles (lower panels of Fig.~\ref{fig:K_3}), the correlations are weaker, as observed for individual PCs in Fig.~\ref{fig:K_1}.
There are larger differences between the LR results for full SBO and the PCR, indicating that the structure-dynamics correlation is spread over more than two PCs.  For the Wahn mixture, the SBO correlation is similar to that of the physically-motivated structural correlators, while the KA mixture shows that the SBO correlation is somewhat weaker.  In this latter case, it appears that bond order misses some aspects of local structure that are important for dynamics.

Overall, the picture that emerges from Figs.~\ref{fig:K_1} and~\ref{fig:K_3} is that the small-particle dynamics have significant correlations with local bond order.  Hence, this aspect of structure has significant predictive power for dynamics, which can be captured through one or two PCs.   For the big particles, the correlations are weaker, and are spread over more PCs.  In all cases, the correlations for the KA  mixture are weaker than those of the  Wahn mixture, consistent with earlier work~\cite{hocky_correlation_2014}.

\begin{figure}[!t]
  \begin{tabular}{cc}
% \includegraphics[width=.5\linewidth]{Wahn/T0.5800/all/config.lammps.xyz.propensity.fit-linear_regression.species-B.propensity-fit.pdf} &
% \hspace{-1em}\includegraphics[width=.52\linewidth]{KA/T0.4500/all/config.lammps.xyz.propensity.fit-linear_regression.species-B.propensity-fit.pdf} \\[-.5em]
% \includegraphics[width=.5\linewidth,clip,trim=0 0 0 5em]{Wahn/T0.5800/all/config.lammps.xyz.propensity.fit-linear_regression.species-A.propensity-fit.pdf} &
% \hspace{-1em}\includegraphics[width=.5\linewidth,clip,trim=0 0 0 5em]{KA/T0.4500/all/config.lammps.xyz.propensity.fit-linear_regression.species-A.propensity-fit.pdf}
\includegraphics[width=.52\linewidth]{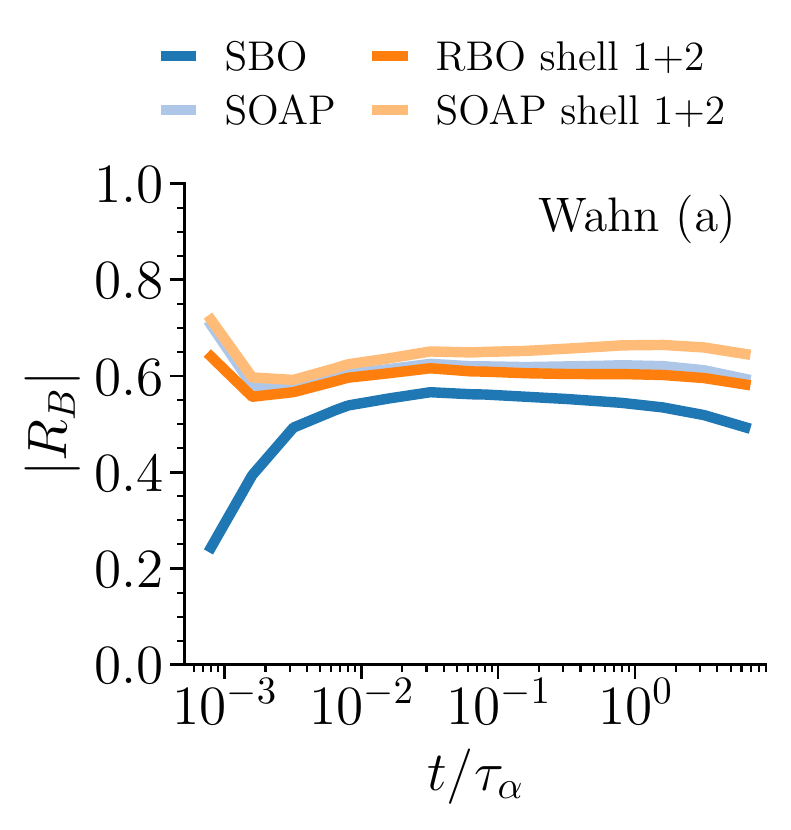} &
\hspace{-1em}\includegraphics[width=.52\linewidth]{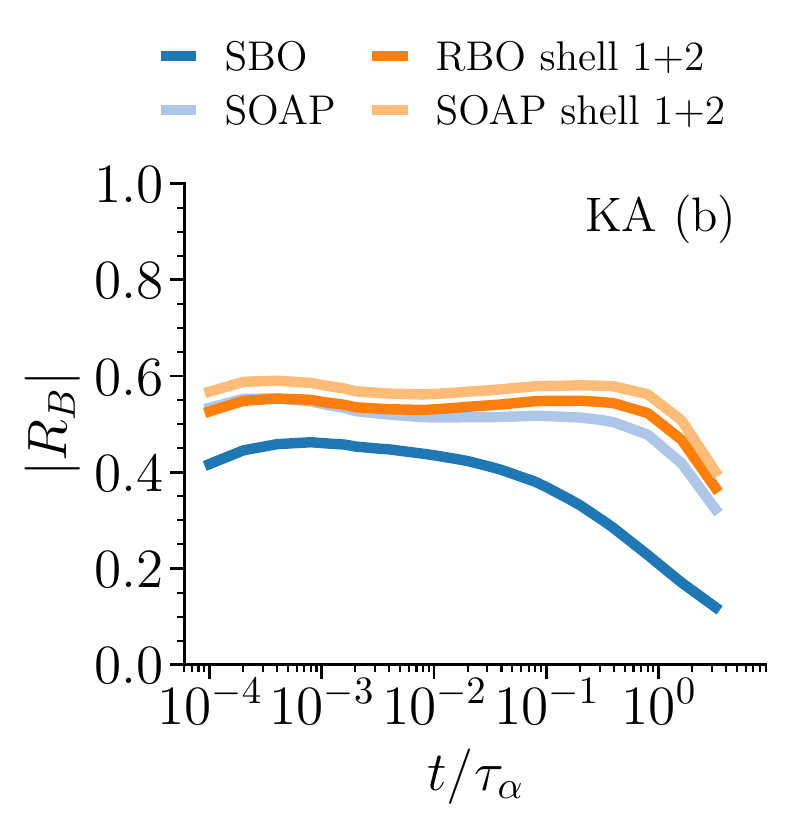} \\[-.5em]
\includegraphics[width=.52\linewidth,clip,trim=0 0 0 5em]{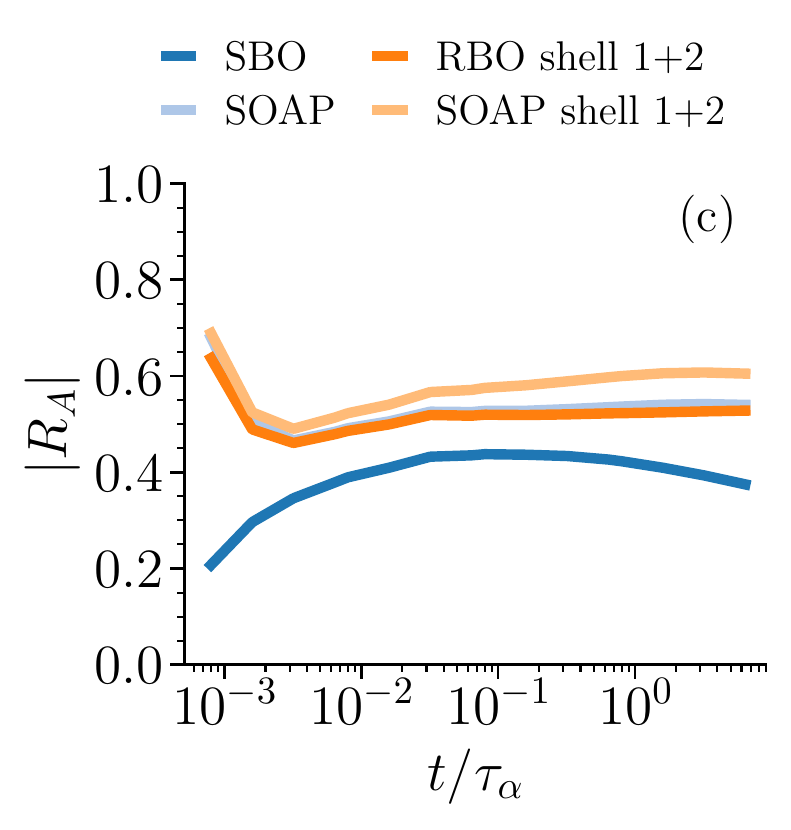} &
\hspace{-1em}\includegraphics[width=.52\linewidth,clip,trim=0 0 0 5em]{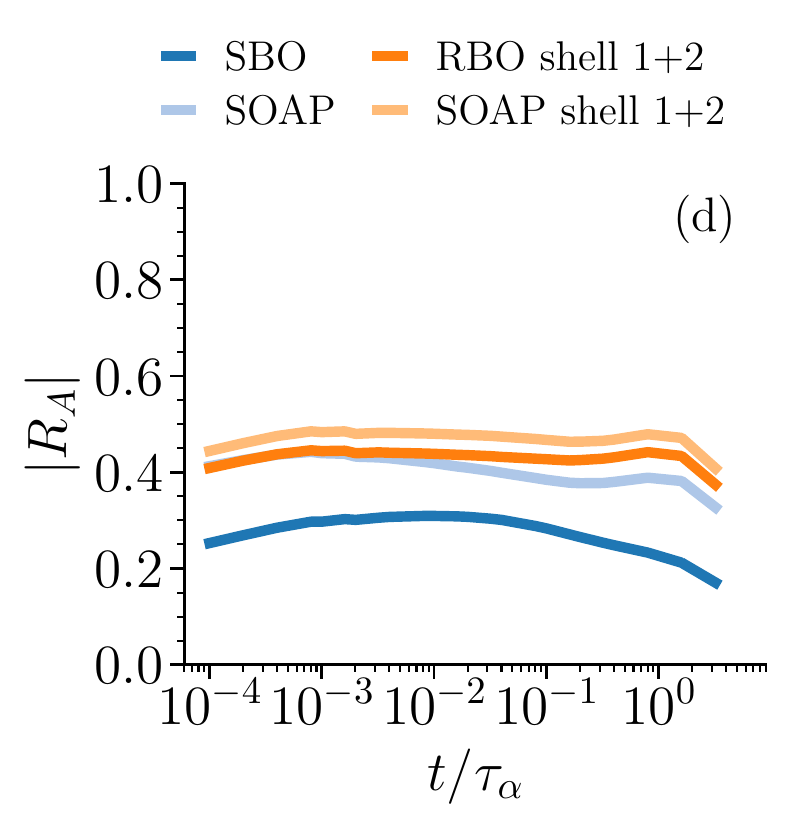}
  \end{tabular}
  \caption{\label{fig:K_4}
Pearson correlation coefficients between the propensity at time $t$ and the linear regression of SOAP (blue) and RBO (red) for the Wahn mixture [(a) and (c)] and the KA mixture [(b) and (d)]. Results are shown separately for the small particles [$|R_B|$, (a) and (b)] and the big particles [$|R_A|$, (c) and (d)]. The light colors are for the first coordination shell only, the dark ones up to the second shell.}
\end{figure}

\subsection{Fitting the propensity of motion using extended structural descriptors: SOAP and RBO}\label{sec:propensity_fit_soap_rbo}

We now resolve the local structure in more detail, using the extended descriptors defined in Sec.~\ref{sec:soap} and~\ref{sec:rbo}.
Adding more structural information in this way will reduce the residual cost $\chi$, corresponding to an improved fit of the dynamics.
We start with the SOAP descriptor focusing on the first coordination shell, choosing $r_\mathrm{cut}$ close to the first minimum of the total $g(r)$ and using $n_\mathrm{max}=6$ radial components, giving 189 independent structural features in total.
Here we only fit the full descriptor, without any dimensionality reduction, because we found that the results of PCA for this descriptor are sensitive to the choice of the hyper-parameters and are anyway difficult to interpret.
We do not expect any overfitting within our simple linear regression model, as the number of datapoints greatly exceeds the number of parameters to fit.

From Fig.~\ref{fig:K_4} we see that the SOAP descriptor does indeed lead to stronger correlations with the dynamics, compared to the SBO descriptor.  This is especially true for short times in the Wahn mixture and longer times in the KA mixture.
The values of the correlation coefficients are also in line with those obtained in supervised learning studies~\cite{Alkemade_Boattini_Filion_Smallenburg_2022}, using support vector machine~\cite{Schoenholz_Cubuk_Sussman_Kaxiras_Liu_2016}, graph neural networks~\cite{bapst_unveiling_2020}, or linear regression of the coarse-grained RBO descriptor with an L2 regularization~\cite{boattini_averaging_2021}.
The improvement over the SBO descriptor, however, comes at the expense of a lack of interpretability of the results: SOAP involves a much larger number of features and it is difficult to understand which are responsible for the correlation with the dynamics.
At any rate, these results indicate that subtle features of the local density within the first coordination shell are responsible for a part of the dynamic fluctuations within the iso-configurational ensemble.

Finally, to capture structural correlations beyond the first coordination shell, we employed the SOAP and RBO descriptors --see Eq.~\eqref{eq:rbo}-- using a larger cutoff distance $r_\mathrm{cut}$.
For the SOAP descriptor, we used $r_\mathrm{cut}=2.2$ and $n_\mathrm{max}=8$, to maintain approximately the same radial resolution, yielding 324 structural features.
For the RBO descriptor, we used Gaussians of width 0.2, centered on a grid $\{ d_n \}$ from $r=0.9$ to 2.2 with spacing 0.1, yielding 104 structural features.
From Fig.~\ref{fig:K_4} we see that extending the descriptors to include the second coordination shell increases slightly the correlations with the propensity.
We do not observe significant differences in the correlations obtained using the SOAP and RBO either, the former leading to marginally higher correlations than the latter.
This indicates that the additional information retained by the SOAP descriptor, which accounts for couplings between different radial orders in the power spectrum (see Sec.~\ref{sec:soap}), is not useful for fitting the propensity.
The simpler RBO descriptor seems to retain all the necessary information that can be exploited to fit the dynamics within a simple linear regression model.
Overall, our results show that including structural correlation beyond the first coordination shell increases the correlation with the dynamics only slightly, and the origin of the increase is difficult to pinpoint.

\section{Comparison of dimensionality reduction by PCA and neural network AE}
\label{sec:pca_vs_ae}

Our analysis so far used PCA, which is possibly the simplest dimensionality reduction method.
As emphasized in Sec.~\ref{sec:dim-redux}, non-linear reduction methods provide superior performance when it comes to mapping complex high-dimensional data on a low-dimensional manifold.
However, this is often achieved at the cost of a lack of interpretability or the need to fine-tune a large number of free parameters.
For instance, the exploration of parameter space and the training process of neural network AEs can quickly become slow and tedious, and the results may not easily generalize.
Moreover, while the explainability of machine learning methods recently started to emerge as a separate research field to tackle the problem of the interpretation of their results~\cite{zdeborova_understanding_2020,linardatos_explainable_2021}, neural networks are still generally regarded as black boxes.

\begin{figure}[!t]
  \includegraphics[width=1\linewidth]{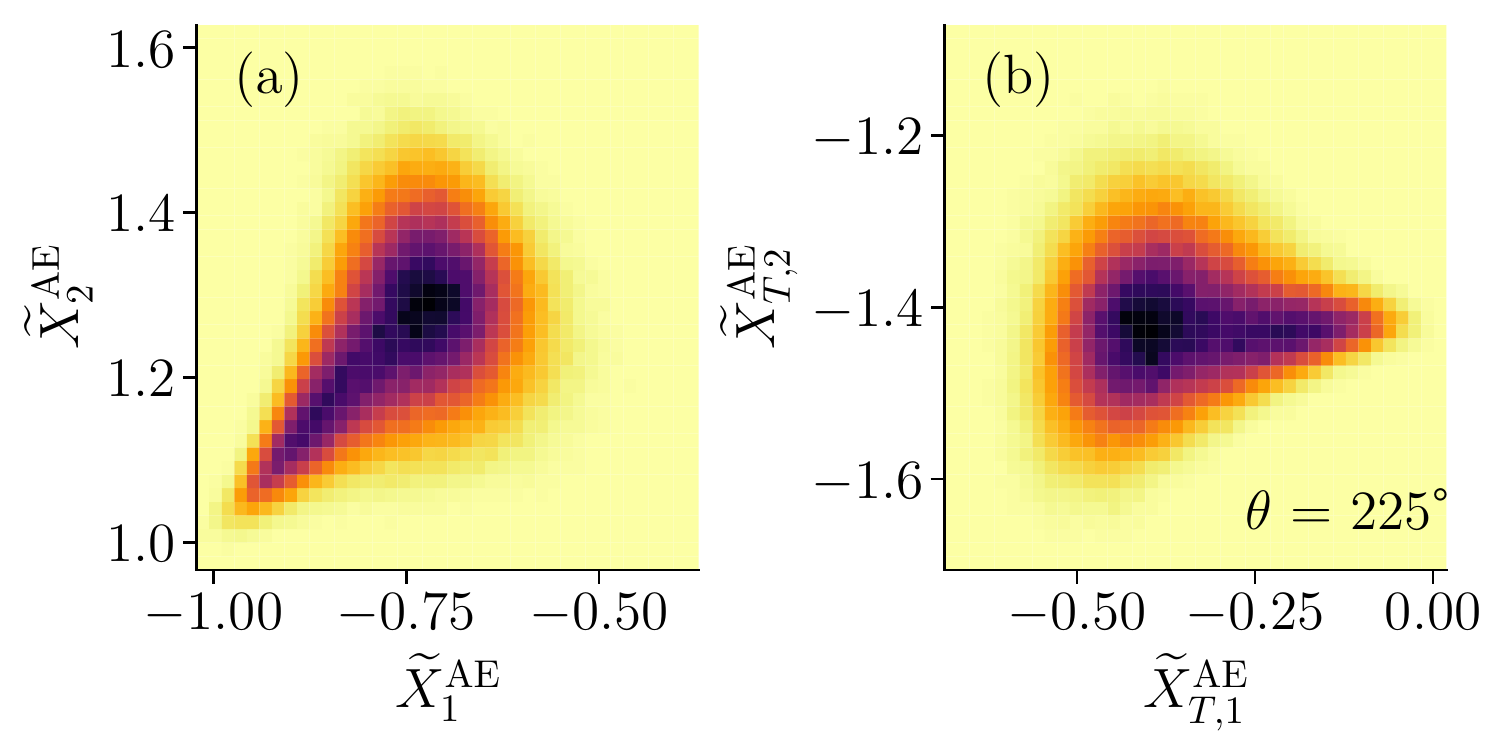}
\caption{(a) Probability density $p(\widetilde{X}_1^\mathrm{AE}, \widetilde{X}_2^\mathrm{AE})$ from a representative solution obtained by the AE with the SBO descriptor in the Wahn model (small particles). (b) Probability density $p(\widetilde{X}_{T,1}^\mathrm{AE}, \widetilde{X}_{T,2}^\mathrm{AE})$ after a rotation of the reduced descriptor by an angle $\theta=225^\circ$, which maximizes $S_R$.}
\label{fig:AE_rotated}
\end{figure}

In this section, we directly compare the results obtained using PCA and the neural network AE, using the SBO descriptor for both the Wahn and KA mixtures.
Our model for the AE was developed using \texttt{TensorFlow}~\cite{tensorflow2015-whitepaper}, following Boattini\textit{ et al.}~\cite{boattini_autonomously_2020} for the choice of free parameters, since the descriptor and the studied physical systems are similar.
Namely, the neural network is composed of three hidden layers with dimensions $(5M, P, 5M)$, where $M$ is the number of features in the SBO descriptor and $P$ is the dimension of the bottleneck that we set to 2.
The input and output layers thus have a dimension of $M$.
Activation functions are the hyperbolic tangent for the encoding and decoding layers, and a linear function for the bottleneck and output layers.
We use mini-batch stochastic gradient descent with momentum, with a batch size of 200 and a momentum of 0.2, and the learning rate is set to $\alpha = 5 \times 10^{-2}$.
We use the mean-squared error as loss function, with an L2 penalty term $\lambda = 10^{-5}$ on the weights for regularization.
Weights and biases are initialized using the Glorot method~\cite{pmlr-v9-glorot10a}.
The network is then trained for 250 iterations (``epochs'' in the data science jargon) for the Wahn mixture and 500 iterations for the KA mixture, on 70\% of the dataset, while the remaining 30\% are used to evaluate the loss function at the end of each iteration.
See Ref.~\onlinecite{zenodo} for full details about the parametrization of the network.

\begin{figure}[!t]
  \includegraphics[width=1\linewidth]{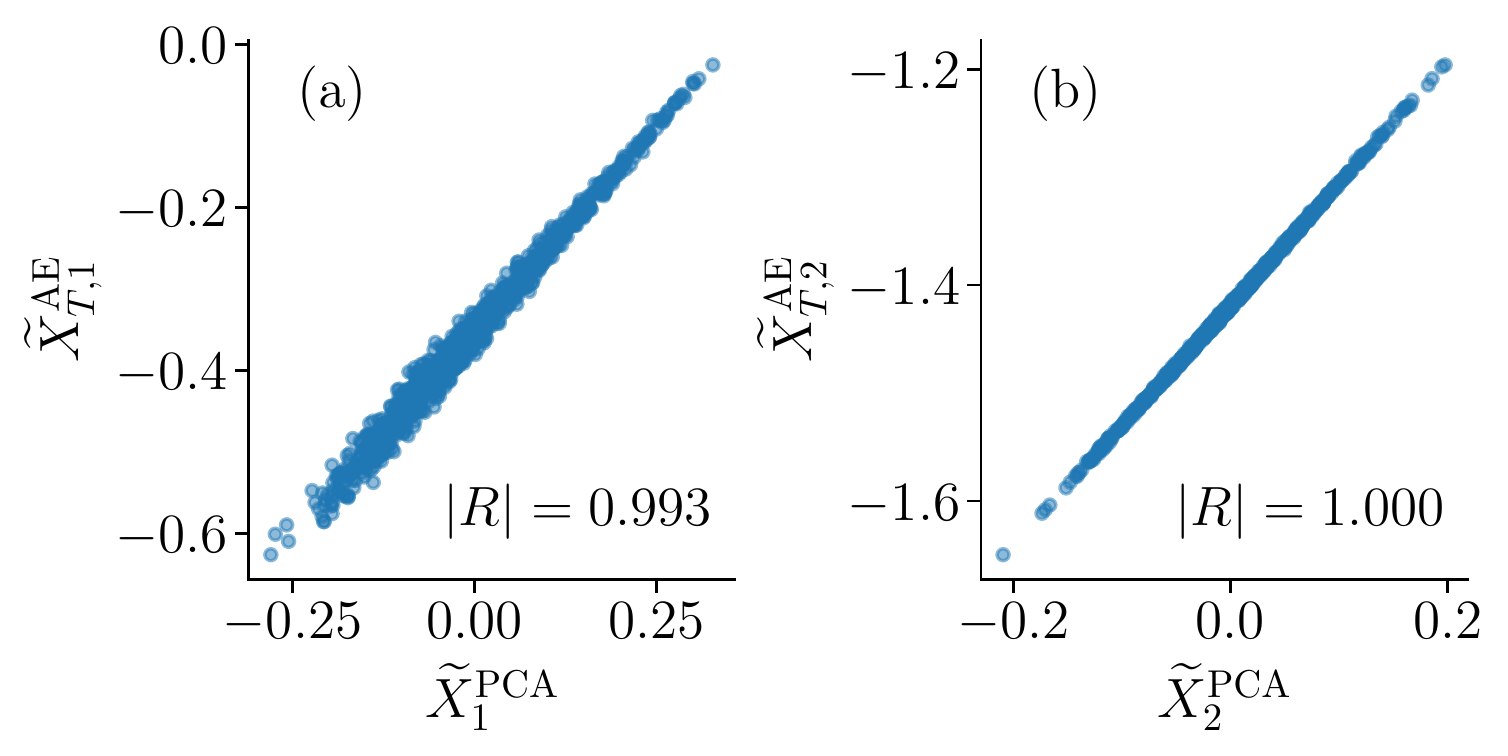}
\caption{Scatter plots of $\widetilde{X}_j^\mathrm{PCA}$ against the reduced features from the AE after rotation, $\widetilde{X}_{T,j}^\mathrm{AE}$ for (a) $j=1$ and (b) $j=2$. The distribution considered for the AE is the same as in in Fig.~\ref{fig:AE_rotated}. Apart from a shift, the two distributions are almost identical, as confirmed by the Pearson correlation coefficients $|R|$. Results are shown for the SBO descriptor in the Wahn model (small particles). For visualization purposes, only a subset of 1000 datapoints are shown.}
\label{fig:pearson_AE_PCA}
\end{figure}

For both models, we process the full datasets $\mathbf{X}$ using both PCA and the AE in order to reduce their dimension to $P=2$ and obtain two datasets $\widetilde{\mathbf X}^\mathrm{PCA}$ and $\widetilde{\mathbf X}^\mathrm{AE}$.
Since there is no constraint on the range and directions of variance of the data in the reduced space found by the AE, we transform $\widetilde{\mathbf X}^\mathrm{AE}$ by rotating the reduced descriptors.  In terms of the data matrix, this corresponds to
\beq
\widetilde{\mathbf X}_T^\mathrm{AE} = \widetilde{\mathbf X}^\mathrm{AE}  \begin{pmatrix}
\cos (\theta) & - \sin (\theta) \\
\sin (\theta) & \cos \theta 
\end{pmatrix} ,
\label{eq:rotation}
\eeq
which yields a new data matrix $\widetilde{\mathbf X}^\mathrm{AE}_T$ (the subscript $T$ indicates the additional transformation).
The angle of rotation $\theta$ is chosen so as to rectify the alignment of the reduced features yielded by the AE to match with those from PCA.
In practice, we iteratively rotated $\mathbf{\widetilde{X}}_T^\mathrm{AE}$ by steps of $\theta = 5^{\circ}$ to maximize
\beq
S_R = |R(\widetilde{X}_{T,1}^\mathrm{AE},\widetilde{X}_{1}^\mathrm{PCA})| + |R(\widetilde{X}_{T,2}^\mathrm{AE},\widetilde{X}_{2}^\mathrm{PCA})| ,
\label{eq:sum_K}
\eeq
where $R(\widetilde{X}_{T,j}^\mathrm{AE},\widetilde{X}_{j}^\mathrm{PCA})$ is the Pearson correlation coefficient between the $j$-th feature of each dataset.
This additional transformation on $\mathbf{\widetilde{X}}^\mathrm{AE}$ does not affect qualitatively the output distribution of the AE but facilitates the comparison with $\mathbf{\widetilde{X}}^\mathrm{PCA}$.

\begin{table*}[!htb]
  \renewcommand{\arraystretch}{1.5}
  \setlength{\tabcolsep}{7pt}
  \begin{tabular}{c||cc||cc}
    \multicolumn{1}{c}{(a) Small particles} & \multicolumn{2}{c}{\textsc{Wahn}} & \multicolumn{2}{c}{\textsc{KA}} \\
    \hline
    \hline
    & $\widetilde{X}_1^\mathrm{PCA}$ & $\widetilde{X}_2^\mathrm{PCA}$ & $\widetilde{X}_1^\mathrm{PCA}$ & $\widetilde{X}_2^\mathrm{PCA}$ \\
    \hline
    \multicolumn{1}{r||}{$\widetilde{X}_{T,1}^\mathrm{AE}$} & 0.998 $\pm$ 0.003 & 0.03 $\pm$ 0.02 & 0.998 $\pm$ 0.003 & 0.05 $\pm$ 0.03 \\
    \multicolumn{1}{r||}{$\widetilde{X}_{T,2}^\mathrm{AE}$} & 0.06 $\pm$ 0.04 & 0.999 $\pm$ 0.001 & 0.05 $\pm$ 0.04 & 0.998 $\pm$ 0.003 \\
    \hline
    \multicolumn{1}{l||}{EVR} & \multicolumn{2}{l||}{0.8016 $\pm$ 0.0006 (0.8009 for PCA)} & \multicolumn{2}{l}{0.680 $\pm$ 0.001 (0.681 for PCA)} \\
    \multicolumn{1}{l||}{Validation loss} & \multicolumn{2}{l||}{0.00073 $\pm$ 0.00001} & \multicolumn{2}{l}{0.00101 $\pm$ 0.00001} \\
	\hline
	\hline
%  \end{tabular}
%  \begin{tabular}{l||ll||ll}
	\multicolumn{1}{c}{} & \multicolumn{1}{c}{} & \multicolumn{1}{c}{} \\
    \multicolumn{1}{c}{(b) Big particles} & \multicolumn{2}{c}{\textsc{}} & \multicolumn{2}{c}{\textsc{}} \\
    \hline
    \hline
    & $\widetilde{X}_1^\mathrm{PCA}$ & $\widetilde{X}_2^\mathrm{PCA}$ & $\widetilde{X}_1^\mathrm{PCA}$ & $\widetilde{X}_2^\mathrm{PCA}$ \\
    \hline
    \multicolumn{1}{r||}{$\widetilde{X}_{T,1}^\mathrm{AE}$} & 0.999 $\pm$ 0.001 & 0.03 $\pm$ 0.01 & 1.000 $\pm$ 0.001 & 0.03 $\pm$ 0.02 \\
    \multicolumn{1}{r||}{$\widetilde{X}_{T,2}^\mathrm{AE}$} & 0.05 $\pm$ 0.02 & 0.999 $\pm$ 0.001 & 0.02 $\pm$ 0.01 & 0.999 $\pm$ 0.001 \\
    \hline
    \multicolumn{1}{l||}{EVR} & \multicolumn{2}{l||}{0.7501 $\pm$ 0.0006 (0.7503 for PCA)} & \multicolumn{2}{l}{0.6993 $\pm$ 0.0008 (0.6995 for PCA)} \\
    \multicolumn{1}{l||}{Validation loss} & \multicolumn{2}{l||}{0.000594 $\pm$ 0.000009} & \multicolumn{2}{l}{0.000551 $\pm$ 0.000004} \\
       \hline
       \hline
  \end{tabular}  	
  	
  \caption{Comparison between the reduced descriptor yielded by PCA, $(\widetilde{X}_1^\mathrm{PCA},\widetilde{X}_2^\mathrm{PCA})$, and the reduced descriptor yielded by the AE, $(\widetilde{X}_{T,2}^\mathrm{AE},\widetilde{X}_{T,2}^\mathrm{AE})$, rotated so as to maximize $S_R$. The first two rows show the absolute value of the Pearson correlation coefficient between the first and second features of both methods, the third row shows the total EVR, and the fourth row shows the loss on the validation set of the AE at the end of the training procedure. Errors are the standard deviation of each quantity over the 10 repetitions of the AE.} %The top table (a) is for the small particles and the bottom table (b) is for the big  particles.}
  	\label{tab:PCA-vs-AE}
\end{table*}

To illustrate this process, Fig.~\ref{fig:AE_rotated} shows an example of the probability densities of the reduced descriptor obtained by the AE for the Wahn mixture, before and after the rotation of the dataset.
We see that the distribution yielded by the AE is qualitatively similar to those shown for PCA in Fig.~\ref{fig:map_red}(c).  The only significant difference is a linear transformation (scale and shift), as shown in Fig.~\ref{fig:pearson_AE_PCA}. This already suggests that the AE does not identify any additional source of structural heterogeneity, beyond what is found in PCA.

Since the training process of the AE has a stochastic nature, we repeated it 10 times.
We identify the ``best'' solution by minimizing the loss on the validation set, among the 10 repetitions.
For each repetition, we also keep track of the EVR and the value of the loss on the validation set at the end of each training, allowing us to assess the variability and robustness of the results.
Table~\ref{tab:PCA-vs-AE}(a,b) shows the absolute value of the average Pearson correlation coefficient between different pairs of features in the reduced descriptors, 
as well as the average EVR and validation loss, for small and big particles, respectively.

For all the models and types of particles, the variability is extremely low, as shown by the small standard deviation on the correlation.
Remarkably, the correlations between ``related'' features ($\widetilde{X}_{T,j}^\mathrm{AE},\widetilde{X}_{j}^\mathrm{PCA}$) are very close to 1, while the correlations between ``opposite'' features ($\widetilde{X}_{T,j}^\mathrm{AE},\widetilde{X}_{k}^\mathrm{PCA}$, $j \neq k$) are close to 0.
Moreover, the average EVR of the AE is identical to that of PCA, which shows that the AE and PCA effectively keep the same amount of information from the original data.
We obtained very similar results for the bare BO descriptor (not shown).
There is almost no difference between the validation losses of the various training iterations, which suggests that they all approach the same minimum of the loss function.
We note that the validation loss is largest for the small particles of the KA mixture, possibly because the number of such particles is 2.5 times lower than the corresponding number for  Wahn mixture.

In conclusion, we found that the output of PCA and AE is almost identical, supporting the idea that the bond-order descriptors do not have a complex distribution that would benefit from a non-linear dimensionality reduction method.
Indeed, the explained variance ratio of the AE is not larger than that of PCA, suggesting that it does not capture any additional information compared to PCA, at least in dimension $P=2$ where the majority of the variance is already restored by both methods.
It appears that PCA is able to capture the key structural features just as accurately as the AE, both in systems with strong and weak structural heterogeneity.
In the context of this study on simple binary mixtures, it appears that the reward of using a complex machine learning method such as the AE does not justify the difficulty in setting it up.
This conclusion is broadly consistent with the findings of recent work on supervised learning of glassy dynamics~\cite{boattini_averaging_2021, Alkemade_Boattini_Filion_Smallenburg_2022}.

\section{Discussion and perspectives}
\label{sec:conc}

We close with a broad view of the context of these results and 
open problems left for future studies.

\subsection{Towards a robust operational definition of locally favored structures}

As noted in the Introduction, a central motivation for unsupervised learning in glassy binary mixtures is 
to pursue a robust definition of locally favored structures, without reference to dynamic properties nor prior knowledge of the relevant symmetries of the particles' arrangements.
Conventional classification methods~\cite{Tanemura_Hiwatari_Matsuda_Ogawa_Ogita_Ueda_1977,gellatly_characterisation_1982,Honeycutt_Andersen_1987,malinsIdentificationStructureCondensed2013,Lazar_Han_Srolovitz_2015} yield a large discrete set of distinct local structures, but many of these are easily transformed into one another by thermal distortions.
Our analysis focused instead on dimensionality reduction of smooth high-dimensional descriptors, as a step towards a coarse-grained notion of LFS.

Among the systems studied, the Wahn and Cu$_{36}$Zr$_{64}$ models have pronounced icosahedral order at low temperatures, which has a clear signature in the probability density of the reduced descriptors. This allows identification of the LFS, as a distinct ``mode'' in the distribution of the reduced descriptor. Indeed, one might use the PC$_1$ defined here as a smooth measure of icosahedral order in these systems.
Similarly, the SiO$_2$ model has directional bonding, which leads to an heterogeneous feature space in which the preferred tetrahedral arrangements are clearly distinguished from other defective structures.
For other systems, namely the KA mixture and Ni$_{33}$Y$_{67}$ model, the structural feature space appears rather homogeneous
and the putative LFS, determined from the largest occurrence of cell signatures in the Voronoi tessellation~\cite{coslovichUnderstandingFragilitySupercooled2007a}, does not stand out from the background of irregular local structures.
Given the range of dimensionality reduction methods and descriptors that we tried, we suspect that this is not a technical limitation, but rather an intrinsic property of these amorphous systems.
Note that the results may be qualitatively different with descriptors that treat separately particles of different species~\cite{Cheng_et_al_2020}, which provide a richer description of compositional fluctuations~\cite{Offei-Danso_Hassanali_Rodriguez_2022}.

Overall, it seems that some glassy mixtures do have well-defined LFS, while others continue to explore a broad spectrum of disordered local structures, even at relatively low temperatures. The latter situation is assumed in mean-field theories of the glass transition~\cite{Kirkpatrick_Thirumalai_2012,Biroli_Bouchaud_2012}, while the first is central to structure-based approaches~\cite{Tarjus_Kivelson_Nussinov_Viot_2005, royallRoleLocalStructure2015}.
It would be interesting to extend this analysis to more complex glassy systems, like polymers and organic glasses, to understand the extent to which LFSs appear there, and whether they can be detected by unsupervised methods.
Finally, a robust operational definition of LFS requires careful consideration of two additional aspects, \textit{viz.}, modality and temperature dependence of the feature space, to which we turn in the next section.

\subsection{Modality and temperature dependence of structural heterogeneity}

While the results for the glassy models with strong icosahedral order in Fig.~\ref{fig:map_red} indicate a population of LFS that is distinct from the rest of the liquid, it would be interesting to understand more precisely if the higher-dimensional structural descriptor can be decomposed into a large number of structural states (or modes).
A related problem arises in studies of polydisperse colloidal suspensions, when looking for an effective description of structure and thermodynamics in terms of a discrete number of families of particles~\cite{Ozawa_Berthier_2017, Patel_Nandi_2021}.
Identifying the number of distinct modes of a high-dimensional distribution is a delicate task in unsupervised learning~\cite{Hastie_Tibshirani_Friedman_2016} and even well-established tests used in cluster analysis (such as the Bayesian information criterion, used for instance in Ref.~\onlinecite{boattini_autonomously_2020}), may fail under some conditions~\cite{siffer_are_2018}.

In this work, we have avoided the question of \textit{how many} modes best fit the distribution of the structural descriptors, but we did attempt a global assessment of the modality of the distribution $p(\widetilde{X})$, to tell whether it is better described by one or multiple modes.
Specifically, we tested a recently proposed statistical criterion called the folding test of unimodality~\cite{siffer_are_2018}, which can also be applied to high-dimensional datasets.
The test outputs a non-negative folding ratio parameter $\Phi$, which is normalized such that $\Phi < 1$ for multimodal distributions.
The normalization considers the approximate folding statistics of the uniform random vector in dimension $d$~\cite{siffer_are_2018}, but we found that 
this normalization leads to a trivial scaling with the dimensionality of our datasets, which makes it unreliable.
It would be interesting to test a suitably modified version of this metric in a future study.

When distinguishing systems that display a well-defined LFS from those with a uniform spectrum of disordered structures, it is important to consider the role of temperature.
One might expect a broad range of structures at high temperature, with LFSs emerging on cooling.
Previous work~\cite{paret_assessing_2020} suggested different degrees of structural fragility in the KA and Wahn model, in that the structural heterogeneity of Wahn has a stronger temperature dependence. We performed a preliminary assessment of the temperature evolution of the EVR, the variance of the PCs, and the non-Gaussianity of their distributions. The results (not shown here) broadly support the idea that the models with icosahedral order studied here have a higher structural fragility than the others.
Finally, we note that while the current unsupervised learning approach can be used to identify structural modes at any given temperature, it is unclear how the corresponding directions in the high-dimensional feature space evolve with temperature.
This should be taken into account by a robust and transparent definition of LFS.

\subsection{Beyond the first coordination shell: how heterogeneous is medium range order?}

In most of the previous sections, we probed structural heterogeneity over a length scale corresponding to the first coordination shell, \textit{i.e.}, we focused on local order.
However, in the analysis carried in Sec.~\ref{sec:propensity_fit_soap_rbo}, we extended the range of the SOAP descriptor to include the second coordination shell.
The results shown in Fig.~\ref{fig:K_4} indicate that this brings only a marginal contribution to the structure-dynamics relationship, hence the dominant source of structural heterogeneity comes from the first coordination shell.

A recent study by Zhang and Kob showed that bond order persists even at large distances in liquids, both under normal and supercooled conditions~\cite{Zhang_Kob_2020}, suggesting that there is interesting structural heterogeneity beyond the first shell.
For instance, the KA mixture is characterized by alternating icosahedral and dodecahedral order in successive coordination shells around big particles. 
To characterize the heterogeneity of such order beyond the first coordination shell,
we analyzed the local structure of the Wahn mixture using the RBO descriptor, restricting ourselves this time only to the second coordination shell.
We used the following two grids of distances for the descriptor: (i) $\{ d_n \} = (0.9, 1.0, 1.1, 1.2, 1.3)$ and (ii) $\{ d_n \} = (1.5, 1.6, 1.7, 1.8, 1.9, 2.0, 2.1, 2.2)$, using Gaussians of width $\delta = 0.2$.
We found that the probability densities $p(\X_1, \X_2)$ changed from bimodal to unimodal, when switching from (i) to (ii), indicating either that our current methodology is not sufficient to capture the heterogeneities in this longer-ranged order, or perhaps that there is little heterogeneity on such scales, even if there is significant ordering.
Similarly, the distributions of the reduced descriptors that characterize the second coordination shell appeared unimodal in the KA mixture.
We defer to a future study a more in-depth analysis of higher order coordination shells, in both closed-packed and network-forming systems, as well as of the temperature dependence of the heterogeneity of medium range order.

\subsection{A critical assessment of unsupervised learning methods for glassy systems}

The words ``unsupervised machine learning'' might suggest a procedure where the user passes some input configurations to an algorithm, which automatically returns a characterization of the local structure, without requiring any assumptions as to the dominant structural motifs. However, this picture is too simplistic, because it neglects at least two places in which the user's assumptions are baked into the method: the choice of the descriptor and of its hyper-parameters. We have emphasized throughout this work that the results depend on these choices: it is essential to keep in mind that they can strongly influence the results.

To illustrate this, we briefly mention two pitfalls that we encountered in this study.  First, it may seem natural to use descriptors that include as much information as possible, to avoid biasing the algorithm towards any particular form of order.
However, in the current context, this hinders interpretability of the results, making it necessary to relate \textit{a posteriori} the reduced variables to other physically-motivated quantities.
Moreover, a high-dimensional descriptor such as SOAP contains much more information than the simple SBO descriptor, but the dimensionality reduction yields qualitatively similar structural features in both cases, compare Fig.~\ref{fig:map_red} and Fig.~\ref{fig:map_soap_1}.  
The reduced descriptors extracted from SOAP also depend sensitively on its hyper-parameters, see Fig.~\ref{fig:sim_soap_1}, whose choice depend on the identification of the relevant length scales in the system.

The second pitfall is that a bad choice of descriptor can lead to detection of erroneous structure, even in a disordered system.  This can be appreciated from Fig.~\ref{fig:map_bare}, which appears to show complex distributions of the first two PCs in all the models considered, hinting at the existence of different local structures.  In fact, these  structures appear because the descriptor  detects neighboring particles according to a fixed cutoff, which splits a continuous family of local structures into distinct sub-populations.  A more reliable representation of the data is obtained by using a smoothed cutoff (Fig.~\ref{fig:map_red}), which reveals the underlying continuous family.

Based on these observations, our conclusion is that unsupervised learning provides a powerful tool in this context, but its application and interpretation still requires care, as with all methods. It should not be surprising that an algorithm requires some guidance to identify the aspects of local structure that are physically relevant in glasses. Providing appropriate guidance requires in turn some physical insight from the user, as well as an understanding of how the data will be used within the algorithm. 

\section*{Acknowledgments}
We thank A. Rodriguez for useful discussions. Part of the simulations were carried out on the CINECA HPC cluster within the CINECA-University of Trieste agreement.

\section*{Author declarations}

\subsection*{Conflict of Interest}

The authors have no conflicts to disclose.

\section*{Data availability}
The data and workflow necessary to reproduce the findings of this study are openly available in the Zenodo data repository at \url{https://doi.org/10.5281/zenodo.7108316}.

\section*{Appendix}
\appendix

\section{Models}\label{app:models}

\subsection{Wahn mixture}

The Wahn model, introduced by Wahnström in Ref.~\onlinecite{wahnstrom_molecular-dynamics_1991}, is a two-component Lennard-Jones mixture composed of type-$A$ (big) and type-$B$ (small) particles with chemical fractions $x_A = x_B = 0.5$.
Particles interact via a Lennard-Jones potential,
\beq
u_{\alpha\beta}(r) = 4 \epsilon_{\alpha\beta} \left[ \left( \frac{\sigma_{\alpha\beta}}{r} \right)^{12} - \left( \frac{\sigma_{\alpha\beta}}{r} \right)^{6} \right] ,
\label{eq:LJ_pot}
\eeq
where $\alpha$ and $\beta$ are species indices.
The values for the interaction parameters are given in the original paper~\cite{wahnstrom_molecular-dynamics_1991}.
Quantities are expressed in the following system of units: the unit of length is $\sigma_{AA}$, the unit of energy is $\epsilon_{AA}$ and the unit of time is $\sqrt{m_A \sigma_{AA}^2 / \epsilon_{AA}}$, this is also valid for the other models presented in the next paragraphs except for the one in Appendix.~\ref{sec:CuZr}.
The number density is $\rho = N/V = 1.297$, where $V$ is the volume of the cubic simulation cell.
% The potential is cut and shifted at a cutoff distance of $2.5 \sigma_{\alpha\beta}$.
We consider molecular dynamics simulation data, produced in the context of Ref.~\onlinecite{paret_assessing_2020}, for a system of $N=20000$ particles.
In the following, we analyze $n_\textrm{conf}=10$ statistically uncorrelated configurations at temperature $T=0.58$.
The estimated mode-coupling crossover temperature is $T_\mathrm{MCT} \approx 0.56$.

\subsection{KA mixture}

The Kob-Andersen (KA) mixture~\cite{kob_scaling_1994} is loosely designed to reproduce the structure of the Ni$_{80}$P$_{20}$ metallic glass former using Lennard-Jones interactions.
It is composed of type-$A$ (big) and type-$B$ (small) particles with chemical fractions $x_A = 0.8$ and $x_B=0.2$, and the number density is set to $\rho = 1.2$. %0397$.
% The potential is cut and shifted in the same way as the Wahn model.
As for the Wahn mixture, we consider molecular dynamics simulation data, produced in the context of Ref.~\onlinecite{paret_assessing_2020}, for a system of $N=20000$ particles.
We consider $n_\textrm{conf}=10$ statistically uncorrelated configurations at temperature $T=0.45$ ($T_\mathrm{MCT} \approx 0.435$).

\subsection{SiO$_2$ model}\label{sec:silica}

We also study a simple binary model, based on short-range interactions, that mimics the structure and dynamics of amorphous silica~\cite{coslovich_dynamics_2009}.
The chemical fractions are $x_\textrm{Si} = 0.33$ and $x_\textrm{O} = 0.67$.
The interaction potential between Si and O particles is of the Lennard-Jones type, as in Eq.~\eqref{eq:LJ_pot}, while the one between identical species is a simple inverse power law,
\begin{equation}
u_{\alpha\alpha} = 4 \epsilon_{\alpha\alpha} \left( \frac{\sigma_{\alpha\alpha}}{r} \right) ^{12} .
\end{equation}
The interaction parameters are given in the original paper.
% The potential is cut off smoothly at a distance $2.2 \sigma_{\alpha\beta}$ by adding a cubic term that ensures continuity up to the second derivative~\cite{grigera_geometric_2002}.
The simulation data analyzed in this work were obtained in Ref.~\onlinecite{Berthier_Biroli_Coslovich_Kob_Toninelli_2012}.
The number of particles is $N=2000$.
We consider $n_\textrm{conf}=20$ statistically uncorrelated configurations at temperature $T=0.3397$ ($T_\mathrm{MCT} \approx 0.31$).

\subsection{Ni$_{33}$Y$_{67}$ model}\label{sec:NiY}

We consider a parametrization of the LJ potential that provides a realistic description of the structure of amorphous alloys of Ni and Y atoms~\cite{dellavalle_microstructural_1994}.
This model is characterized by a single energy scale and diameters $\sigma_{\alpha\beta}$, determined by fitting structural data on experimental Ni-Y alloys at several compositions.
As in Ref.~\onlinecite{dellavalle_microstructural_1994}, we use chemical fractions $x_\textrm{Ni}=0.33$ and $x_\textrm{Y}=0.67$.
In the following, we will refer to Ni and Y atoms in this mixture as small and big particles, respectively.
We carried out molecular dynamics simulations for a system composed of 4000 particles, which we cooled at constant density $\rho=1.5$.
This corresponds approximately to the highest densities reached along an isobaric path at $P=10$ in Ref.~\onlinecite{coslovichUnderstandingFragilitySupercooled2007a}.
The system is then equilibrated and simulated at $T=0.55$ ($T_\textrm{MCT} \approx 0.52$), and we collect $n_\textrm{conf}=40$ statistically uncorrelated configurations.

\subsection{Cu$_{64}$Zr$_{36}$ model}\label{sec:CuZr}

We simulate with LAMMPS~\cite{plimpton_fast_1995} an embedded atom model for CuZr alloys, using the interatomic potentials developed in Ref.~\onlinecite{Cheng_Sheng_Ma_2008}.
We use a model system of size $N=16000$ with chemical fractions $x_\textrm{Cu}=0.64$ and $x_\textrm{Zr}=0.36$, which displays a pronounced icosahedral local order at low temperatures~\cite{Soklaski_Nussinov_Markow_Kelton_Yang_2013, soklaski_locally_2016}.
Note that such local order develops prominently around the (small) Cu particles.
The model is also prone to crystallization into a CuMg$_2$ Laves phase~\cite{Ryltsev_Klumov_Chtchelkatchev_Shunyaev_2016} and we confirmed this kind of instability at zero pressure for temperatures in the range between $800 K$ and $900 K$.
We could however equilibrate and simulate the system for 0.12 $ns$ at zero pressure and temperature $T=800K$, while observing only a modest minor drift of the potential energy.
At this temperature, the structural relaxation time $\tau_\alpha$ is about 3 $ps$ and we consider $n_\textrm{conf}=11$ statistically uncorrelated configurations.
Given that it is difficult to avoid crystallization for temperatures above $800 K$, we do not have a precise estimate of the MCT crossover temperature for this model. However, inspection of the data reported in Refs.~\onlinecite{Soklaski_Nussinov_Markow_Kelton_Yang_2013, soklaski_locally_2016} suggests that it may lie in the range 750-800$K$.

\section{Additional results on the SBO descriptor}
\label{app:sbo}

\begin{figure*}[!t]
  \begin{tabular}{ccccc}
 %             \includegraphics[width=0.2\linewidth]{KA/T0.4500/config.xyz.sbo.scaling-False.redux-True.heatmap_marginal.species-1.pdf} &
 % \hskip-1em  \includegraphics[width=0.2\linewidth]{Ni33Y67/T0.55/config.xyz.sbo.scaling-False.redux-True.heatmap_marginal.species-1.pdf} &
 % \hskip-1em  \includegraphics[width=0.2\linewidth]{Wahn/T0.5800/config.xyz.sbo.scaling-False.redux-True.heatmap_marginal.species-1.pdf} &
 % \hskip-1em  \includegraphics[width=0.2\linewidth]{Cu64Zr36/T800/config.xyz.sbo.scaling-False.redux-True.heatmap_marginal.species-1.pdf} &
 % %  \hskip-1em  \includegraphics[width=0.2\linewidth]{SiO2/T0.3397/config.xyz.sbo.scaling-False.redux-True.heatmap_marginal.species-1.pdf} \\
\includegraphics[height=.27\linewidth]{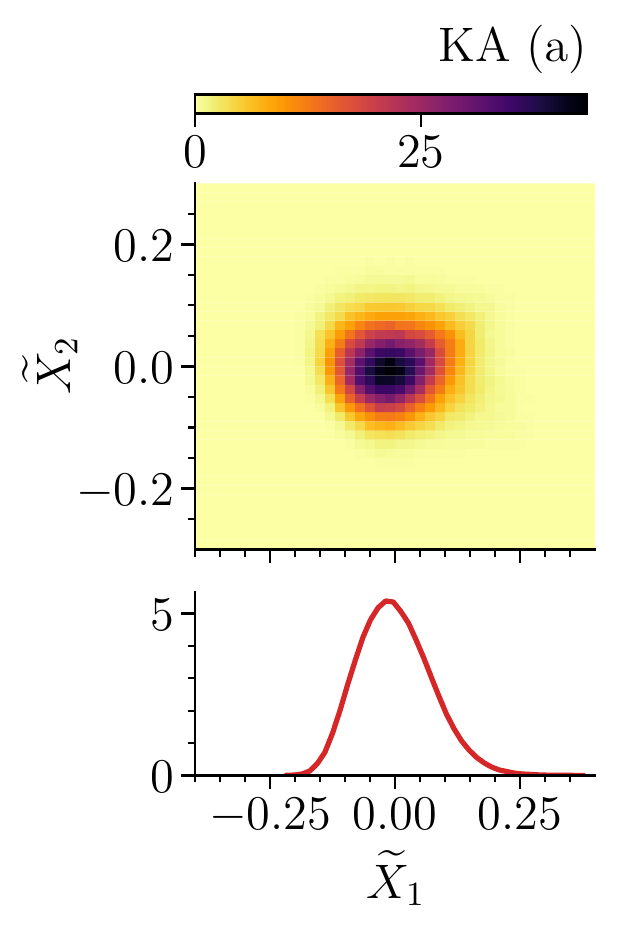}
\includegraphics[height=.27\linewidth]{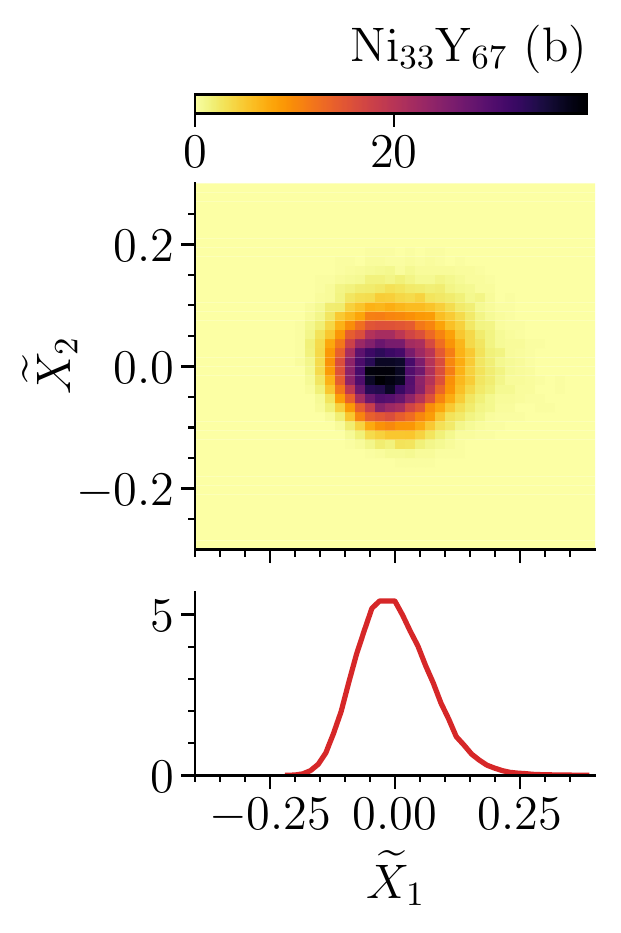}
\includegraphics[height=.27\linewidth]{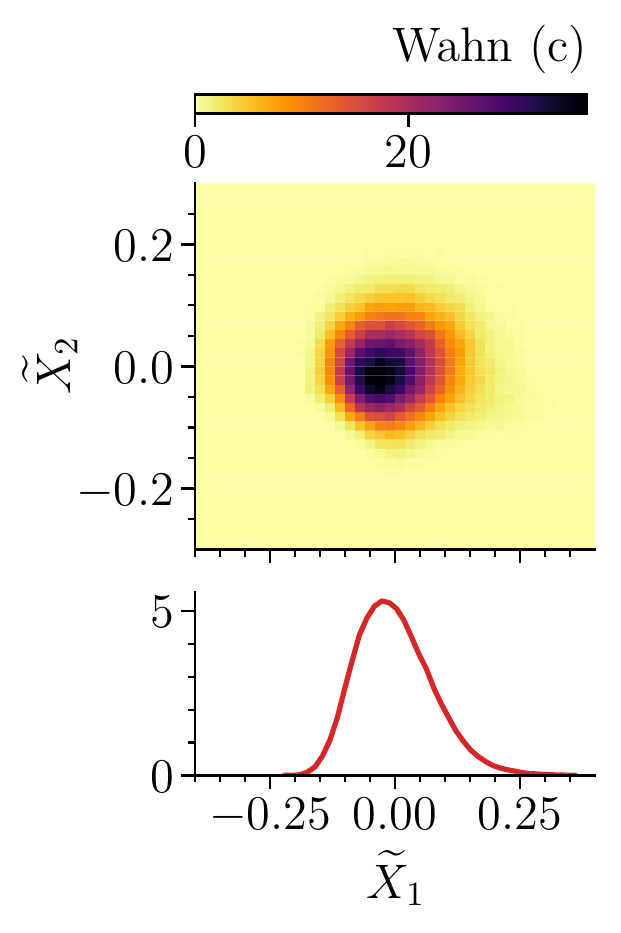}
\includegraphics[height=.27\linewidth]{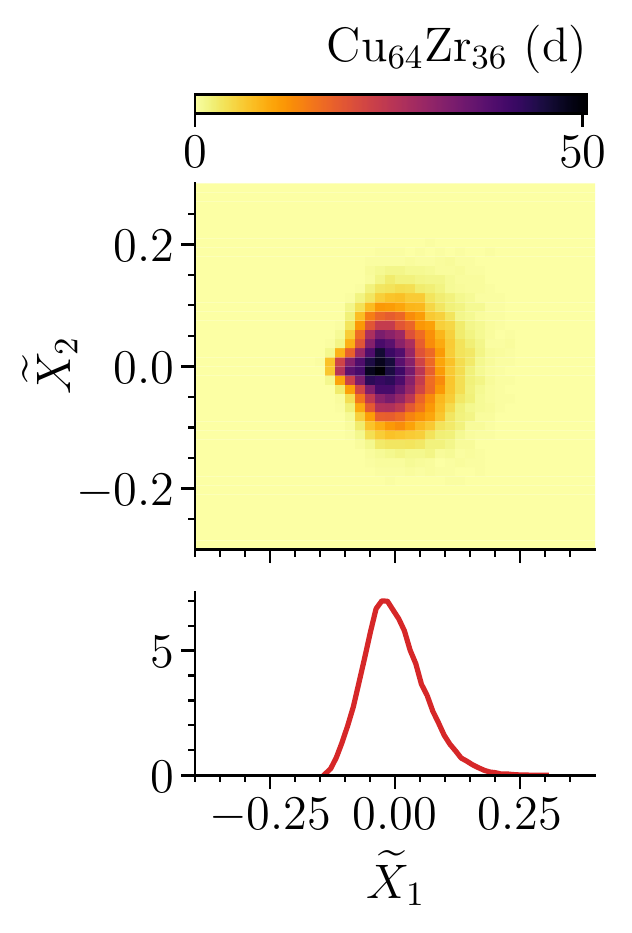}
  \end{tabular}
\caption{\label{fig:map_red_big} Same as Fig.~\ref{fig:map_red} but for the large particles of the close-packed mixtures: (a) KA, (b) \NiY, (c) Wahn, (d) \CuZr.}
\end{figure*}

\begin{figure*}[!t]
  \begin{tabular}{cccc}
 \hskip-.5em\includegraphics[height=.258\linewidth]{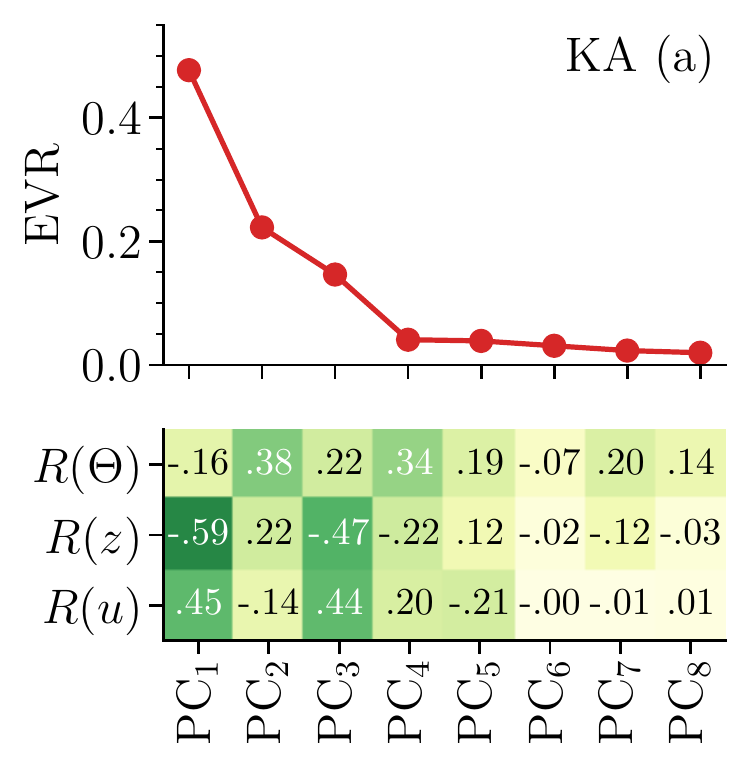} &
 \hskip-1.em\includegraphics[height=.258\linewidth]{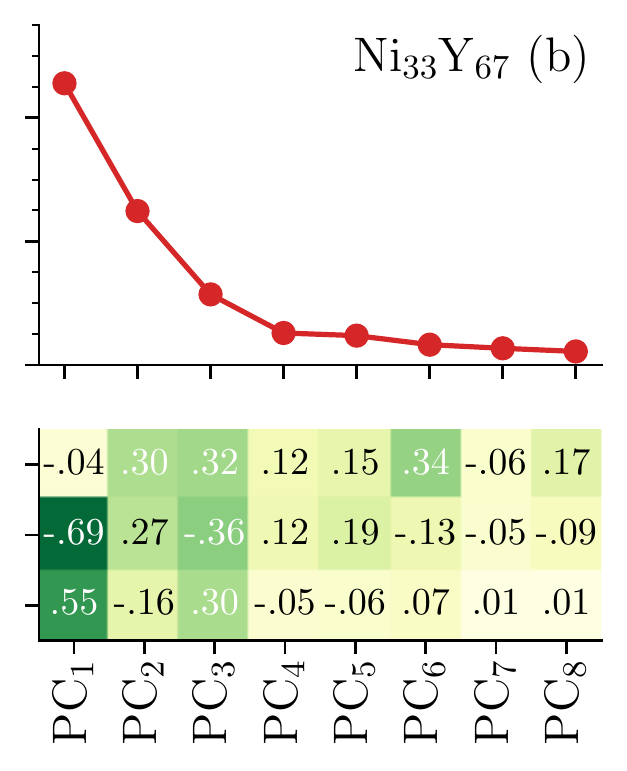} &
 \hskip-1.em\includegraphics[height=.258\linewidth]{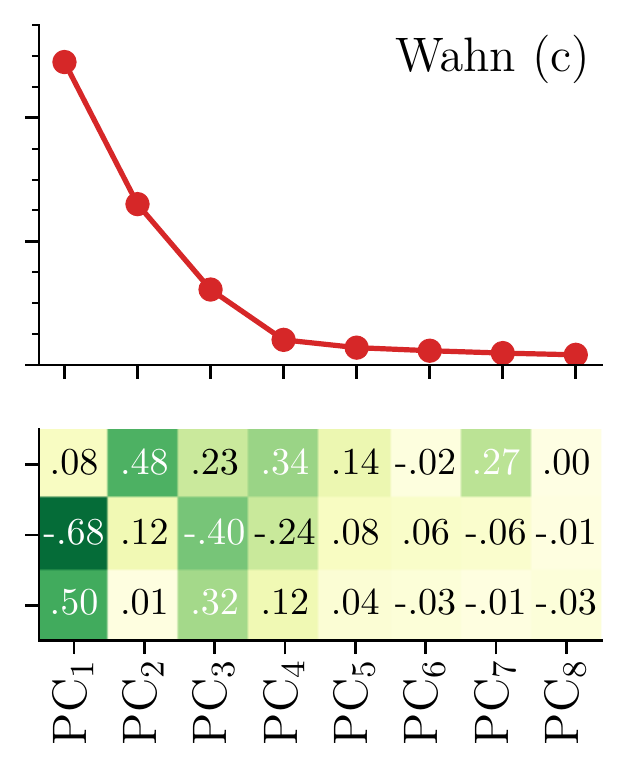} &
 \hskip-1.em\includegraphics[height=.258\linewidth]{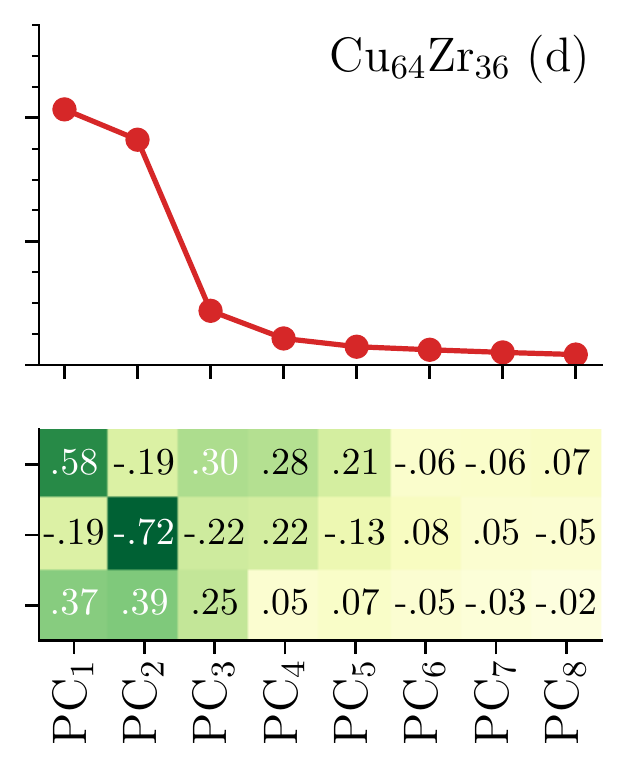} \\
  \end{tabular}
\caption{\label{fig:overview_structure_big} Same as Fig.~\ref{fig:overview_redux} but for the large particles of the close-packed mixtures: (a) KA, (b) \NiY, (c) Wahn, (d) \CuZr. Note that no well-defined LFS can be identified from the Voronoi tessellation around these particles, therefore the correlation with $\ell$ is not considered.}
\end{figure*}

We provide here supplementary information and results on the PCA of the SBO descriptor in the studied models. Namely, Tables~\ref{table:eigenvectors} and \ref{table:eigenvectors_big} show the eigenvectors $V^{(1)}$ and $V^{(2)}$, \textit{i.e.}, the first two PC directions. We also show additional results for the big particles of the close-packed mixtures: Fig.~\ref{fig:map_red_big} presents the distributions $p(\X_1, \X_2)$ obtained from PCA of the SBO descriptor, while Fig.~\ref{fig:overview_structure_big} shows the EVR and the correlation of each PC with the physically motivated structural measures introduced in Sec.~\ref{sec:pc_vs_ql}. Note the little bump in the distribution $p(\X_1, \X_2)$ for \CuZr in Fig.~\ref{fig:map_red_big}(d), which could be due to the presence of a small crystallite in the sample.

\begin{table*}[!tb]
  \begin{tabular}{l|r|rrrrrrrr}
    \hline
    \hline
    \multicolumn{10}{c}{KA}  \\
    \hline
    % \ExpandableInput{./sprouts/plots/data/KA/T0.4500/config.xyz.sbo.scaling-False.species-2.pca-eigenvectors.tex}
        & EVR &$Q_1^\mathrm{S}$    & $Q_2^\mathrm{S}$    & $Q_3^\mathrm{S}$    & $Q_4^\mathrm{S}$    & $Q_5^\mathrm{S}$    & $Q_6^\mathrm{S}$    & $Q_7^\mathrm{S}$    & $Q_8^\mathrm{S}$    \\ 
\hline 
PC$_1$ & 37\% & +0.06& +0.07& +0.19& +0.66& -0.51& +0.05& +0.49& +0.12\\ 
PC$_2$ & 31\% & +0.07& +0.01& +0.02& -0.39& -0.50& +0.65& +0.01& -0.40\\ 

    \hline
    \multicolumn{10}{c}{\NiY} \\
    \hline
    % \ExpandableInput{./sprouts/plots/data/Ni33Y67/T0.55/config.xyz.sbo.scaling-False.species-2.pca-eigenvectors.tex}
        & EVR &$Q_1^\mathrm{S}$    & $Q_2^\mathrm{S}$    & $Q_3^\mathrm{S}$    & $Q_4^\mathrm{S}$    & $Q_5^\mathrm{S}$    & $Q_6^\mathrm{S}$    & $Q_7^\mathrm{S}$    & $Q_8^\mathrm{S}$    \\ 
\hline 
PC$_1$ & 38\% & +0.09& +0.06& +0.06& +0.25& -0.82& +0.31& +0.40& -0.03\\ 
PC$_2$ & 29\% & -0.02& -0.00& +0.10& +0.78& +0.12& -0.41& +0.10& +0.43\\ 
    \hline
    \multicolumn{10}{c}{Wahn} \\
    \hline
    % \ExpandableInput{./sprouts/plots/data/Wahn/T0.5800/config.xyz.sbo.scaling-False.species-2.pca-eigenvectors.tex}
    & EVR &$Q_1^\mathrm{S}$    & $Q_2^\mathrm{S}$    & $Q_3^\mathrm{S}$    & $Q_4^\mathrm{S}$    & $Q_5^\mathrm{S}$    & $Q_6^\mathrm{S}$    & $Q_7^\mathrm{S}$    & $Q_8^\mathrm{S}$    \\ 
\hline 
PC$_1$ & 61\% & -0.03& -0.04& -0.06& -0.13& -0.41& +0.68& -0.50& -0.31\\ 
PC$_2$ & 20\% & -0.05& +0.01& +0.03& +0.12& +0.72& +0.09& -0.64& +0.22\\ 
    \hline
    \multicolumn{10}{c}{\CuZr} \\
    \hline
    % \ExpandableInput{./sprouts/plots/data/Cu64Zr36/T800/config.xyz.sbo.scaling-False.species-2.pca-eigenvectors.tex}
        & EVR &$Q_1^\mathrm{S}$    & $Q_2^\mathrm{S}$    & $Q_3^\mathrm{S}$    & $Q_4^\mathrm{S}$    & $Q_5^\mathrm{S}$    & $Q_6^\mathrm{S}$    & $Q_7^\mathrm{S}$    & $Q_8^\mathrm{S}$    \\ 
\hline 
PC$_1$ & 65\% & -0.03& -0.04& -0.05& -0.16& -0.53& +0.65& -0.42& -0.31\\ 
PC$_2$ & 18\% & +0.03& -0.01& -0.01& -0.07& -0.69& -0.16& +0.70& -0.09\\ 
    \hline
    \multicolumn{10}{c}{SiO$_2$} \\
    \hline
    % \ExpandableInput{./sprouts/plots/data/SiO2/T0.3397/config.xyz.sbo.scaling-False.species-1.pca-eigenvectors.tex}
    & EVR &$Q_1^\mathrm{S}$    & $Q_2^\mathrm{S}$    & $Q_3^\mathrm{S}$    & $Q_4^\mathrm{S}$    & $Q_5^\mathrm{S}$    & $Q_6^\mathrm{S}$    & $Q_7^\mathrm{S}$    & $Q_8^\mathrm{S}$    \\ 
\hline 
PC$_1$ & 65\% & +0.25& +0.61& -0.05& +0.04& +0.60& -0.08& -0.02& +0.44\\ 
PC$_2$ & 16\% & +0.11& -0.41& -0.34& +0.29& +0.04& -0.71& +0.17& +0.28\\ 
    \hline
    \hline
  \end{tabular}
  \caption{\label{table:eigenvectors}Eigenvectors $V^{(1)}$ and $V^{(2)}$ of the covariance matrix, corresponding to the two largest eigenvalues of the SBO descriptor for the small particles of the close-packed mixtures and for the Si particles in the SiO$_2$ models.}
\end{table*}

\begin{table*}[!tb]
  \begin{tabular}{l|r|rrrrrrrr}
    \hline
    \hline
    \multicolumn{10}{c}{KA} \\
    \hline
    % \ExpandableInput{./sprouts/plots/data/KA/T0.4500/config.xyz.sbo.scaling-False.species-1.pca-eigenvectors.tex}
        & EVR &$Q_1^\mathrm{S}$    & $Q_2^\mathrm{S}$    & $Q_3^\mathrm{S}$    & $Q_4^\mathrm{S}$    & $Q_5^\mathrm{S}$    & $Q_6^\mathrm{S}$    & $Q_7^\mathrm{S}$    & $Q_8^\mathrm{S}$    \\ 
\hline 
PC$_1$ & 48\% & -0.01& -0.01& +0.00& +0.04& +0.12& +0.75& -0.64& +0.07\\ 
PC$_2$ & 22\% & +0.06& +0.03& +0.03& +0.06& +0.06& -0.43& -0.40& +0.80\\ 
    \hline
    \multicolumn{10}{c}{\NiY} \\
    \hline
    % \ExpandableInput{./sprouts/plots/data/Ni33Y67/T0.55/config.xyz.sbo.scaling-False.species-1.pca-eigenvectors.tex}
    & EVR &$Q_1^\mathrm{S}$    & $Q_2^\mathrm{S}$    & $Q_3^\mathrm{S}$    & $Q_4^\mathrm{S}$    & $Q_5^\mathrm{S}$    & $Q_6^\mathrm{S}$    & $Q_7^\mathrm{S}$    & $Q_8^\mathrm{S}$    \\ 
\hline 
PC$_1$ & 46\% & -0.02& -0.02& -0.01& +0.04& +0.19& +0.87& -0.42& -0.12\\ 
PC$_2$ & 25\% & +0.07& +0.02& +0.03& +0.02& -0.01& -0.21& -0.65& +0.72\\ 
    \hline
    \multicolumn{10}{c}{Wahn} \\
    \hline
    % \ExpandableInput{./sprouts/plots/data/Wahn/T0.5800/config.xyz.sbo.scaling-False.species-1.pca-eigenvectors.tex}
        & EVR &$Q_1^\mathrm{S}$    & $Q_2^\mathrm{S}$    & $Q_3^\mathrm{S}$    & $Q_4^\mathrm{S}$    & $Q_5^\mathrm{S}$    & $Q_6^\mathrm{S}$    & $Q_7^\mathrm{S}$    & $Q_8^\mathrm{S}$    \\ 
\hline 
PC$_1$ & 49\% & -0.00& -0.00& +0.02& +0.07& +0.20& +0.74& -0.63& -0.06\\ 
PC$_2$ & 26\% & +0.04& +0.01& +0.04& +0.09& +0.22& -0.39& -0.44& +0.77\\ 
    \hline
    \multicolumn{10}{c}{\CuZr} \\
    \hline
    % \ExpandableInput{./sprouts/plots/data/Cu64Zr36/T800/config.xyz.sbo.scaling-False.species-1.pca-eigenvectors.tex}
    & EVR &$Q_1^\mathrm{S}$    & $Q_2^\mathrm{S}$    & $Q_3^\mathrm{S}$    & $Q_4^\mathrm{S}$    & $Q_5^\mathrm{S}$    & $Q_6^\mathrm{S}$    & $Q_7^\mathrm{S}$    & $Q_8^\mathrm{S}$    \\ 
\hline 
PC$_1$ & 41\% & +0.07& +0.04& +0.01& +0.08& +0.24& +0.41& -0.82& +0.31\\ 
PC$_2$ & 36\% & +0.00& +0.02& -0.01& +0.01& +0.07& +0.73& +0.13& -0.67\\     
    \hline
  \end{tabular}
  \caption{\label{table:eigenvectors_big}Eigenvectors $V^{(1)}$ and $V^{(2)}$ of the covariance matrix, corresponding to the two largest eigenvalues of the SBO descriptor for the big particles of the close-packed mixtures.}
\end{table*}

% \bibliography{main}
% \bibliographystyle{apsrev4-1}


\end{document}